\documentclass[aps,prx,reprint,preprintnumbers,superscriptaddress,nofootinbib,longbibliography,floatfix]{revtex4-2}
\pdfoutput=1
\usepackage{rotating}
\usepackage{array}
\usepackage{amsmath}
\usepackage[normalem]{ulem}
\usepackage{slashed}
\usepackage{booktabs}
\usepackage[pdftex,table]{xcolor}
\usepackage{units}
\usepackage{xfrac}
\usepackage{mathtools}
\usepackage{empheq}
\usepackage[]{units}
\usepackage{multirow}
\usepackage{amssymb}
\usepackage{url}
\usepackage{comment}
\usepackage{physics}
\usepackage{color,soul}
\usepackage{bbm}
\usepackage[caption=false]{subfig}

\usepackage{hyperref}
\hypersetup{
  colorlinks=true,
  citecolor=blue,
  linkcolor=blue,
  urlcolor=blue
}

\newcommand{\geant}{\textsc{Geant}}
\newcommand{\calosc}{\textsc{CaloScore}}

\begin{document}

\title{Score-based Generative Models for Calorimeter Shower Simulation}

\author{Vinicius Mikuni}
\email{vmikuni@lbl.gov}
\affiliation{National Energy Research Scientific Computing Center, Berkeley Lab, Berkeley, CA 94720, USA}

\author{Benjamin Nachman}
\email{bpnachman@lbl.gov}
\affiliation{Physics Division, Lawrence Berkeley National Laboratory, Berkeley, CA 94720, USA}
\affiliation{Berkeley Institute for Data Science, University of California, Berkeley, CA 94720, USA}

\begin{abstract}
    Score-based generative models are a new class of generative algorithms that have been shown to produce realistic images even in high dimensional spaces, currently surpassing other state-of-the-art models for different benchmark categories and applications. In this work we introduce \calosc, a score-based generative model for collider physics applied to calorimeter shower generation. Three different diffusion models are investigated using the Fast Calorimeter Simulation Challenge 2022 dataset. \calosc~is the first application of a score-based generative model in collider physics and is able to produce high-fidelity calorimeter images for all datasets, providing an alternative paradigm for calorimeter shower simulation.
\end{abstract}

\maketitle


\section{Introduction}
\label{sec:intro}

Detailed detector simulations are an essential component of data analysis in particle and nuclear physics.  These simulations are used to fold particle-level predictions for comparisons with data and are used to unfold data for detector-effects in order to compare with predictions and other experiments.  Simulations are also a critical input for the design of future experiments.  

The most widely used detector simulations are built on the program \geant~\cite{geant4,geant4-add1,geant4-add2}.  Achieving precision requires significant computing time as propagating particles through materials results in a large number of secondary particles each undergoing electromagnetic and/or nuclear interactions.  For this reason, the most complex detectors to emulate are calorimeters, whose purpose is to stop particles and measure the deposited energy.  An $\mathcal{O}(1)$ fraction of all computing in HEP goes towards simulating particle propagation inside dense materials with \geant.

The experiments at the Large Hadron Collider (LHC) generate billions of events per run, each of which has hundreds to thousands of individual calorimeter showers.  Within the experiments' computing budgets, it is not possible to run \geant-based (`full') simulation for all events.  Therefore, all of the experiments have developed fast simulation methods that replace physics models with simpler parametric models that are tuned to the full simulation.  The fast simulation models are constructed with relatively few parameters in order to facilitate efficient optimization and validation. This fundamentally limits their precision, in particular in the ability to model complex correlations in high-dimensions.  These correlations may also not be explicitly part of an optimization that uses only a relatively small number of one-dimensional observables.

Deep learning offers a complementary approach to engineered parameteric models.  Flexible neural networks are used to transform random numbers (called a \textit{latent space} in machine learning) into structured data.  So far, there have been three main strategies for deep generative modeling. Generative adversarial networks (GANs)~\cite{Goodfellow:2014upx} optimize the generator network by means of an auxiliary network (`discriminator') that tries to classify generated examples from real examples. Variational autoencoders (VAEs)~\cite{kingma2014autoencoding} learn a stochastic map from the data space to a latent space and back, preserving the statistics of the latent space and data space.  Normalizing flows (NFs)~\cite{pmlr-v37-rezende15} use invertible transformations so that the probability density can be computed and the generator is optimized using the log likelihood.  A number of deep generative models have been proposed for emulating calorimeter showers and other particle detectors~\cite{deOliveira:2017pjk,Paganini:2017hrr,Paganini:2017dwg,Vallecorsa:2019ked,SHiP:2019gcl,Chekalina:2018hxi,ATL-SOFT-PUB-2018-001,Carminati:2018khv,Vallecorsa:2018zco,Musella:2018rdi,Erdmann:2018kuh,Deja:2019vcv,Derkach:2019qfk,Erdmann:2018jxd,Oliveira:DLPS2017,deOliveira:2017rwa,Hooberman:DLPS2017,Belayneh:2019vyx,buhmann2020getting,2009.03796,Kansal:2020svm,Maevskiy:2020ank,Rehm:2021zow,Rehm:2021zoz,Rehm:2021qwm,Kansal:2021cqp,Khattak:2021ndw,Anderlini:2021qpm,Buhmann:2021caf,Bieringer:2022cbs,Monk:2018zsb,1816035,Buhmann:2021lxj,Hariri:2021clz,Fanelli:2019qaq,Orzari:2021suh,Touranakou:2022qrp,Krause:2021ilc,Krause:2021wez}.

The first proposals for generating calorimeter showers with deep generative models used GANs (starting with \textsc{CaloGAN}~\cite{deOliveira:2017pjk,Paganini:2017hrr,Paganini:2017dwg}).  The evaluation of GANs is fast and there are no constraints on the form of the generator function.  However, GAN optimization is challenging as it is a minimax problem due to the competition between the generator and discriminator networks. Furthermore, GANs are known to suffer from \textit{mode collapse} where the generator learns to produce only a subset of the possible showers.  Despite these challenges, the ATLAS Collaboration become the first experiment to replace part of their fast simulation with a GAN~\cite{ATLAS:2021pzo} and are already using it to generate billions of events for Run~3.  Other experiments are also exploring the use of GANs for detector simulation~\cite{Erdmann:2018jxd,Chekalina:2018hxi,Deja:2019vcv}. 

Since the first GAN studies, there have been a number of innovations to improve the precision of deep generative calorimeter simulation.  This includes variations/modifications to the GAN setup including Wasserstein GANs~\cite{https://doi.org/10.48550/arxiv.1701.07875,Erdmann:2018jxd} to help with training stability and mode collapse and refining networks~\cite{buhmann2020getting,2009.03796} to correct the spectra of an initial generative model.  Beyond GANs, recent work with NFs has shown great promise~\cite{Krause:2021ilc,Krause:2021wez}.  Normalizing flows tend to be robust to mode collapse and are minimized with a convex loss function.  Furthermore, the normalization of the resulting generative model seems to have beneficial regularization properties for the training.  The authors of the \textsc{CaloFlow} model~\cite{Krause:2021ilc,Krause:2021wez} showed that a post-hoc classifier struggled to distinguish showers generated from their neural network and showers from \textsc{Geant}~4, a critical milestone in this field.

While NFs show great promise for fast calorimeter simulation, they have difficulty scaling to higher-dimensional datasets.  Such datasets are important for the upgraded CMS forward calorimeter~\cite{CERN-LHCC-2017-023} as well as ultra-fine calorimeters proposed for future detectors~\cite{CALICE:2018ibt}. Two of the three datasets in the recent \textit{Fast Calorimeter Simulation Challenge 2022}~\cite{faucci_giannelli_michele_2022_6366324,michele_faucci_giannelli_2022_6368338,faucci_giannelli_michele_2022_6366271,ATLAS:2021pzo} have dimensionality at least an order of magnitude beyond what has been studied with NFs.  The NF training time for these data is prohibitive and would require significant research and development.

In this paper, we examine the potential of a new class of deep generative algorithms named score-based generative models~\cite{https://doi.org/10.48550/arxiv.1907.05600,ho2020denoising,Song2021ScoreBasedGM}. Like NFs, score-based models minimize a convex loss function with a single generator network that also provides access to the full data likelihood after training (see App.~\ref{app:cnf}).  However, unlike NFs, the gradient of the data density (the `score') is learned instead of the density. This choice introduces more flexibility to the network architecture, since the Jacobian of the transformation does not need to be computed during training. This additional flexibility, contrary to normalizing flows, also allows the introduction of bottleneck layers (layers with fewer neurons than the previous layer), greatly reducing the number of trainable parameters and improving the scalability of the model.  We demonstrate this explicitly on the Fast Calorimeter Simulation Challenge 2022 datasets\footnote{Scripts used to reproduce the results shown in this document are available at: https://github.com/ViniciusMikuni/CaloScore}.  


This paper is organized as follows. Sec.~\ref{sec:score} introduces how deep generative models can be constructed using the likelihood gradients instead of likelihoods directly.  Different choices of drift and diffusion functions investigated in this work are introduced in Sec.~\ref{sec:drift}. Section~\ref{sec:sampling} describes how new samples are generated from the trained model. Description of the Fast Calorimeter Simulation Challenge 2022 datasets and network architecture are presented in Secs.~\ref{sec:dataset} and \ref{sec:model}  as well as a GAN model used for comparison. Finally, numerical results are discussed in Sec~\ref{sec:results}. The paper ends with conclusions and outlook in Sec.~\ref{sec:conclusions}.

\section{Generative models using gradients of the data}
\label{sec:score}

\begin{figure*}[ht]
\centering
\includegraphics[width=\textwidth]{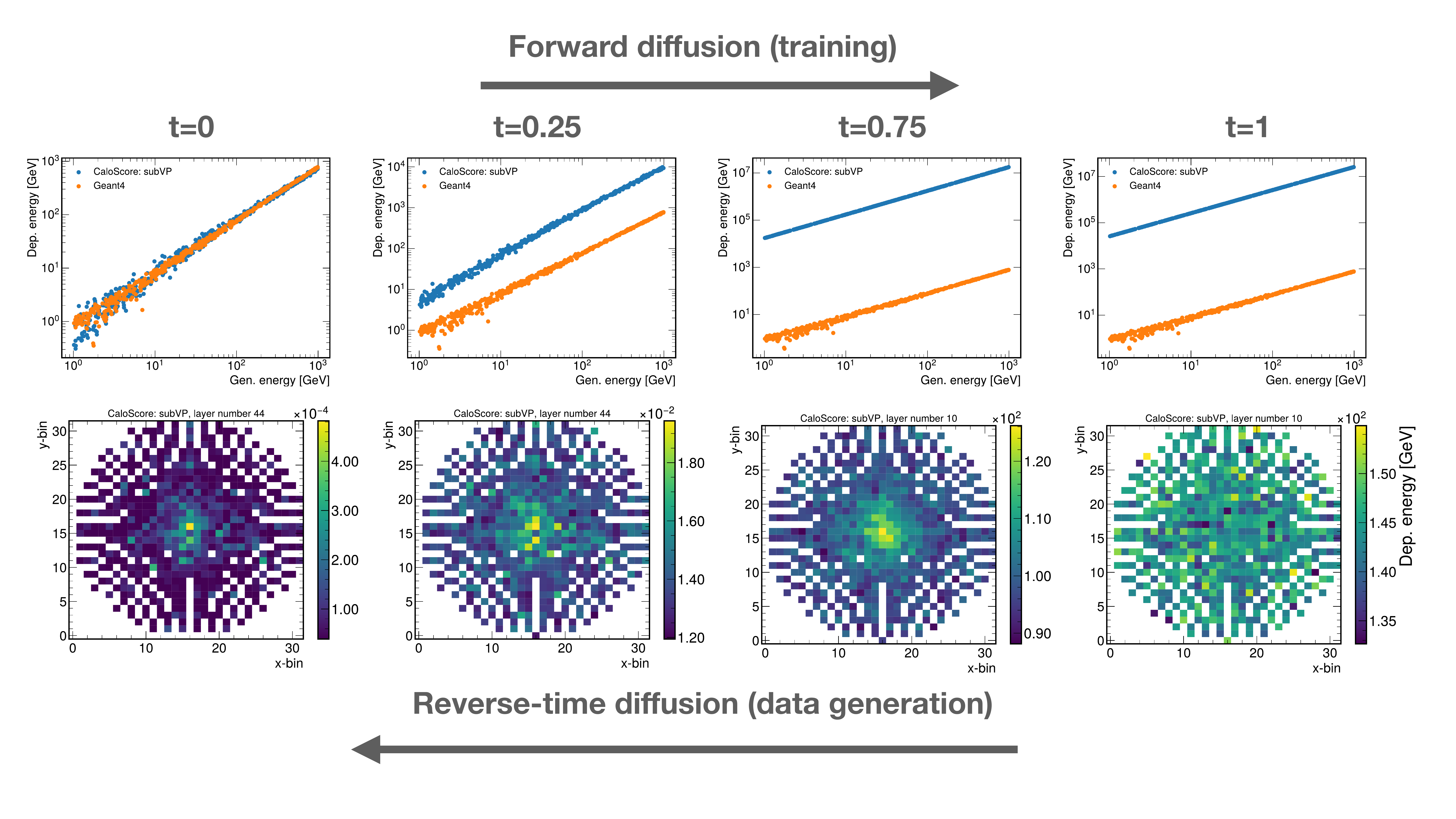}
\caption{The score-based generative model is trained using a diffusion process that slowly perturbs the data. Generation of new samples is carried out by reversing the diffusion process using the learned score-function, or the gradient of the data density. For different time-steps, we show the distribution of deposited energies versus generated particle energies (top) and the energy deposition in a single layer of a calorimeter (bottom), generated with our proposed \calosc~ model.}
\label{fig:scheme}
\end{figure*}

Our approach most closely follows Ref.~\cite{Song2021ScoreBasedGM}.  Before providing the technical details, we briefly describe the key method components.  First, a neural network is learned to approximate the data score, $\nabla_x\log p_\text{data}$ for some high-dimensional distribution $x\in\mathbb{R}^\mathrm{D}$ described by the probability density $p_\text{data}$.  In many high energy physics examples, including calorimeter simulation, we do not have access to $p_\text{data}$ analytically - we can only sample from it.  Without access to $p_\text{data}$, we cannot use a loss function like the mean squared error to directly approximate (`match') the score.  NFs circumvent this problem by maximizing the likelihood of the data.  There is no analog of maximum likelihood for score matching and so a different innovation is required.  

For data that are smeared, it can be shown that matching the score of the smeared data is equivalent to matching the score of the smearing function~\cite{6795935}.  For data purposefully smeared by an analytically tractable smearing function (e.g. a Gaussian), this means that all of components required to compute the loss function are known.  The methods explored in this paper make use of purposeful smearing, where the amount of smearing is increased/decreased to estimate the density or generate samples, respectively. A schematic of the idea is shown in Fig.~\ref{fig:scheme}.

A stochastic process that continuously corrupts data is described by the following stochastic differential equation (SDE):
\begin{equation}
    \mathrm{d}x = f(x,t)\mathrm{d}t + g(t)\mathrm{dw}.
\end{equation}
The initial data $x(t=0) := x_0\in \mathbb{R}^d$ sampled from the distribution $p_\mathrm{data}$ 
evolve over time given a set of drift and diffusion coefficients  $f(x,t):\mathbb{R}^{d}\rightarrow\mathbb{R}^d$ and $g(t):\mathbb{R}\rightarrow\mathbb{R}$, respectively. The Wiener process, or Brownian motion,  $w(t):\mathbb{R}\rightarrow\mathbb{R}$ is indexed by the time parameter $t\in[0,1]$. The goal of the generative model is to reverse this process, generating new observations starting from a noise distribution and solving the reverse SDE defined as:
\begin{equation}
    \mathrm{d}x = [f(x,t)-g(t)^2\nabla_x\log p_t(x)]\mathrm{d}t + g(t)\mathrm{d\bar{w}},
    \label{eq:rsde}
\end{equation}
where $\mathrm{\bar{w}}$ is the Wiener process in the reverse time direction \cite{ANDERSON1982313} with the gradient $\nabla_x\log p_t(x)$ of the time-dependent density $p_t(x)$  named the \textit{data score}. For fixed choices of the drift and diffusion coefficients, the only term in Eq.~\ref{eq:rsde} that needs to be estimated is the time-dependent score function. However, since the score-function is not tractable, we use neural networks that aim to minimize the Frobenius norm of the difference
\begin{equation}
    \frac{1}{2}\mathbb{E}_{p_\mathrm{data}(x)}\left [ \left\| s_\theta(x) - \nabla_x\log p_\mathrm{data}\right\| ^2_2\right ],
    \label{eq:score_mse}
\end{equation}
with $s_\theta(x)$ the output of a network with trainable parameters $\theta$. Since the true score function is not known, Eq.~\ref{eq:score_mse} cannot be readily computed. Instead, a denoising score matching \cite{6795935} strategy is used. In this strategy, instead of learning the score function of the data, we aim to learn the score function of data that have been perturbed with a known perturbation function since it is sufficient to match the score of the perturbation function. Note that while matching the score of the smeared data requires computing an expectation value over the smeared data, matching the score of the smearing function requires computing the smeared data and an expectation value over the smeared data.

Given a Gaussian perturbation kernel $p_\sigma(\tilde{x}|x):=\mathcal{N}(x,\sigma^2)$ and $p_\sigma(\tilde{x}):=\int p_{\mathrm{data}}(x)p_\sigma(\tilde{x}|x)\mathrm{d}x$, the probability density of the perturbed data, the loss function minimized during training is:
\begin{equation}
    \frac{1}{2}\mathbb{E}_{p_\sigma(\tilde{x}|x)p_{\mathrm{data}}}\left [ \left\| s_\theta(\tilde{x}) -  \nabla_{\tilde{x}} \log p_\sigma(\tilde{x}|x)\right\| ^2_2\right ].
    \label{eq:loss_score}
\end{equation}
The advantage of this strategy is that we can directly estimate the last term in Eq.~\ref{eq:loss_score}, since:
\begin{equation}
    \nabla_{\tilde{x}} \log p_\sigma(\tilde{x}|x) = \frac{x-\tilde{x}}{\sigma^2} \sim \frac{\mathcal{N}(0,1)}{\sigma}
\end{equation}

The time component can be made explicit by rewriting the loss function in Eq.~\ref{eq:loss_score} as:

\begin{equation}
    \frac{1}{2}\mathbb{E}_{t}\mathbb{E}_{p(x_t|x_0)p(x_0)}\left [ \lambda(t)\left \| s_\theta(x,t) -  \nabla_{x_t} \log p_t(x_t|x_0)\right\| ^2_2\right ].
    \label{eq:loss_score_time}
\end{equation}

The weighting function $\lambda(t):\mathbb{R}\rightarrow\mathbb{R}$ ensures the loss function has the same order of magnitude at all times and is chosen to be inversely proportional to $\mathbb{E}\left [ \left\| \nabla_{x_t}\log p_t(x_t|x_0)\right\|^2_2 \right ]$. When the drift coefficient $f(x,t)$ is chosen to be an affine function of $x$, the resulting perturbation kernel is always Gaussian \cite{sarkka2019applied} and can be chosen such that both mean and variance are known in closed form, making Eq.~\ref{eq:loss_score_time} efficient to compute during training. 

\section{Choice of drift and diffusion coefficients}
\label{sec:drift}

In this work we investigate three different choices of drift and diffusion coefficients that result in perturbation kernels that are easy to calculate in closed form. The first SDE, initially proposed in \cite{https://doi.org/10.48550/arxiv.1907.05600}, is defined as:
\begin{equation}
    \mathrm{d}x = \sqrt{\frac{\mathrm{d} [\sigma^2(t)]}{\mathrm{d} t}}\mathrm{dw}.
    \label{eq:ve_sde}
\end{equation}
The parameter $\sigma(t) =\sigma_{\textrm{min}}\left(\frac{\sigma_{\textrm{max}}}{\sigma_{\textrm{min}}}\right)^t$ is defined with $\sigma_{\textrm{min}} = 0.01$ and $\sigma_{\textrm{max}} = 50$ to ensure $x(1)\sim\mathcal{N}(0, \sigma_{\textsc{max}}^2)$ is independent from $x(0)$. Since the time-dependent variance of the resulting perturbation explodes when $t\rightarrow\infty$, this SDE is often referred to variance exploding (VE) SDE.

The second SDE is a continuous version of the discrete perturbation introduced in \cite{ho2020denoising}, defined as:
\begin{equation}
    \mathrm{d}x =  -\frac{1}{2}\beta(t)x\mathrm{d}t + \sqrt{\beta(t)}  \mathrm{dw}.
    \label{eq:vp_sde}
\end{equation}
 The parameter $\beta(t) = \beta_{\textrm{min}} + t \left( \beta_{\textrm{max}} - \beta_{\textrm{min}} \right)$ with $\beta_{\textrm{min}} = 0.1$ and $\beta_{\textrm{max}}=20$ is used, resulting in $x(1)\sim\mathcal{N}(0,1)$. The variance of this process is fixed to one when the initial distribution also has unit variance, hence its is referred to as variance preserving (VP) SDE.

The last SDE, introduced in \cite{Song2021ScoreBasedGM}, is defined as:
\begin{equation}
    \mathrm{d}x =  -\frac{1}{2}\beta(t)x\mathrm{d}t + \sqrt{\beta(t)(1-e^{-2\int_0^t\beta(s)\mathrm{d}s})}  \mathrm{dw}.
    \label{eq:subvp_sde}
\end{equation}
When the parameter $\beta(t)$ is chosen to be identical as the one used in Eq.~\ref{eq:vp_sde}, the variance of the stochastic process is always smaller than the variance of the VP SDE, hence receiving the name subVP. Similarly to Eq.~\ref{eq:vp_sde}, $x(1)$ also follows a normal distribution. 

The resulting perturbation kernels induced by each SDE are listed in Tab.~\ref{tab:perturbation}.

\begin{table}[ht]
    \centering
	\small
    \caption{Perturbation kernel induced by different SDE choices.}
    \label{tab:perturbation}
	\begin{tabular}{lcc}
        SDE & Perturbation kernel\\
        \hline
        VE & $\mathcal{N}(x(0), \sigma^2(t) - \sigma^2(0))$\\
        VP & $\mathcal{N}(x(0)e^{-\frac{1}{2}\int_0^t \beta(s) \mathrm{d}s}, 1 - e^{-\int_0^t \beta(s) \mathrm{d}s})$\\
        subVP & $\mathcal{N}(x(0)e^{-\frac{1}{2}\int_0^t \beta(s) \mathrm{d}s}, (1 -  e^{-\int_0^t \beta(s) \mathrm{d}s})^2)$\\

	\end{tabular}
\end{table}

\section{Sample generation}
\label{sec:sampling}
New samples are generated by solving the reverse diffusion process defined in Eq.~\ref{eq:rsde} using a numerical SDE solver. In this work, we use the  Euler-Maruyama algorithm~\cite{kloeden1992stochastic} followed by an additional corrector step that uses the Langevin MCMC approach \cite{Parisi:1980nh,Grenander1994REPRESENTATIONSOK} to increase the sampling quality. For each time decrement $\Delta t$, the updated state of the system is described as:
\begin{equation}
    x_{t-\Delta t} = x_t + [f(x_t,t) -g^2(t)s_\theta(x_t,t)]\Delta t + g(t)\sqrt{|\Delta t|}z,
    \label{eq:EM}
\end{equation}
where $z\sim\mathcal{N}(0,1)$ is sampled at each time step. The corrector step takes the updated state from Eq.~\ref{eq:EM} and applies the correction
\begin{equation}
    x_{t-\Delta t}' = x_{t-\Delta t} + \epsilon s_\theta(x_{t-\Delta t},t) g(t) + \sqrt{2\epsilon}z,
    \label{eq:corrector}
\end{equation}
where $\epsilon$ is a tunable parameter that determines the strength of the correction applied. A dimension-independent expression for $\epsilon$ is defined as a function of a signal-to-noise ratio $r$  parameter as:
\begin{equation}
    \epsilon = 2r^2\frac{\left\| z\right\|_2^2}{\left\| s_\theta \right\|_2^2},
    \label{eq:snr}
\end{equation}
where $\left\| z\right\|_2$ and $\left\| s_\theta \right\|_2$ are the batch-average norms of the Gaussian noise and trained score function.

High fidelity samples require the time interval $\Delta t$ to be small, possibly leading to hundreds of iterative steps and consequently hundreds of network evaluations. We decrease the number of function evaluations by reusing the score function evaluated in Eq.~\ref{eq:EM} during the calculation of the corrector step in Eq.~\ref{eq:corrector}. This approach effectively decreases the number of function evaluations by a factor two while no decrease in generation quality is observed.

\section{Fast Calorimeter Simulation Challenge 2022}
\label{sec:dataset}
\calosc~is trained on the datasets created by the Fast Calorimeter Simulation Challenge 2022~\cite{faucci_giannelli_michele_2022_6366324,michele_faucci_giannelli_2022_6368338,faucci_giannelli_michele_2022_6366271,ATLAS:2021pzo}. A total of three datasets are provided, representing calorimeter shower simulations with \textsc{Geant} of different geometries and granularities. Dataset 1 \cite{michele_faucci_giannelli_2022_6368338} is based on the ATLAS  open dataset \cite{ATL-SOFT-PUB-2020-006,ATLAS:2021pzo} and is similar to the current ATLAS detector calorimeter geometry. Showers are generated at the calorimeter surface in the pseudo-rapidity range $\eta\in[0.20,0.25]$. While samples consisting of both photons and pions are provided, we evaluate our model  using only the photon dataset. The voxelization procedure is defined such that it minimizes the amount of empty voxels, while maintaining high fidelity compared to the full simulation. This strategy results in different number of voxels per calorimeter layer and a total of 368 voxels to represent the full detector slice. Photon energies are provided in this configuration for 15 incident energies from 256 MeV up to 4 TeV in steps given by powers of two. For each generated energy, 10k samples are provided with this number reduced at higher energies due to long simulation times, resulting in a total of 121k used during training. 

Datasets 2 \cite{faucci_giannelli_michele_2022_6366271} and 3 \cite{faucci_giannelli_michele_2022_6366324} contain each 100k examples and are simulated using a common detector layout but with different voxelization granularity. The detector simulated has a concentric cylinder geometry with 45 layers, where each layer consists of active (silicon) and passive (tungsten) material, simulated with \textsc{Geant4}. Electrons are generated at the detector surface with initial energy sampled from a log-uniform distribution ranging from 1~GeV to 1~TeV. In dataset 2, each layer consists of 144 readout cells, with 9 in the radial and 16 in the angular directions. Dataset 3 is more granular, consisting of 900 readout cells in each layer, with 18 in the radial and 50 in the angular directions.

Even though the initial voxelization is provided in Cylindrical coordinates, we found it beneficial to convert the voxels in datasets 2 and 3 to Cartesian coordinates. This change allows the effective usage of 3D convolutional neural networks (CNNs) to build the score model. While CNNs are applicable in polar coordinates, they struggle to learn periodic boundary conditions. Convolutional operations are also less effective, since the majority of the energy depositions are located near $r=0$, or near the corners of the image. Since the datasets are only available after voxelization, the transformation from Cylindrical to Cartesian coordinates inevitably leads to loss of information\footnote{A one-to-one assignment between the two sets of coordinates is possible, but requires the distance interval in Cartesian coordinates to follow a non-linear function since the transformation of coordinates is itself non-linear.}. Nevertheless, we observe improved generation quality after changing the coordinate system. See App.~\ref{app:transform} for more details of the transformation. The total number of voxels after the transformation is chosen to be similar to the original number, resulting in $12\times12=14$4 and $32\times 32=1024$ voxels per layer for datasets 2 and 3, respectively.

Energies deposited in each voxel span multiple orders of magnitude, motivating yet another transformation of the inputs before training the generative model. First, each voxel energy $E_v$ is normalized by the value of the generated energy of the particle $E_0$ times a factor $f$ that ensures the normalized voxel energy $E'_v = \frac{E_v}{fE_0}$ lies between 0 and 1. Naively, the factor $f$ could be taken as 1, since energy conservation should ensure the sum of deposited energies to not exceed the initial particle's energy. However, the sampling fraction of the calorimeter may lead to a mismatch between these numbers. This effect is particularly important in dataset 1, when particle energies as low as 256 MeV are considered. We take $f$ as the highest deviation between total deposited and generated energies, fixed to $f=3.1$ for dataset 1 and $f=2$ for datasets 2 and 3. 

The normalized energy depositions are then transformed to log-space, similarly to the strategy used in \textsc{CaloFlow}. The log-transformed value $u_v$ is defined as:
\begin{equation}
    u_v=\log \left ( \frac{x}{1-x} \right ), x=\alpha + (1-2\alpha)E'_v.
    \label{eq:log_transform}
\end{equation}
The value $\alpha$ is set to $10^{-6}$ and avoids a discontinuity when $E'_v=0$. The generated particle energy, used as a conditional input to the model, is also transformed before training. The transformed conditional energy $u_0$ is defined as:
\begin{equation}
    u_0 = \frac{e_0-e_{\textrm{min}}}{e_{\textrm{max}}-e_{\textrm{min}}},
\end{equation}
where $e_{\textrm{min}}$ and $e_{\textrm{max}}$ are the minimum and maximum energies available in the dataset. 

Finally, dataset 1 is also modified by adding an extra dimension that encodes an overall energy normalization. Before applying the log-transformation in Eq.~\ref{eq:log_transform}, the total deposited energy is calculated. Instead of normalizing each voxel by the generated energy, the total deposited energy is used, ensuring the sum of all voxels is equal to 1. The additional entry is then defined as the total deposited energy normalized by the initial particle energy times the factor $f$. This strategy improves the estimation of the total energy deposition during training, now encoded as a single entry rather than the sum of all voxels.

\begin{figure*}[ht]
\centering
\includegraphics[width=\textwidth]{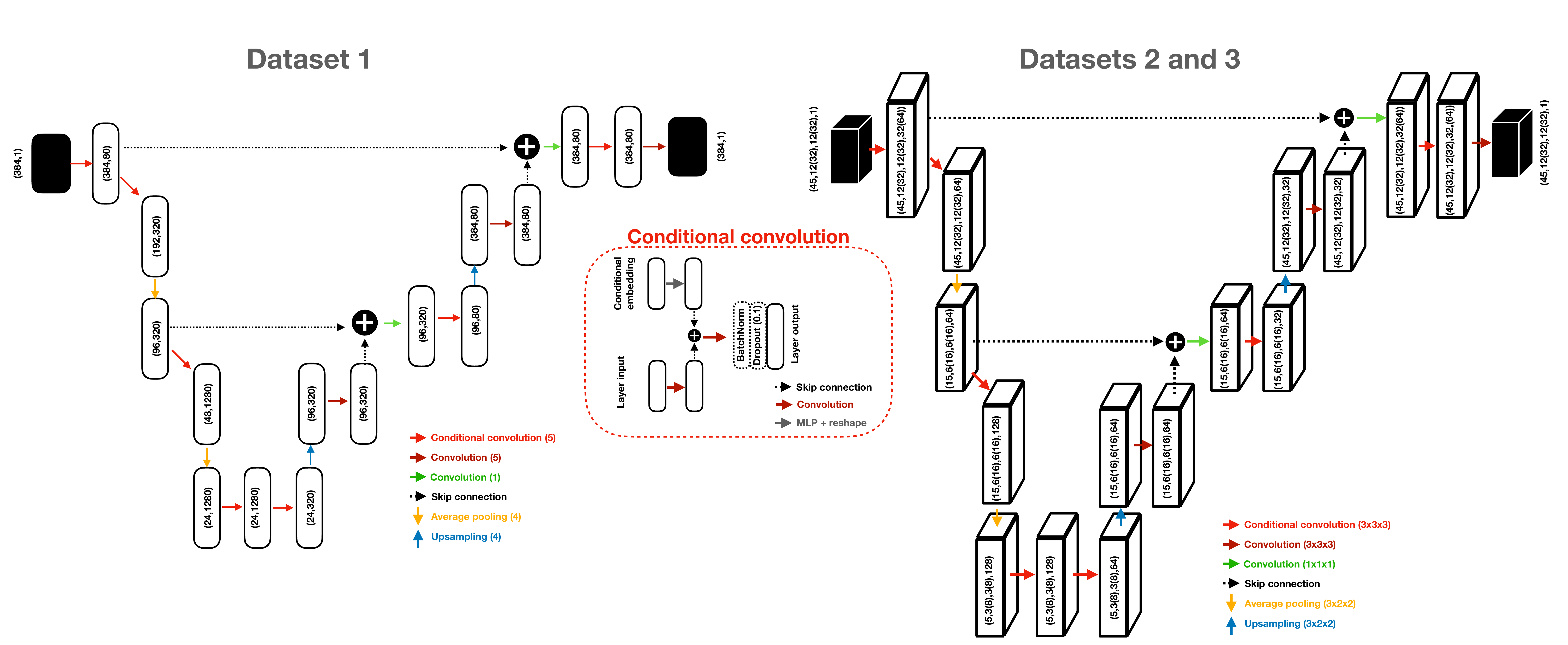}
\caption{Network architectures used for datasets 1 (left), 2 and 3 (right) based on the U-Net architecture. Values in parentheses represent the dimensionality of the data at each layer of the network. Parenthesis after convolutional operations represent the kernel size used. The swish non-linear activation function is used after each convolution operation. Conditional inputs (time and energy) are used multiple times in the model to define conditional convolution operations described in the middle.}
\label{fig:arch}
\end{figure*}

\section{Model architecture and training details}
\label{sec:model}
The score function is built from a modified version of the U-net model \cite{ronneberger2015u} where and encoder-decoder architecture with skip connections is used. 3D convolution operations are used as the basic layers in \calosc~for datasets 2 and 3, leveraging the regular geometry. Dataset 1, on the other hand, is irregular and consists of different number of voxels per detector layer. In this case the score function is built based on 1D convolutional operations. This approach leads to dataset 1 requiring bigger kernel and layer sizes compared to datasets 2 and 3, resulting in a bigger model architecture.

Each convolutional operation uses the swish \cite{ramachandran2017searching} non-linear activation function, with a kernel size of 5 (dataset 1) and 3 (datasets 2 and 3). The number of dimensions is reduced in the encoder section of the network through average pooling operations, reducing the total number of dimensions by a factor 4 for dataset 1 and a factor $3\times 2\times2=12$ for datasets 2 and 3 after each pooling layer. In the opposite direction, upsampling layers are used to increase the dimensionality by repeating entries multiple times. 

The different network architectures used for each dataset are shown in Fig.~\ref{fig:arch}.

Conditional inputs used to train the model, namely the time component and generated energy, are first transformed using random Fourier features \cite{tancik2020fourier}. The transformed features for each conditional input are then concatenated and passed over 2 fully connected layers of sizes 256 (dataset 1) or 128 (datasets 2 and 3), both followed by a swish activation function. This set of \textit{conditional embeddings} are used during multiple stages of the model architecture. In particular, conditional convolutional operations are created by adding the conditional embeddings as an additional bias at the output of convolutional layers.

Additionally, we train a Wasserstein GAN (WGAN) \cite{arjovsky2017wasserstein} using an additional gradient penalty (GP) term \cite{gulrajani2017improved} to enforce the Lipschitz constraint of the critic model. WGAN-GP is a well-established model for calorimeter detector simulation and currently used in the ATLAS Collaboration~\cite{ATLAS:2021pzo} and studied in the context of different detector geometries applied to different experimental collaborations \cite{Erdmann:2018jxd,Chekalina:2018hxi,Buhmann:2020pm}.

A similar network architecture to \calosc~is used to facilitate the comparison between approaches. The generator network takes as inputs an uniform distribution with dimensions 50, 100, 150 for datasets 1, 2, and 3 respectively. The noise vector is then concatenated with the conditional energy before passing through a fully connected layer with swish activation function. The output size of the fully connected layer is set to be the same size as last bottleneck layer of the U-Net model used in \calosc. After reshaping, the same architecture that follows the last bottleneck layer in the right-hand side of the U-net is used. Similarly, the critic network takes as inputs real or generated samples and uses the same convolutional layers as the ones used in left-hand side of the U-Net model before the last bottleneck layer. From the last bottleneck layer, an additional fully connected layer is used with output size of 1 and no activation function. This choice results in all \calosc~models and WGAN-GP with similar number of trainable weights. 

The implementation of all models is carried out with \textsc{Tensorflow}~\cite{tensorflow} optimized with \textsc{Adam}~\cite{adam} in the case of \calosc and \textsc{RMSProp}  in the WGAN-GP implementation, all with initial learning rate set to $10^{-4}$. In \calosc, the learning rate is reduced by a factor 2 if the loss function does not improve for a period of 100 consecutive epochs, evaluated using an independent dataset. The evaluation dataset is taken as 20\% of the total amount of available training events. The models are trained for a total of 2000 epochs. The epoch with lowest evaluation loss is saved for further inspection. For all models, 16 NVIDIA A100 GPUs are used simultaneously interfaced with the Horovod package \cite{sergeev2018horovod} on the Perlmutter supercomputer. The batch size in each GPU is set to 128 (datasets 1 and 2) and 64 (dataset 3). The WGAN-GP model is trained for 10000 epochs with fixed learning rate. The last training epoch is saved for further evaluation, but others epochs before the last were checked to produce similar results as the ones presented. 

Sample generation is performed in \calosc as described in Sec.~\ref{sec:sampling}, with signal-to-noise ratio fixed to 0.2 and total number of time steps set to 100, in dataset 1, and 200, in datasets 2 and 3. See App.~\ref{app:corrector} for differences in generation quality for other parameter choices. The total number of trainable weights and the time required to generate 100 calorimeter showers in a single GPU with batch size fixed to 100 are listed in Tab.~\ref{tab:resources}. 
\begin{table}[ht]
    \centering
	\small
    \caption{Number of dimensions, trainable parameter, and time to generate 100 new calorimeter showers for each dataset studied in this work. Generation times for \geant~are based on the average time required to generate samples over the energy range provided.}
    \label{tab:resources}
	\begin{tabular}{l|c|c|c|c|ccc}
        Dataset & N. of & N. of & \multicolumn{3}{c}{Time to 100 showers [s]}\\
        & voxels & weights & {\scriptsize \calosc} & {\scriptsize WGAN-GP} & {\scriptsize \geant}\\
        \hline
        dataset 1 & 384 & 32M & 4.0 & 1.3 & $\mathcal{O}(10^2-10^3)$\\
        dataset 2 & 6480 & 1.4M & 5.8 & 1.33 & $\mathcal{O}(10^4)$\\
        dataset 3 & 46080 &1.7M & 33.4 & 2.06 & $\mathcal{O}(10^4)$\\
	\end{tabular}
\end{table}

The lack of a regular geometry resulted in \calosc~requiring almost 20 times more trainable parameters for dataset 1 compared to datasets 2. On the other hand, dataset 3 has only $\sim$20\% more trainable weights than dataset 2, even though dataset 3 has 7 times more voxels. This observation suggests that the model complexity is mostly determined by the network architecture rather than the number of voxels present in the dataset, contrary to normalizing flows where the complexity from the Jacobian determinant calculation increases at least linearly with the number of voxels. 

Although the ultimate goal is for the generation time to be significantly faster than for \geant, the time to generate new calorimeter showers with \calosc~is comparatively slow to other machine learning-based models as a result of the hundreds of function evaluations.  In this paper, we have focused on modeling the complex data distributions and we leave explorations of accelerating inference to future studies.

\section{Results}
\label{sec:results}
\begin{figure*}[ht]
\centering
\includegraphics[width=0.3\textwidth]{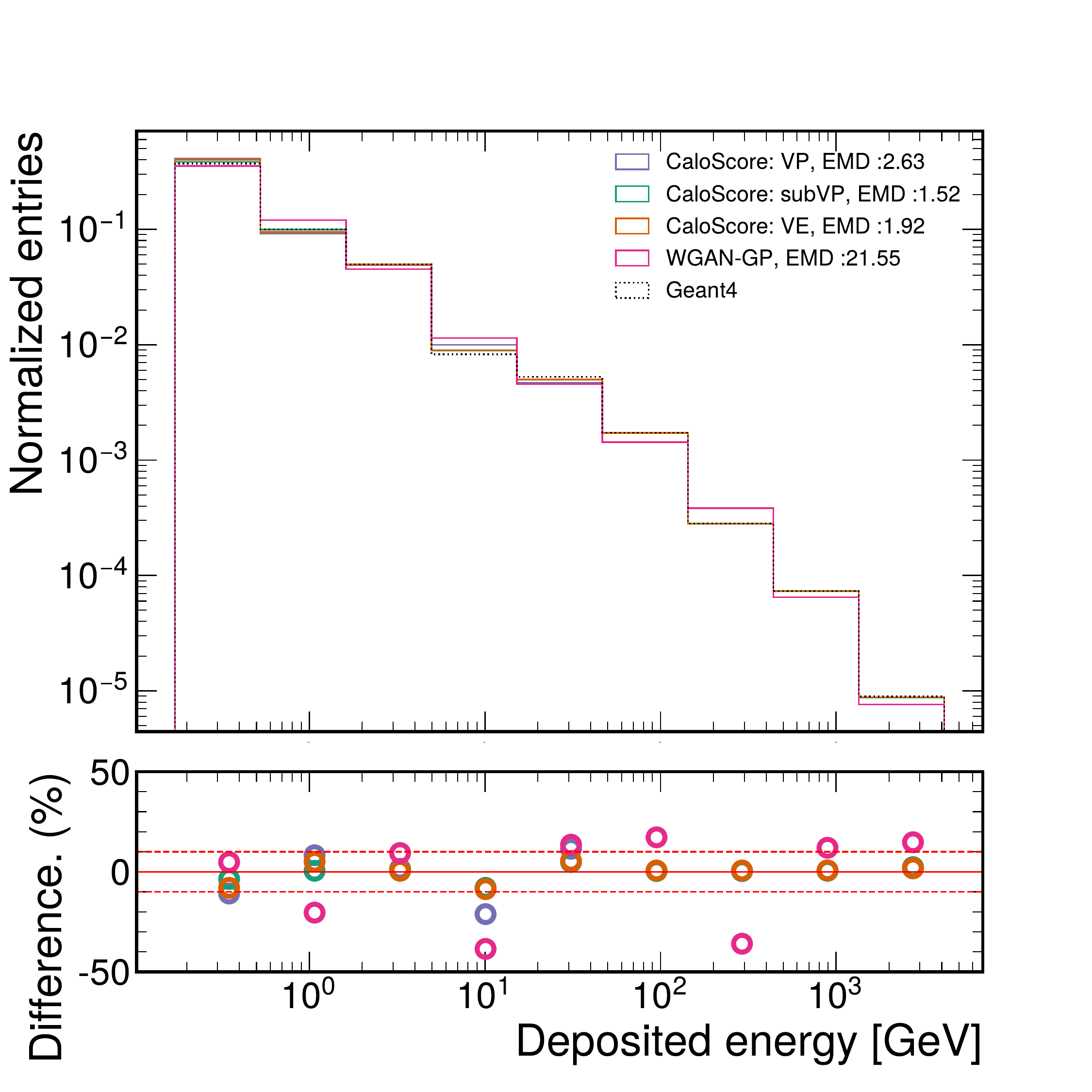}
\includegraphics[width=0.3\textwidth]{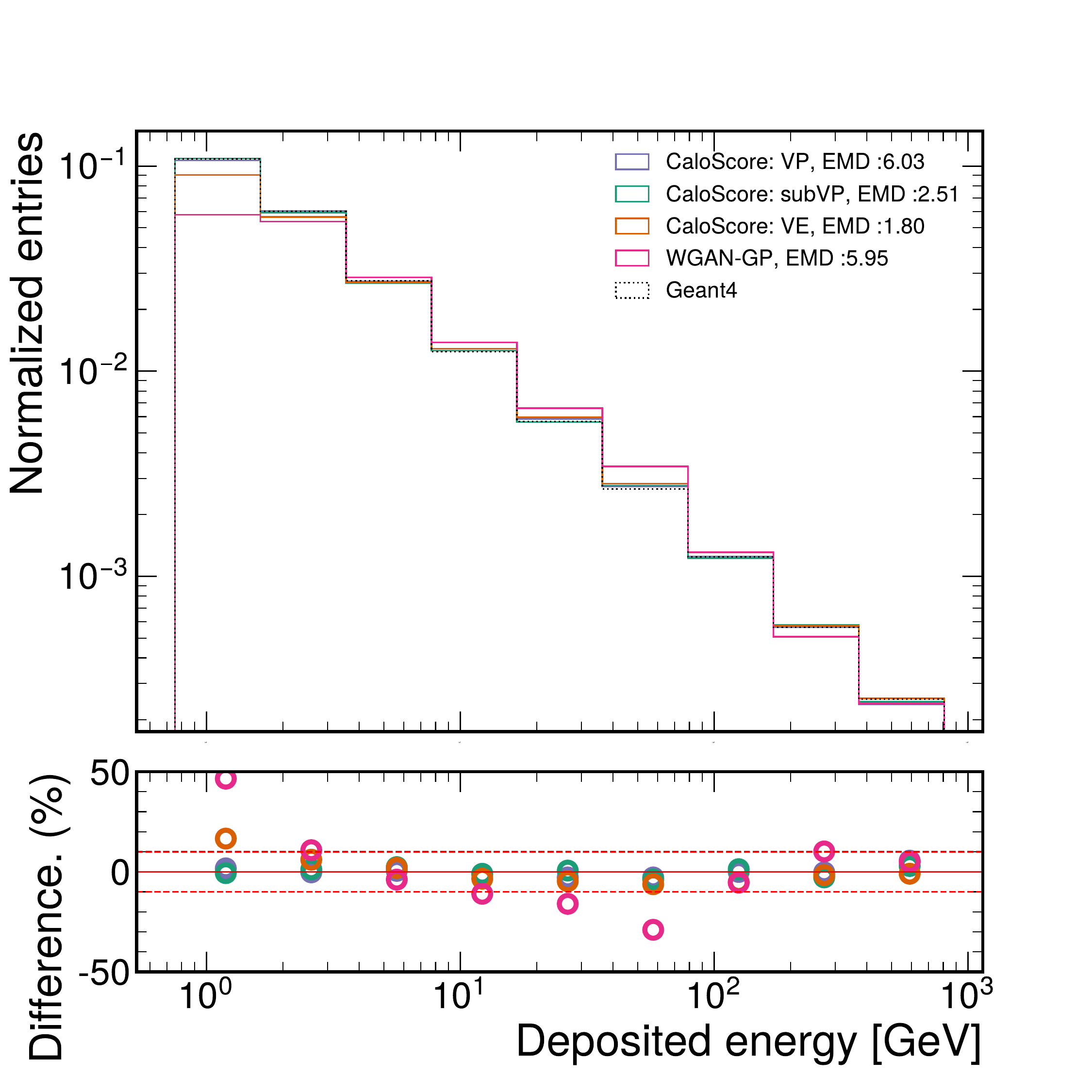}
\includegraphics[width=0.3\textwidth]{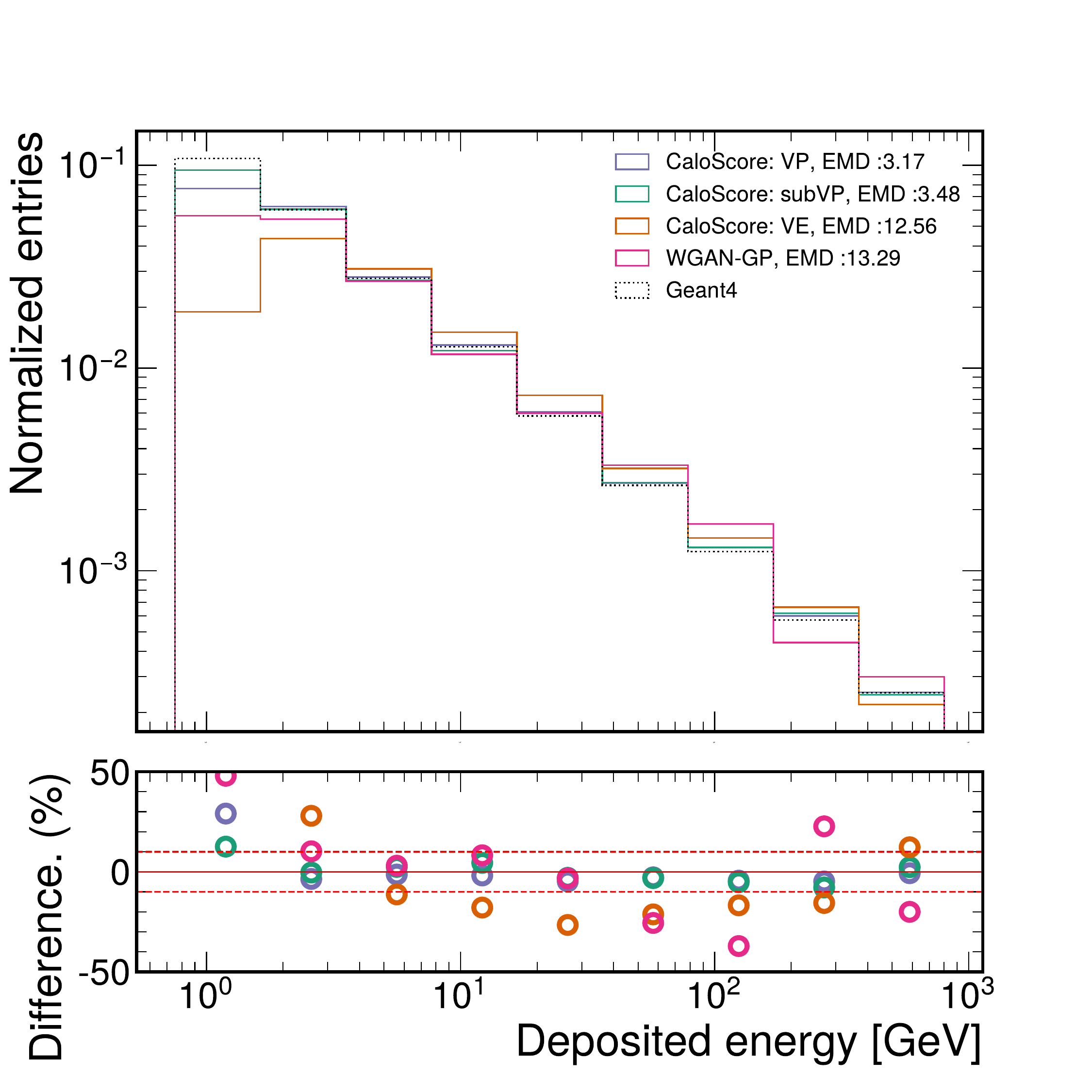}
\includegraphics[width=0.3\textwidth]{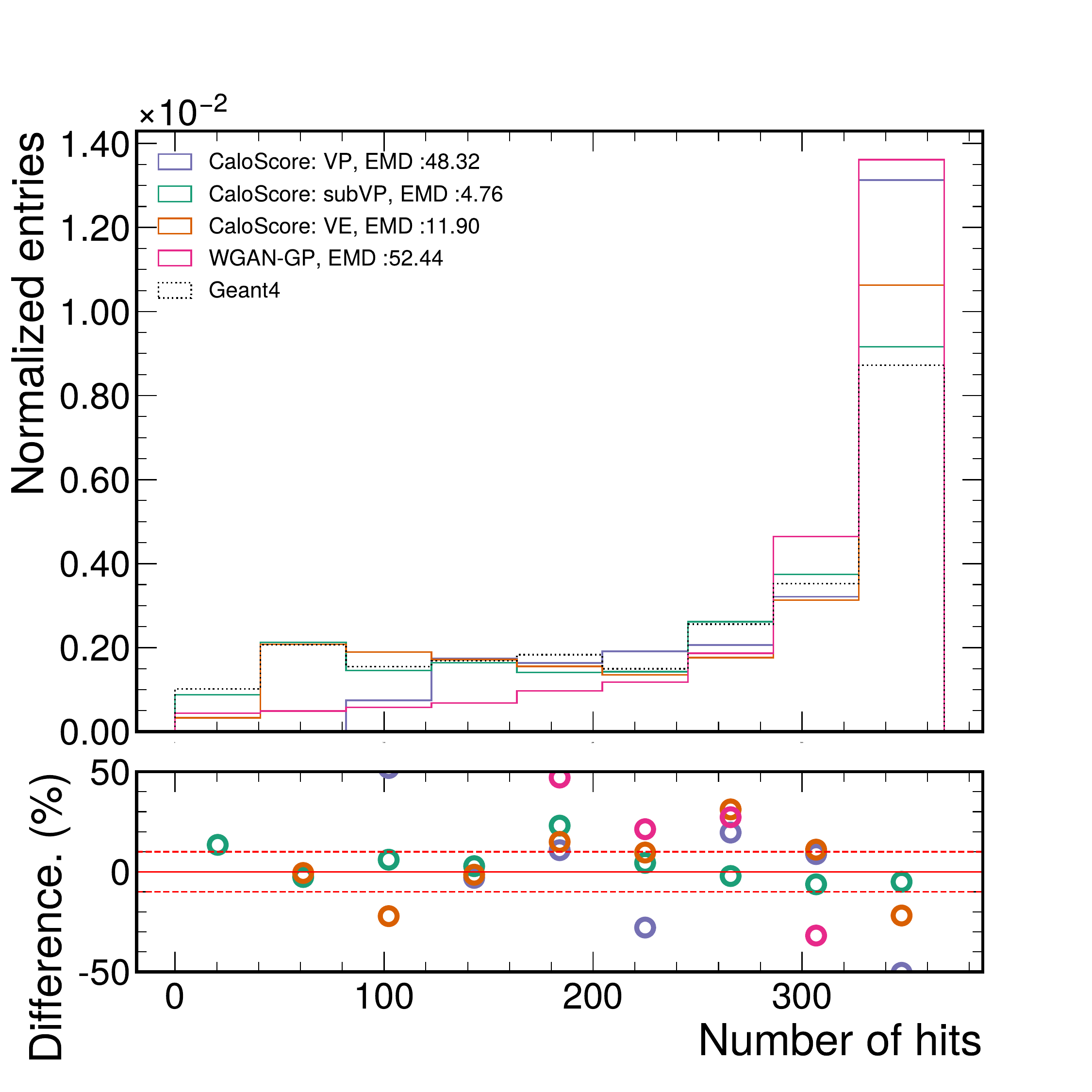}
\includegraphics[width=0.3\textwidth]{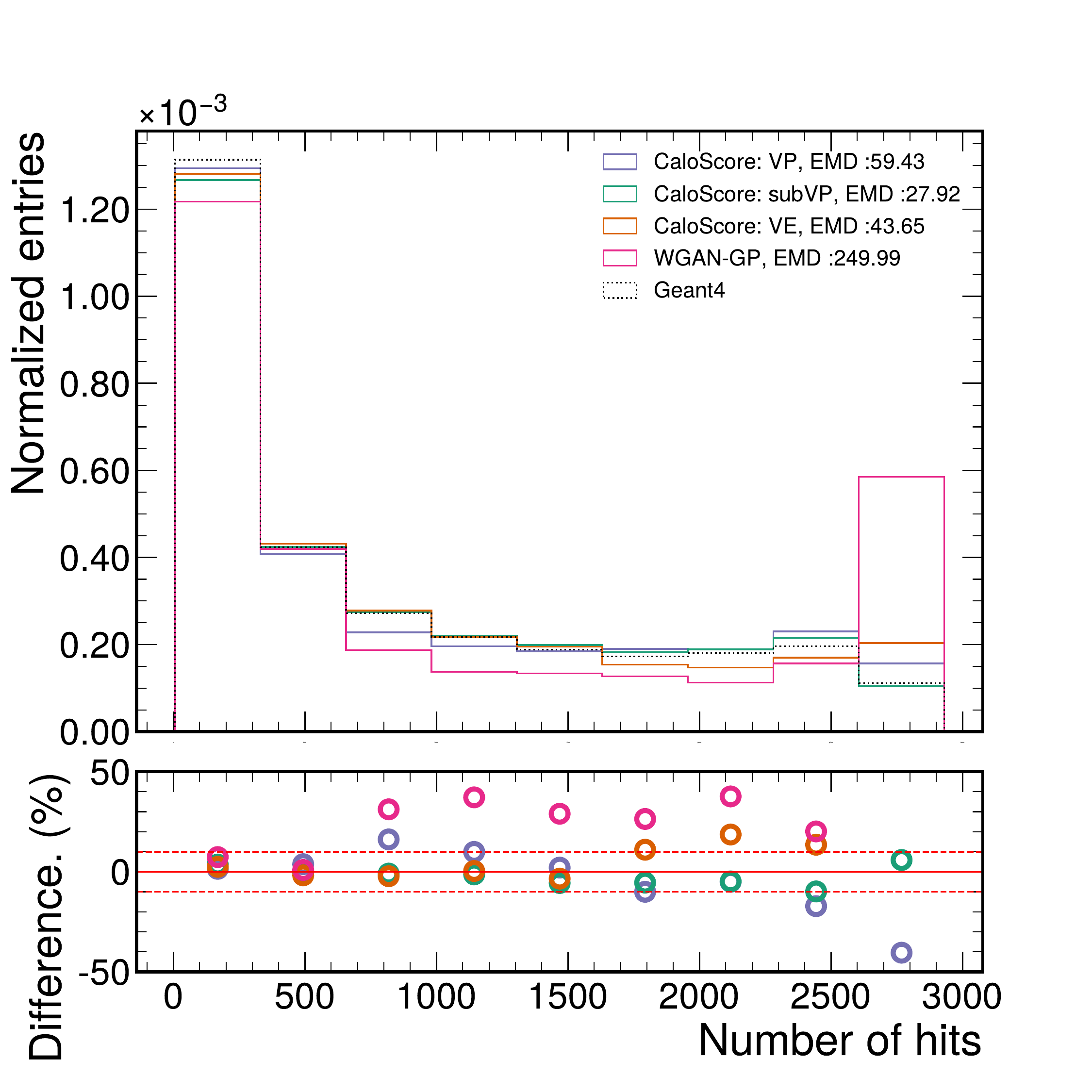}
\includegraphics[width=0.3\textwidth]{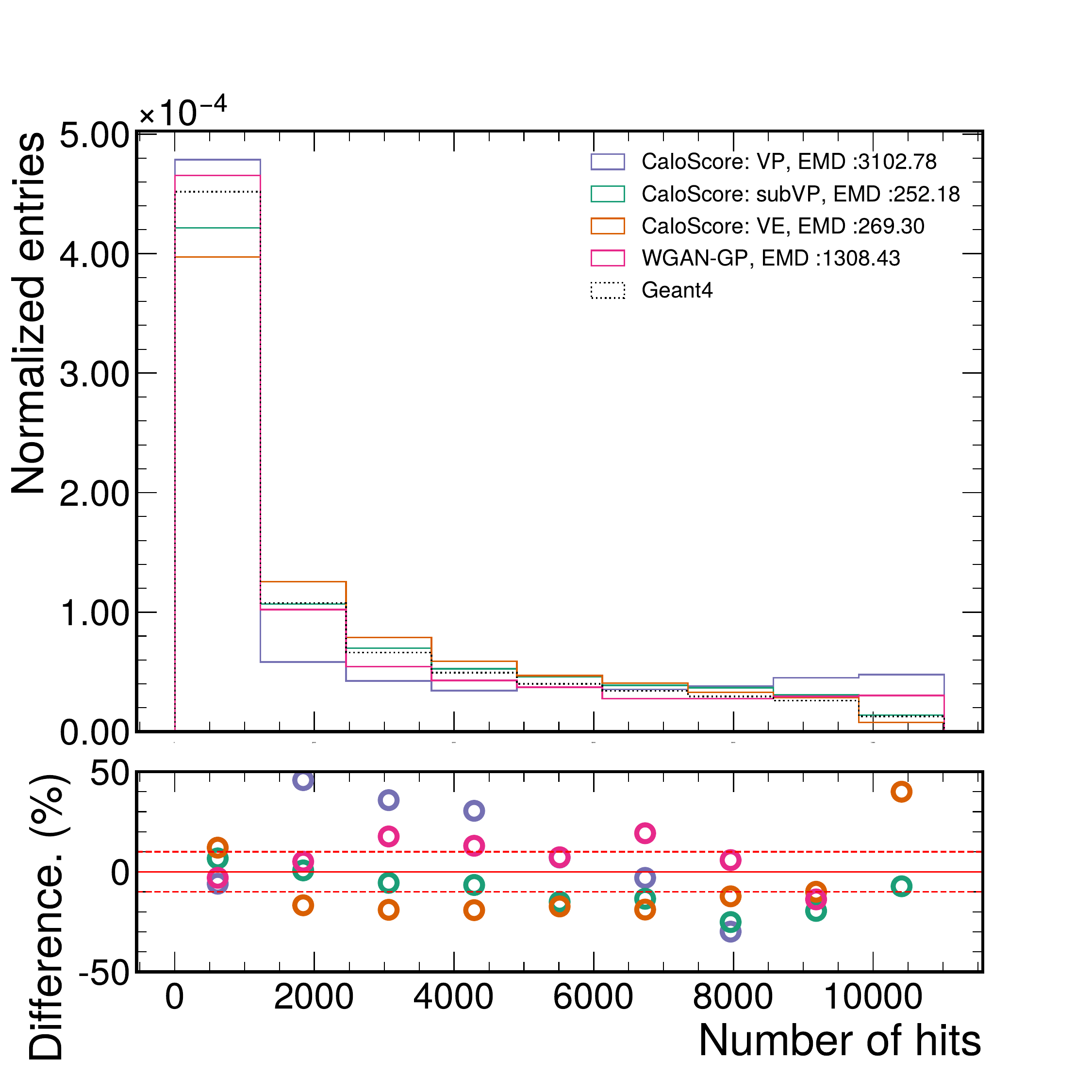}

\caption{Comparison of the sum of all voxel energies (top) and number of hits (bottom) for datasets 1 (left), 2 (middle), and 3 (right). Dashed red bands represent the 10\% deviation interval of the generated samples when compared to \geant~predictions. The earth mover's distance (EMD) between each distribution and the \geant~distribution is also provided.}
\label{fig:etot_nhits}
\end{figure*}

Multiple distributions are used to evaluate the quality of generated samples using different \calosc~SDE implementations and additional WGAN-GP model.  The total energy deposited in the detector and the number of calorimeter hits are shown in Fig.~\ref{fig:etot_nhits}. A hit is defined as any voxel with energy deposition above a certain energy threshold. The energy thresholds are taken as the minimum energy observed in each challenge dataset, set to 0.01 keV for dataset 1 and 15.1 keV for datasets 2 and 3. The 1-Wasserstein distance between distributions, referred as the earth mover's distance (EMD), is also calculated between the  \geant~and different generative model implementations.

A good agreement between the \geant~and generated samples is observed for all diffusion models and datasets. At low deposited energies, the difference between \calosc~and \geant~increase and is most noticeable for dataset 3. The subVP implementation shows a better agreement overall, followed by VP and VE, indicating that bounding the variance of the diffusion process is beneficial, specially as the number of voxels increase. Conversely, the WGAN-GP implementation shows a higher EMD value compared to \calosc. 

\begin{figure*}[ht]
\centering
\includegraphics[width=0.3\textwidth]{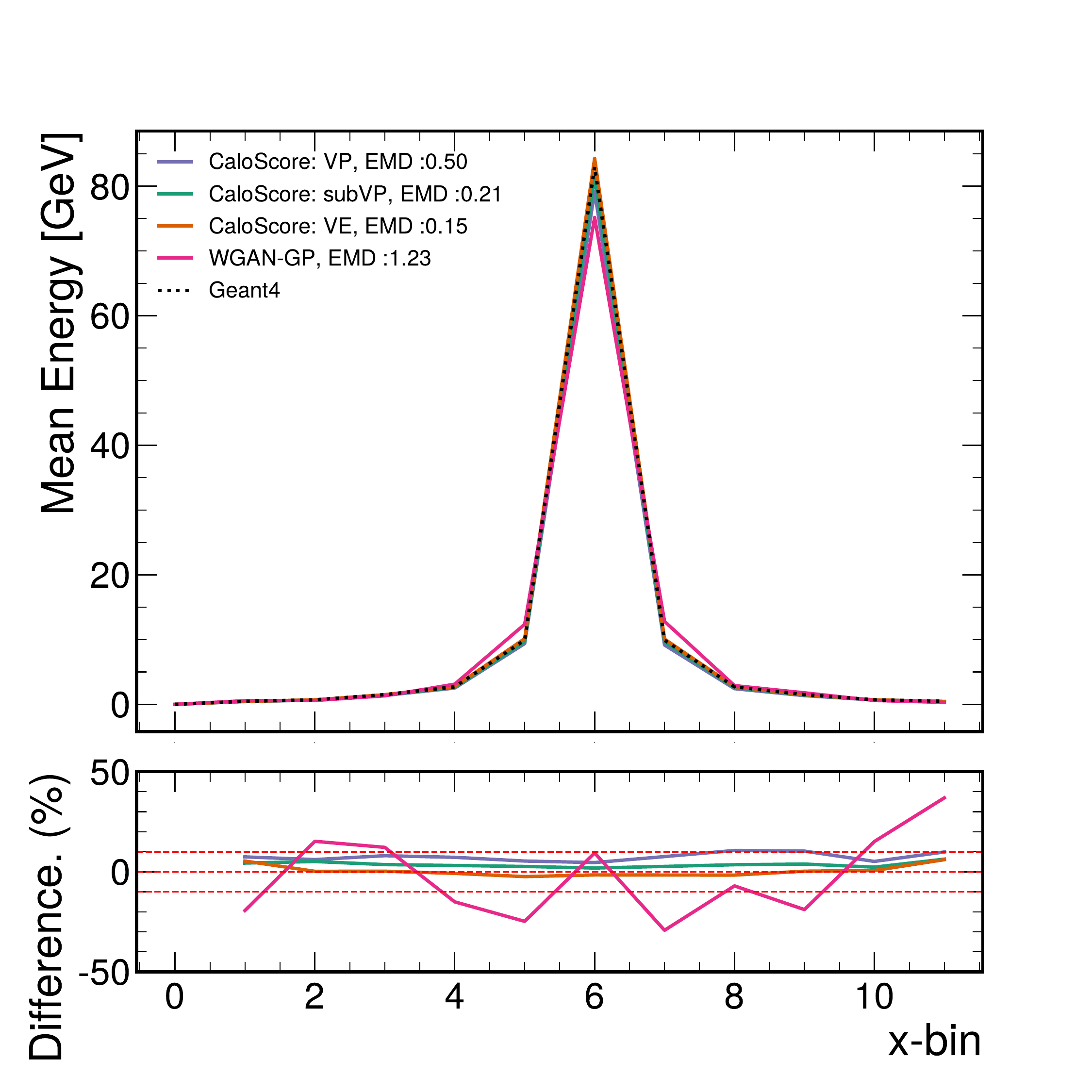}
\includegraphics[width=0.3\textwidth]{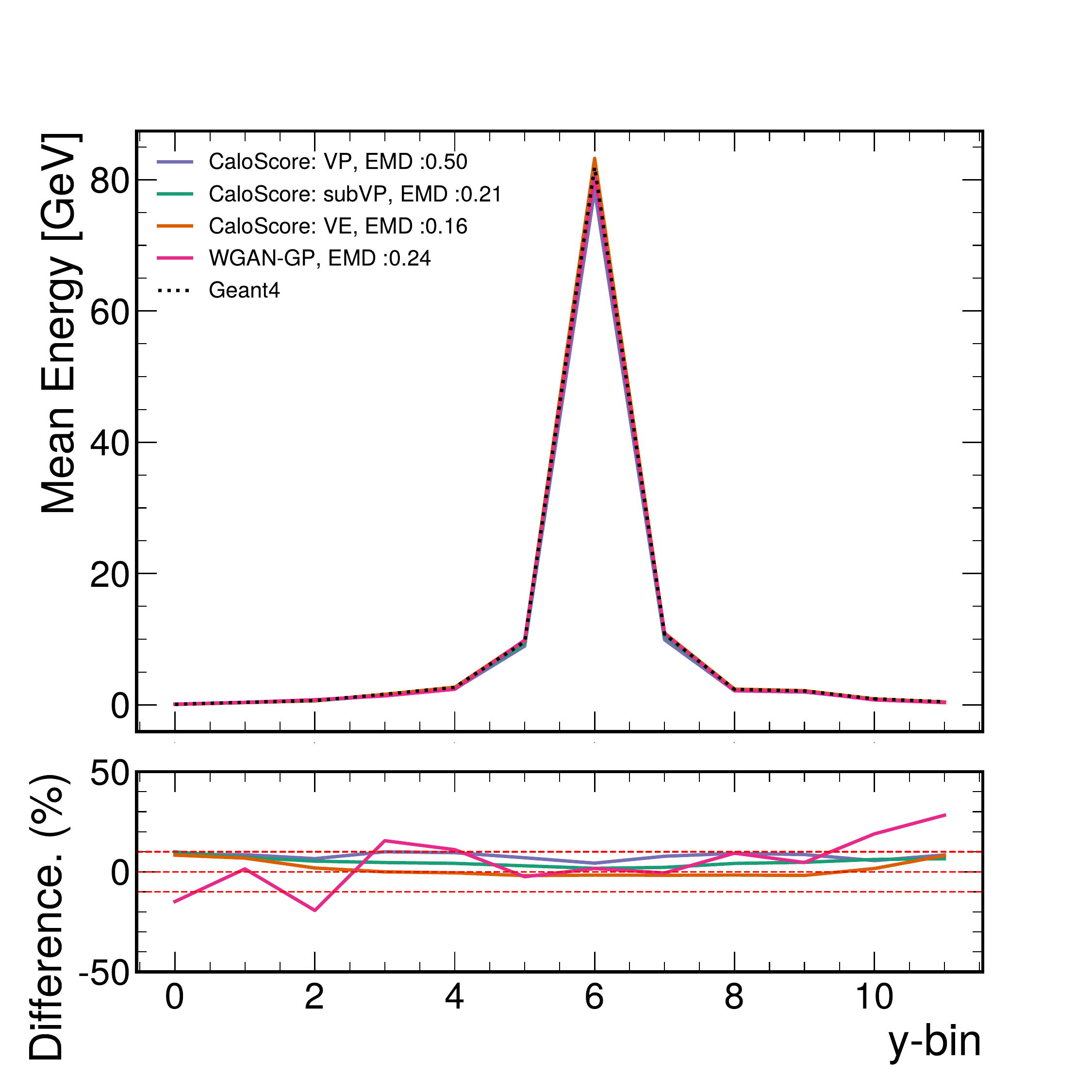}
\includegraphics[width=0.3\textwidth]{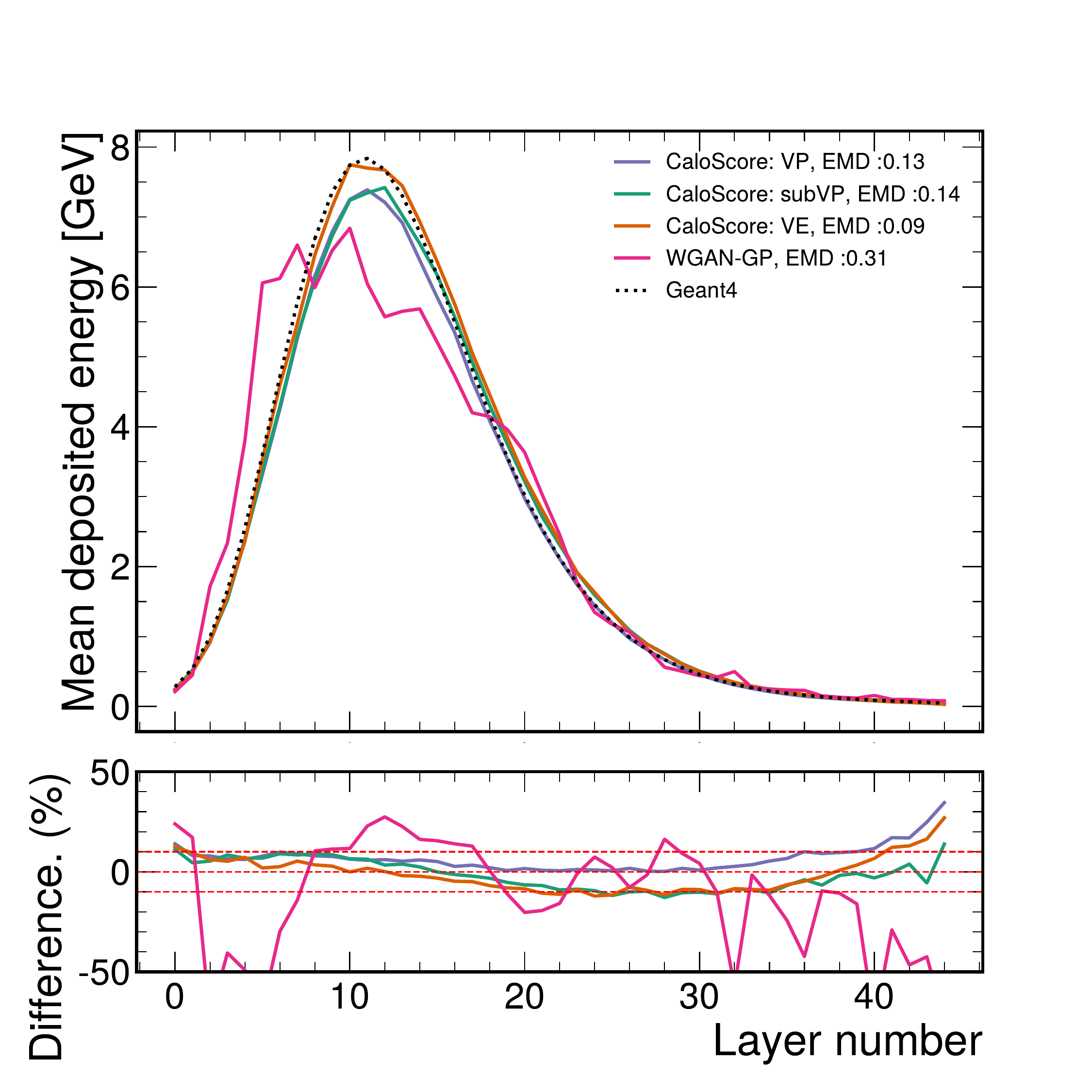}
\includegraphics[width=0.3\textwidth]{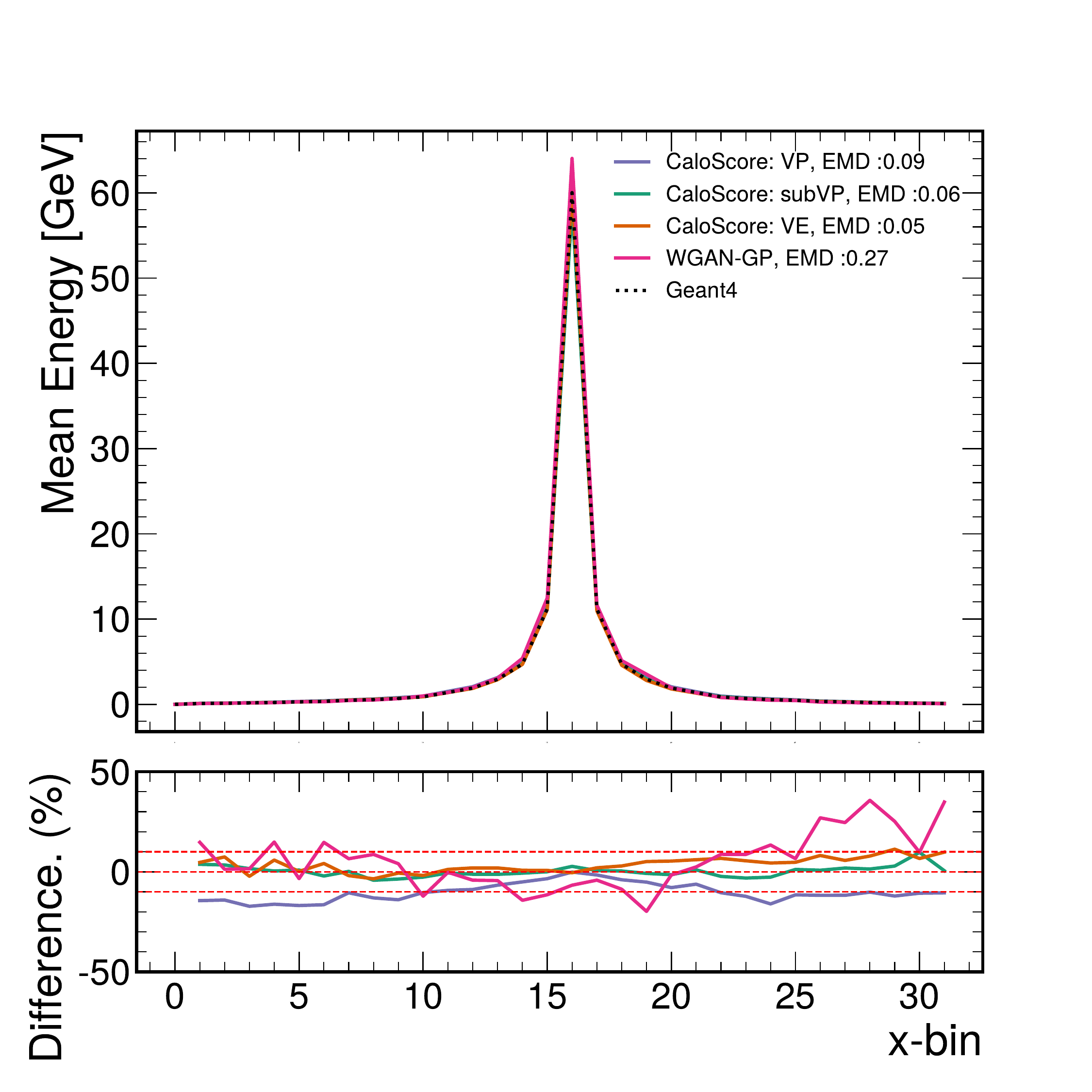}
\includegraphics[width=0.3\textwidth]{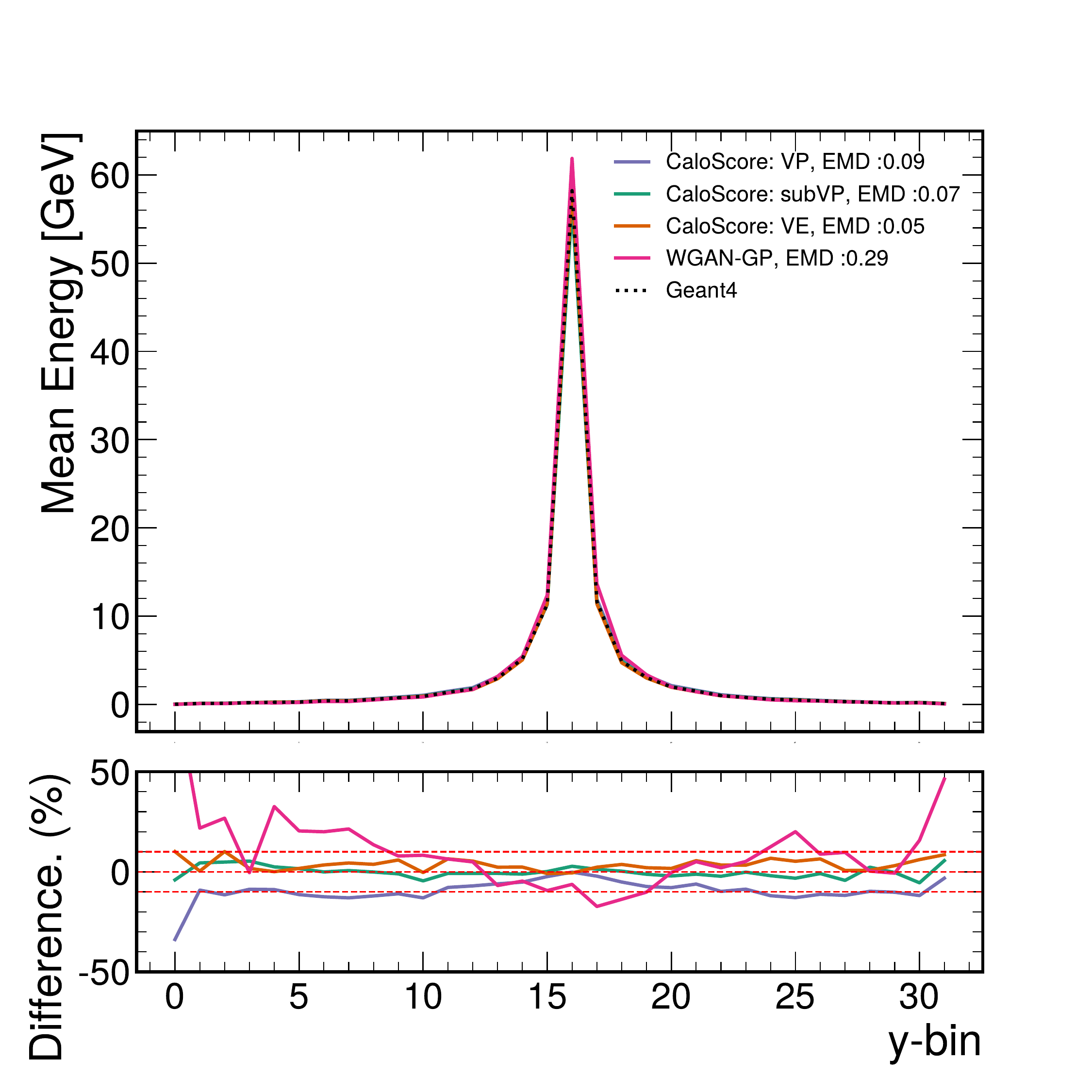}
\includegraphics[width=0.3\textwidth]{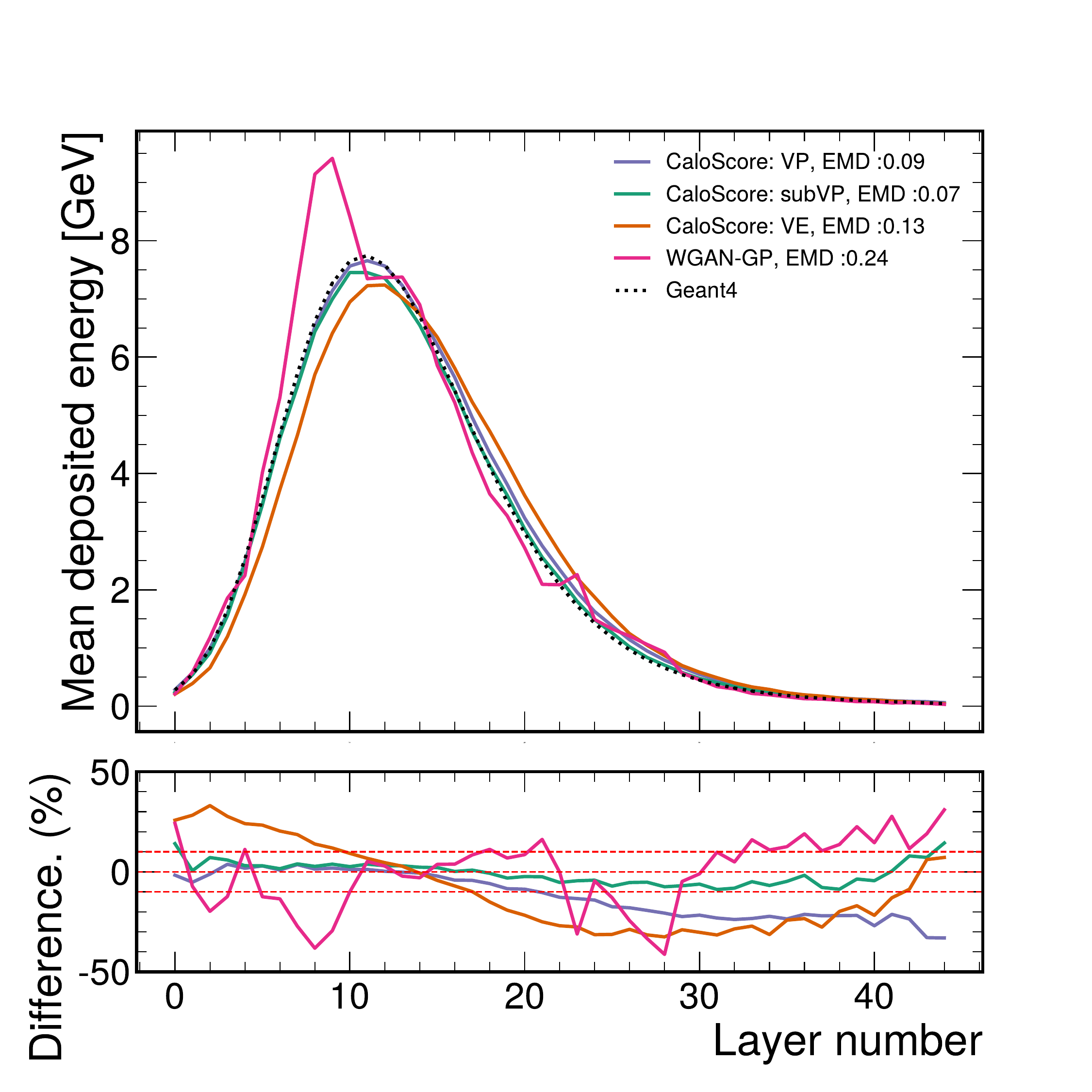}

\caption{Comparison of the average deposited energies in the x- (left), y- (middle), and z-coordinates (right) for datasets 2 (top) and 3 (bottom). Dashed red bands represent the 10\% deviation interval of the generated samples when compared to \geant~predictions. The earth mover's distance (EMD) between each distribution and the \geant~distribution is also provided.}
\label{fig:emean}
\end{figure*}

A similar conclusion is derived from the average  energy deposition as a function of the layer number and Cartesian coordinates $x$ and $y$ shown in Fig.~\ref{fig:emean}.

While the VE implementation agrees with the simulation response in dataset 2, the prediction for dataset 3 is shifted as seen from the layer-dependent distribution. In the case of the WGAN-GP, the predictions are also shifted, but in the opposite direction, predicting a higher energy fraction at the initial layers of the detector.

\begin{figure*}[ht]
\centering
\includegraphics[width=0.3\textwidth]{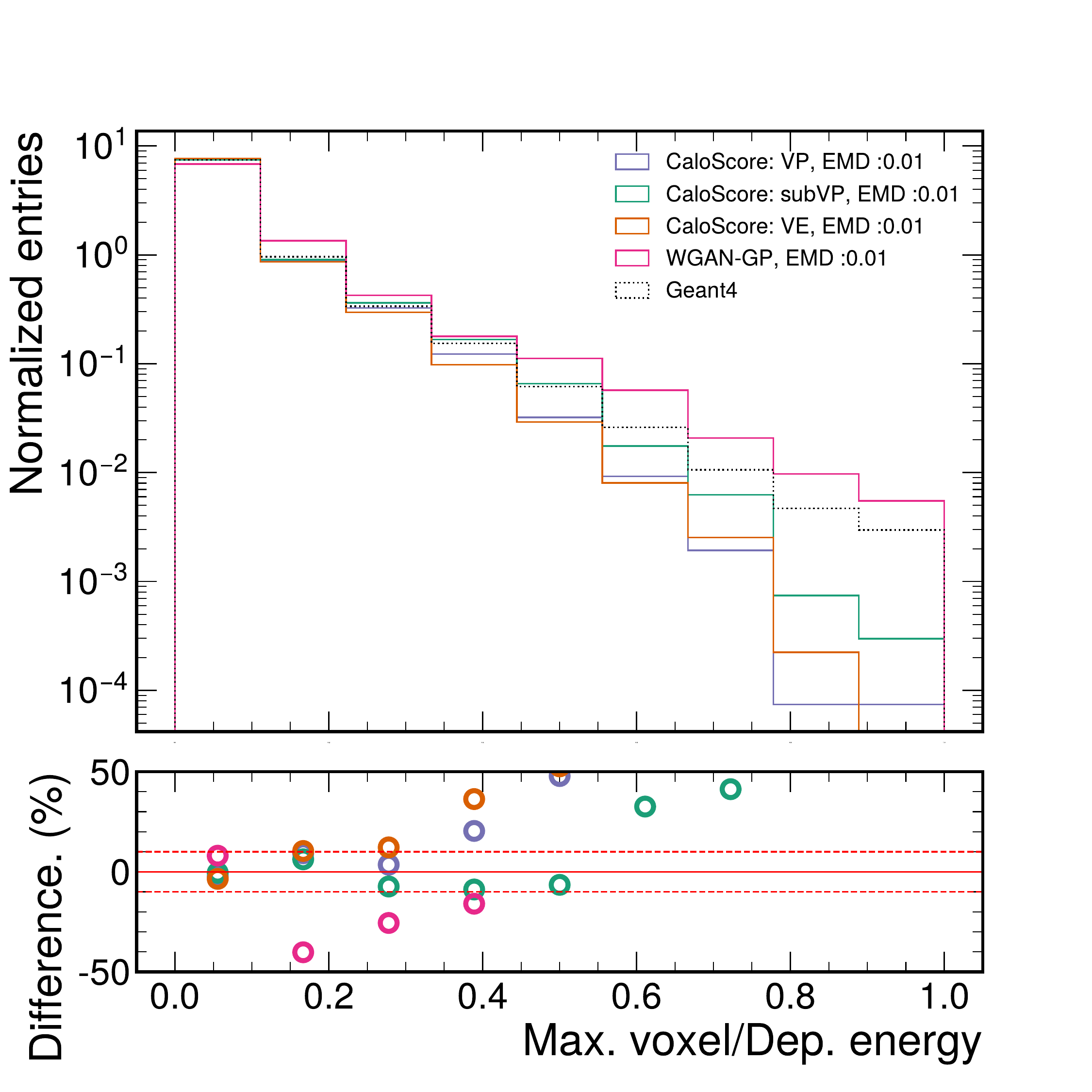}
\includegraphics[width=0.3\textwidth]{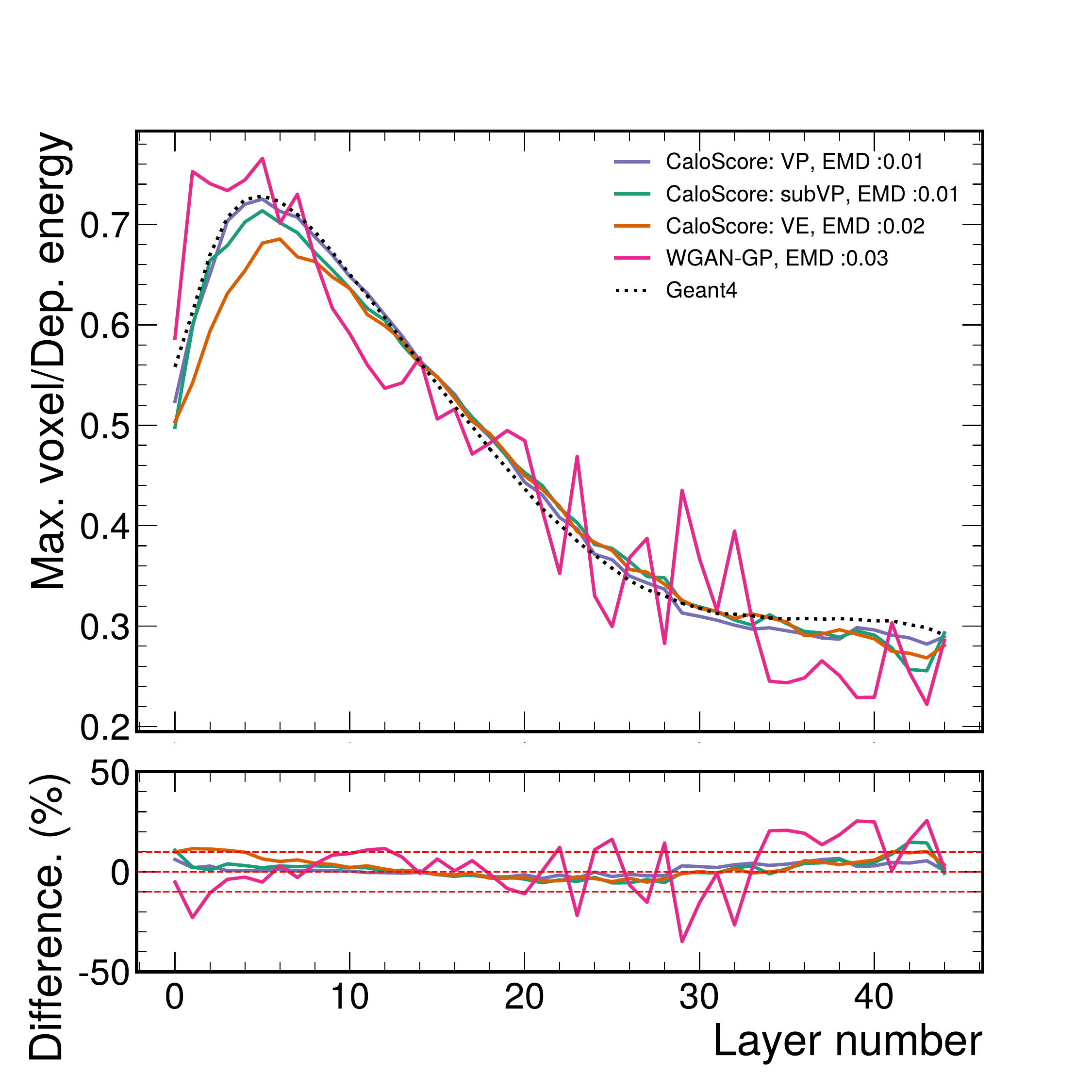}
\includegraphics[width=0.3\textwidth]{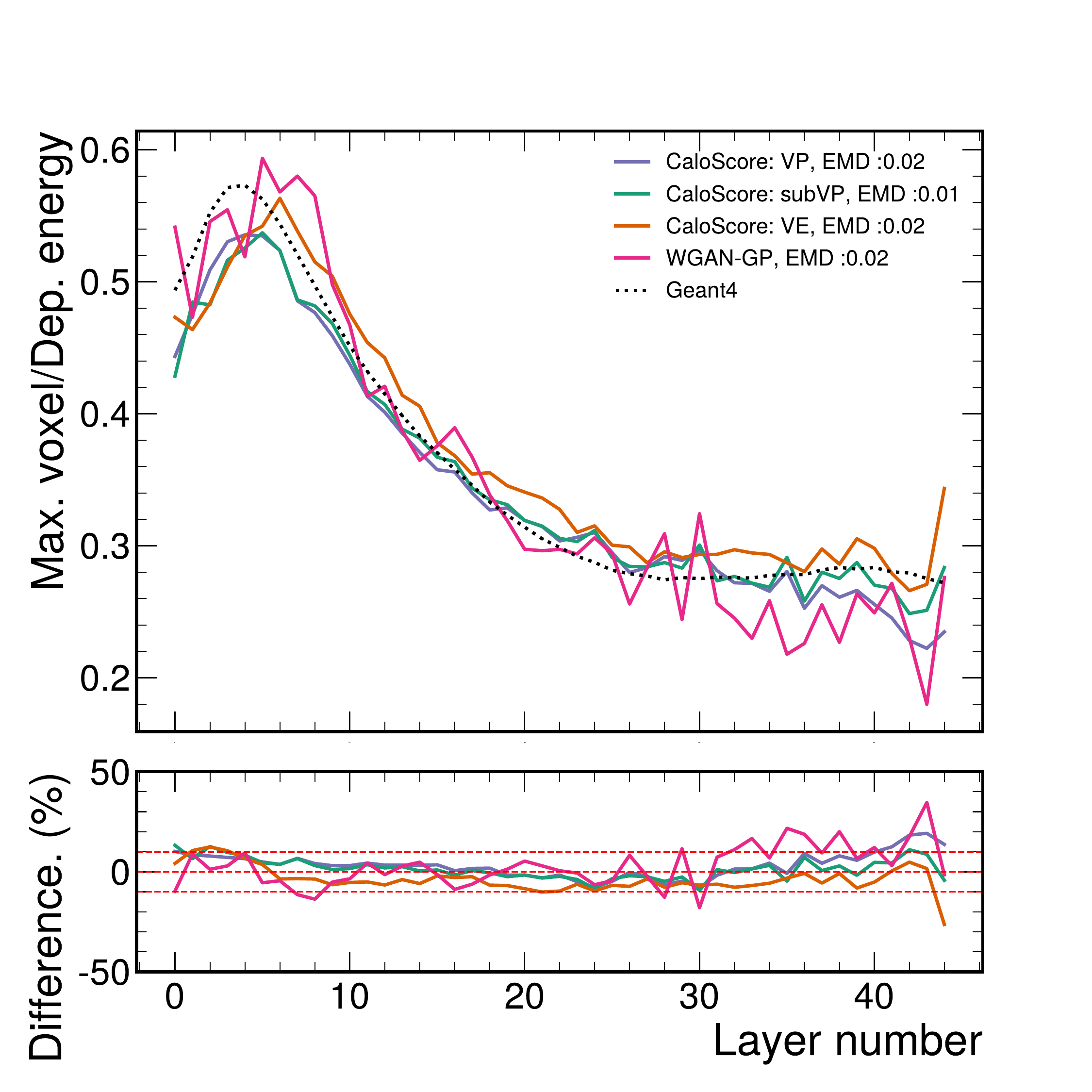}

\caption{Comparison of the maximum deposited energy in a single voxel divided by the sum of deposited energies in datasets 1 (left), 2 (middle), and 3 (right). Dashed red bands represent the 10\% deviation interval of the generated samples when compared to \geant~predictions. The earth mover's distance (EMD) between each distribution and the \geant~distribution is also provided.}
\label{fig:max_e}
\end{figure*}

Similarly, the maximum energy deposited in a single voxel normalized to the total deposited energy is shifted in the VE implementation for dataset 3 as shown in Fig. \ref{fig:max_e}. While the low energy fraction region for dataset 1 is well described by all \calosc~implementations, the high energy fraction region starts to show deviations from the \geant~predictions and is best described by the WGAN-GP model. The maximum energy fraction as a function of the layer number for datasets 2 and 3 shows a good agreement between the different \calosc~implementations, with most of the distributions showing deviations within the 10\% interval. 

\begin{figure*}[ht]
\centering
\includegraphics[width=0.22\textwidth]{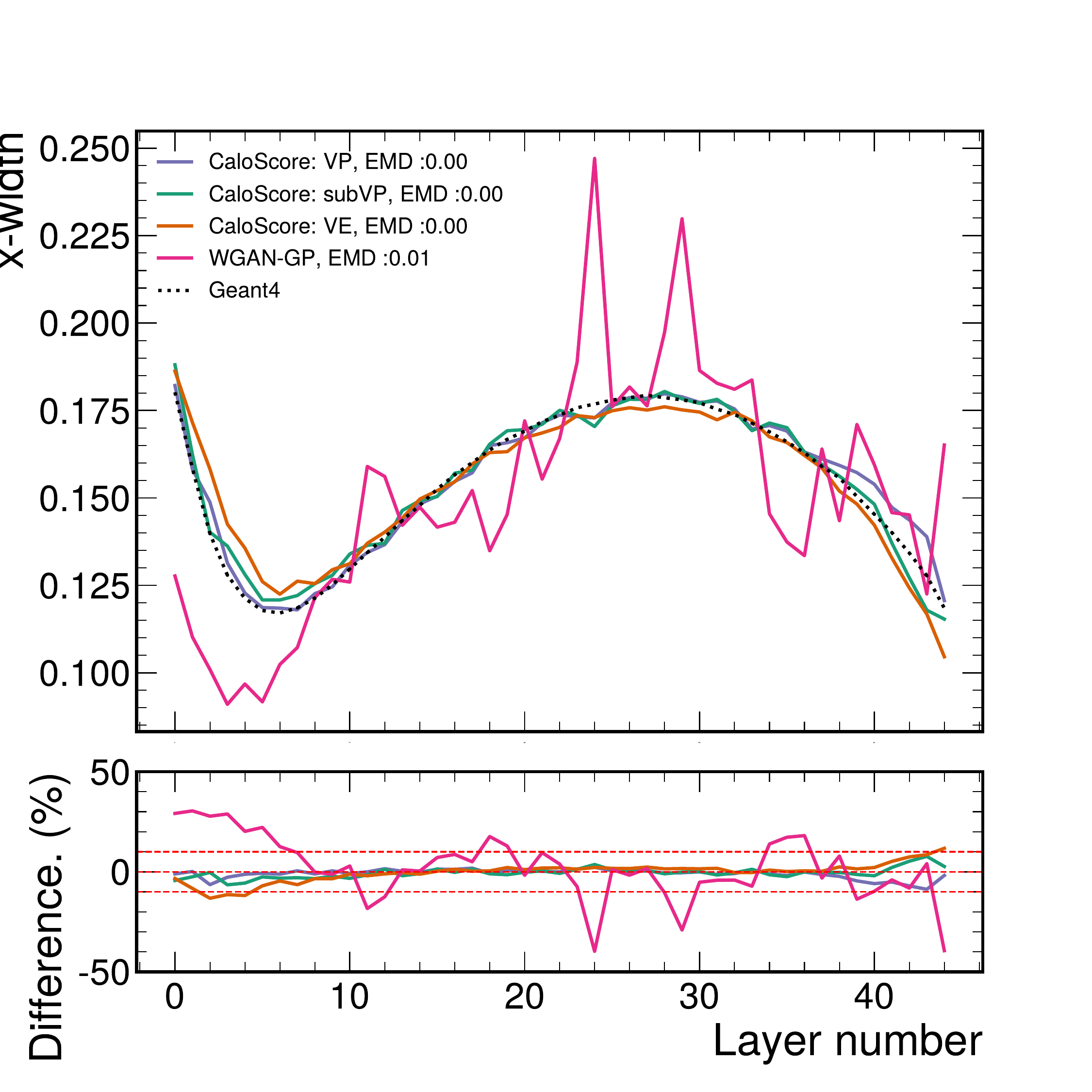}
\includegraphics[width=0.22\textwidth]{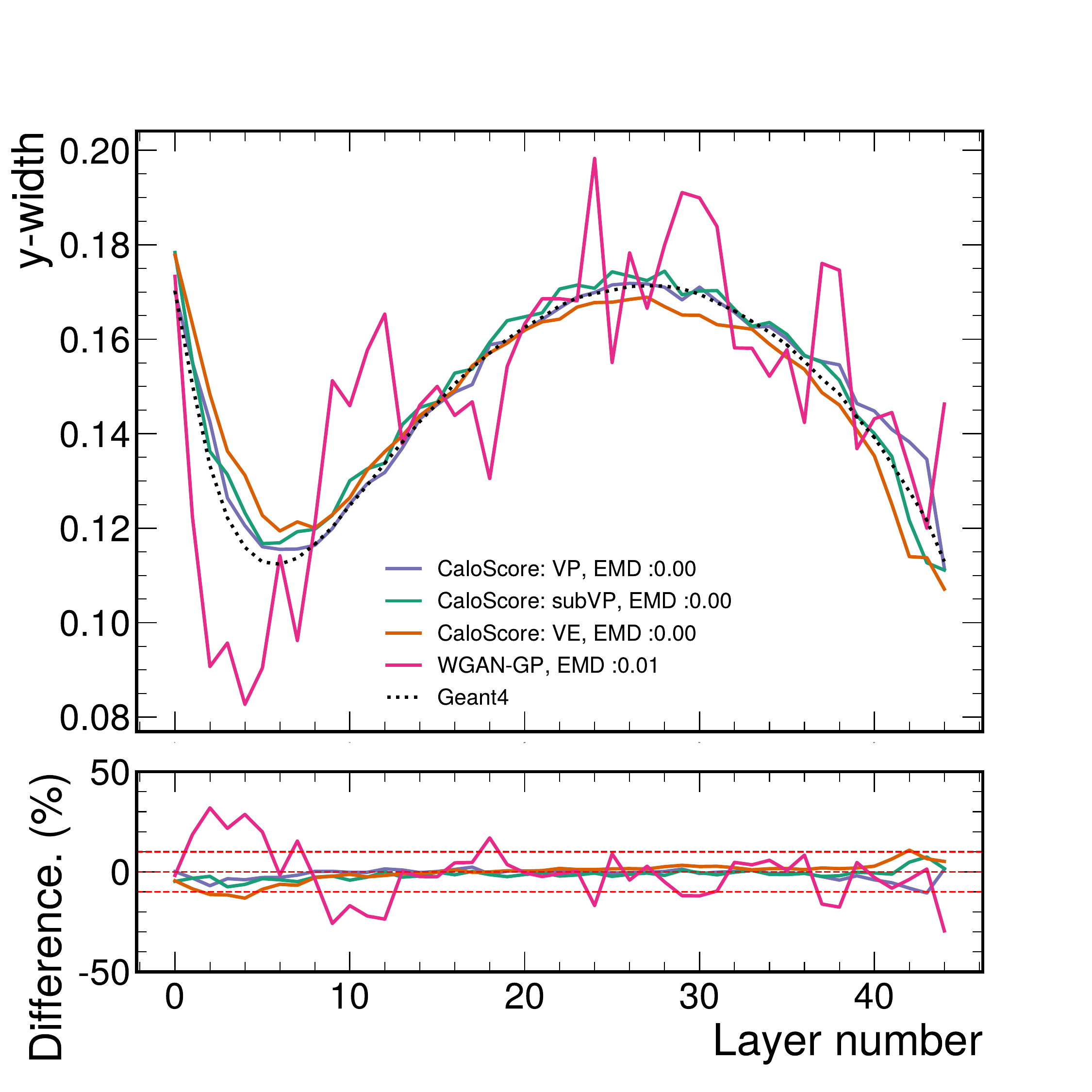}
\includegraphics[width=0.22\textwidth]{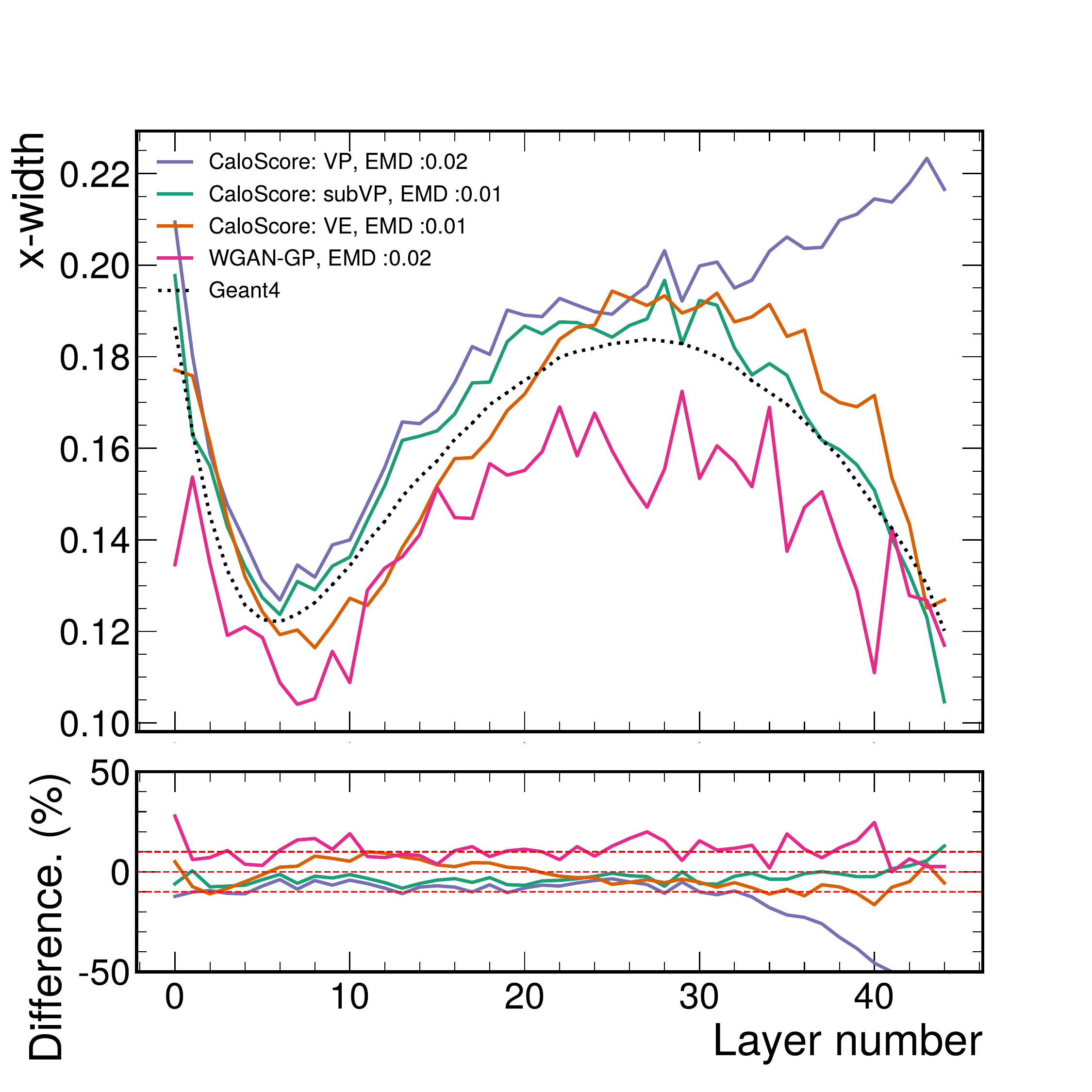}
\includegraphics[width=0.22\textwidth]{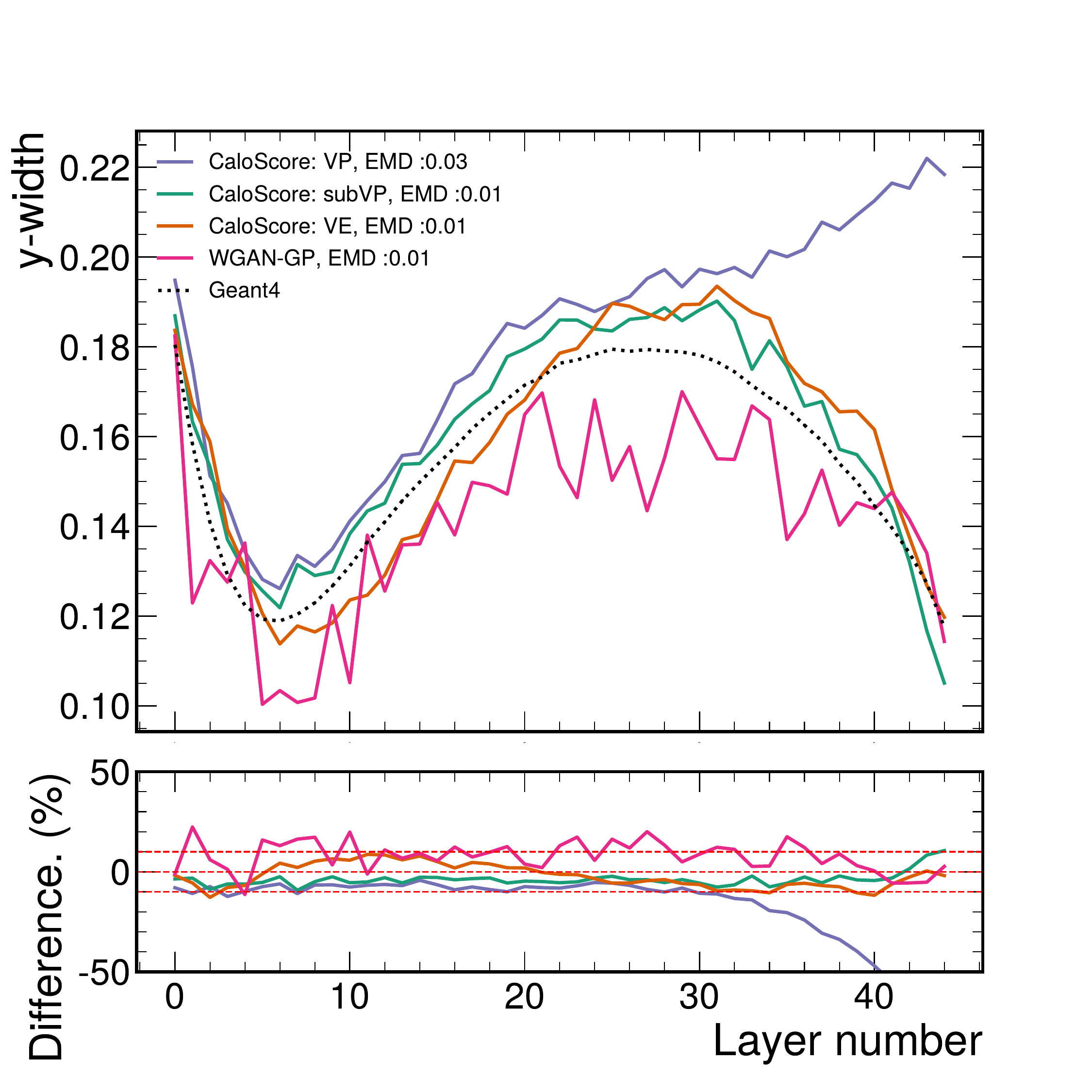}

\caption{Comparison of the particle shower width in the x-  and y-  directions in datasets 2 (first two figures from the left) and 3 (last two figures from the left). Dashed red bands represent the 10\% deviation interval of the generated samples when compared to \geant~predictions. The earth mover's distance (EMD) between each distribution and the \geant~distribution is also provided.}
\label{fig:shower_w}
\end{figure*}

The angular distribution of the calorimeter shower is investigated in datasets 2 and 3 in terms of the shower width, shown in Fig.~\ref{fig:shower_w}.  The shower width $\sigma_i$ with $x_i, i\in [1,2]$ representing the x- and y- coordinates is calculated as: 
\begin{equation}
    \sigma_i = \sqrt{\left< x_i^2 \right > - \left< x_i \right >^2 },
\end{equation}
with energy-weighted mean defined as
\begin{equation}
    \left< x_i \right > =  \frac{\sum_j x_{i,j} E_j}{\sum_j E_j}.
\end{equation}

A good agreement between all \calosc~implementations and  \geant~predictions is observed in dataset 2, with all implementations showing less than 10\% deviation in all calorimeter layers. However, for dataset 3, the VP implementations shows a disagreement at the last layers of the detector while the shift observed in Fig.~\ref{fig:emean} for the VE implementation leads to a similar shift in the shower width. Nevertheless, the subVP implementation maintains the same level of agreement as observed in dataset 2. The WGAN-GP implementation however shows larger fluctuations compared to \calosc and consistently smaller shower widths in dataset 3.

\begin{figure*}[ht]
\centering
\includegraphics[width=0.18\textwidth]{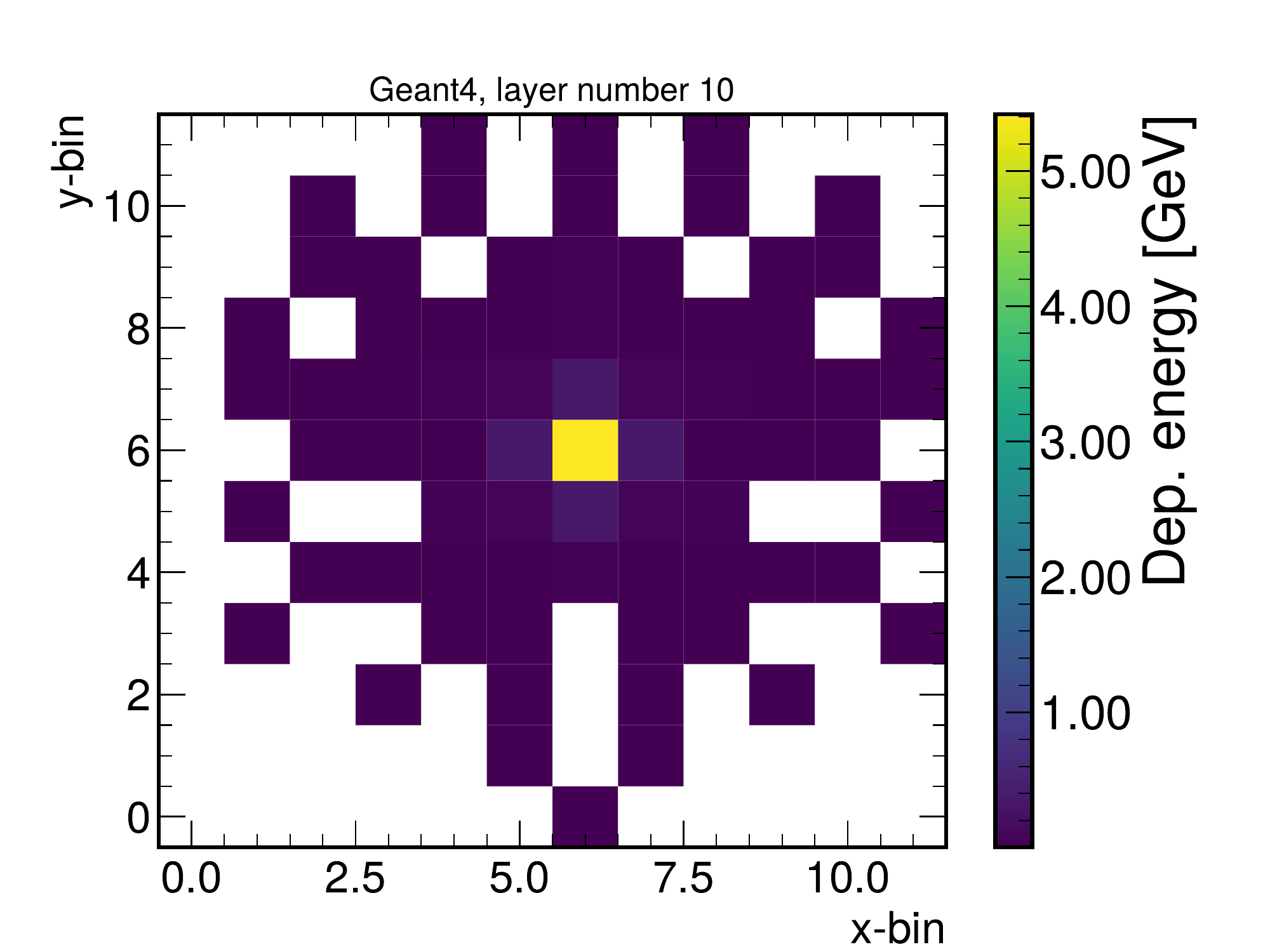}
\includegraphics[width=0.18\textwidth]{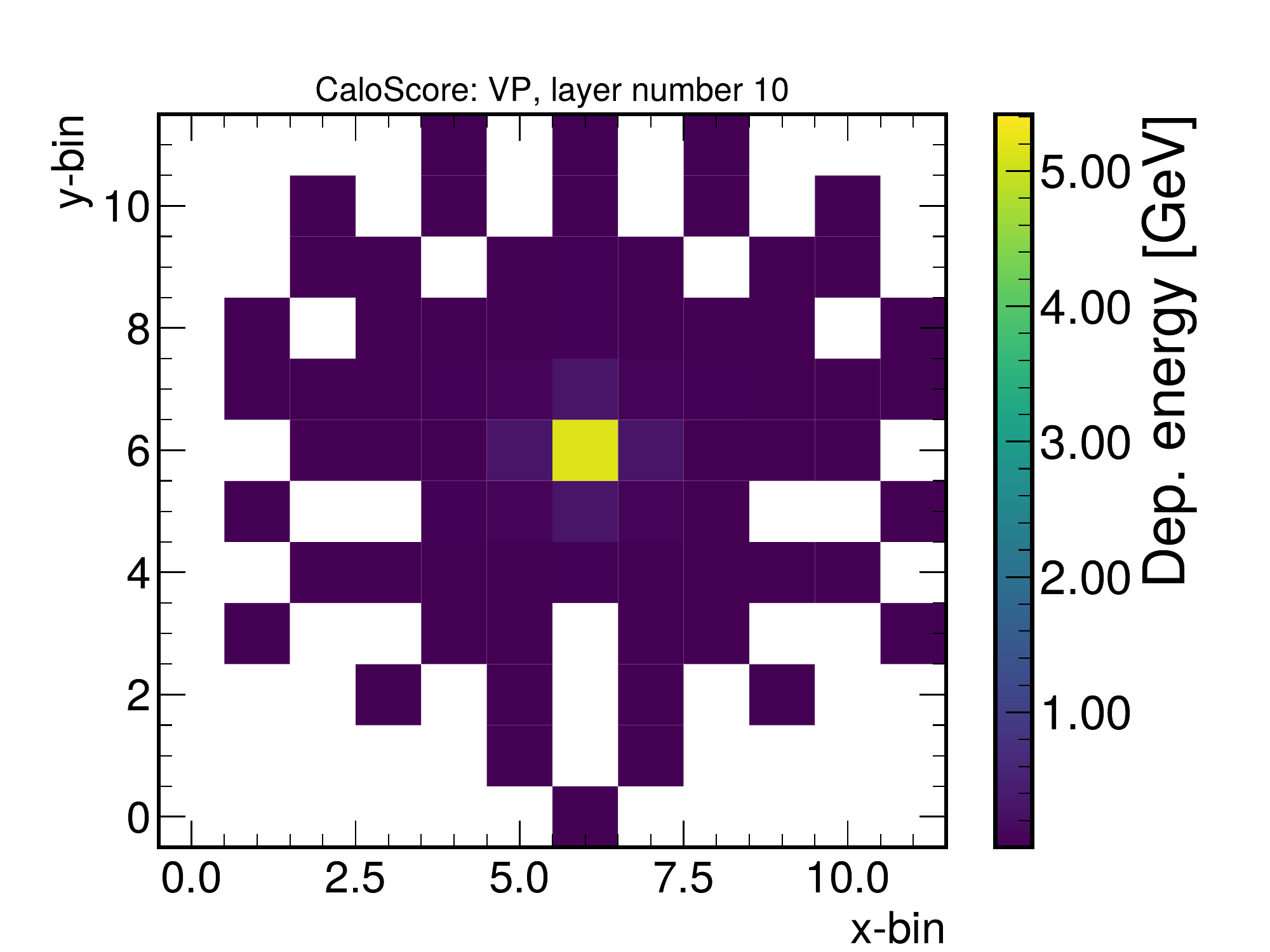}
\includegraphics[width=0.18\textwidth]{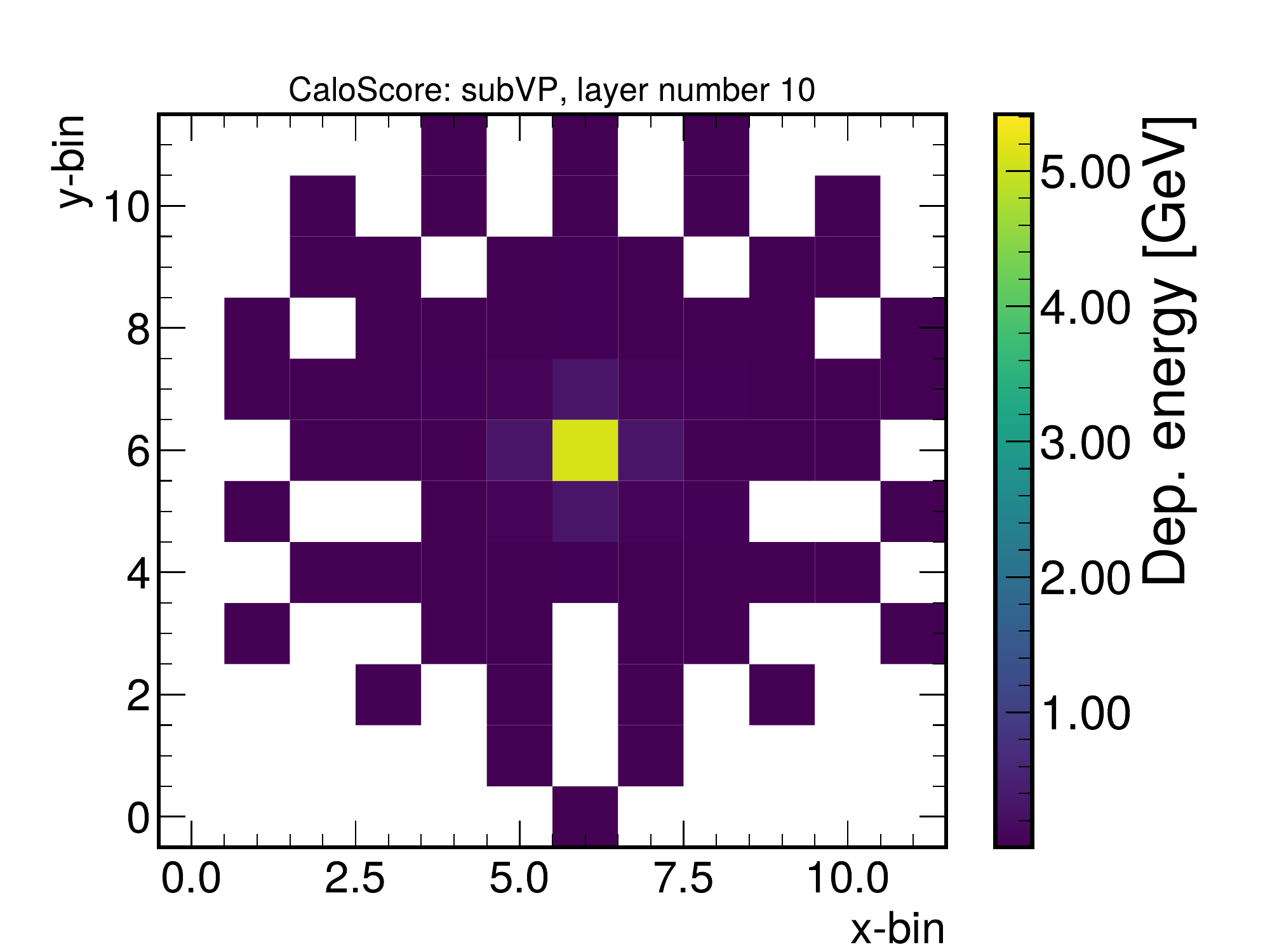}
\includegraphics[width=0.18\textwidth]{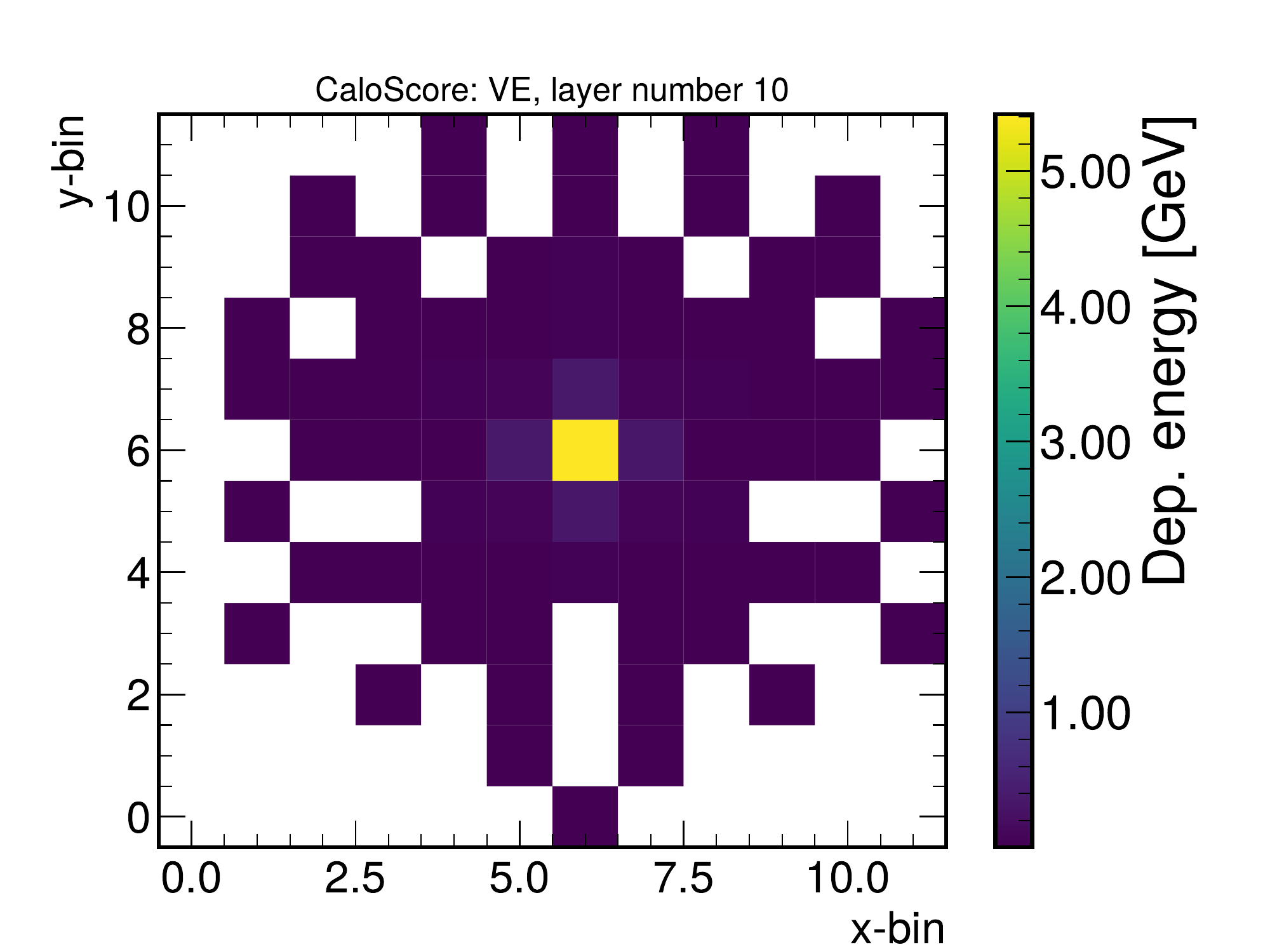}
\includegraphics[width=0.18\textwidth]{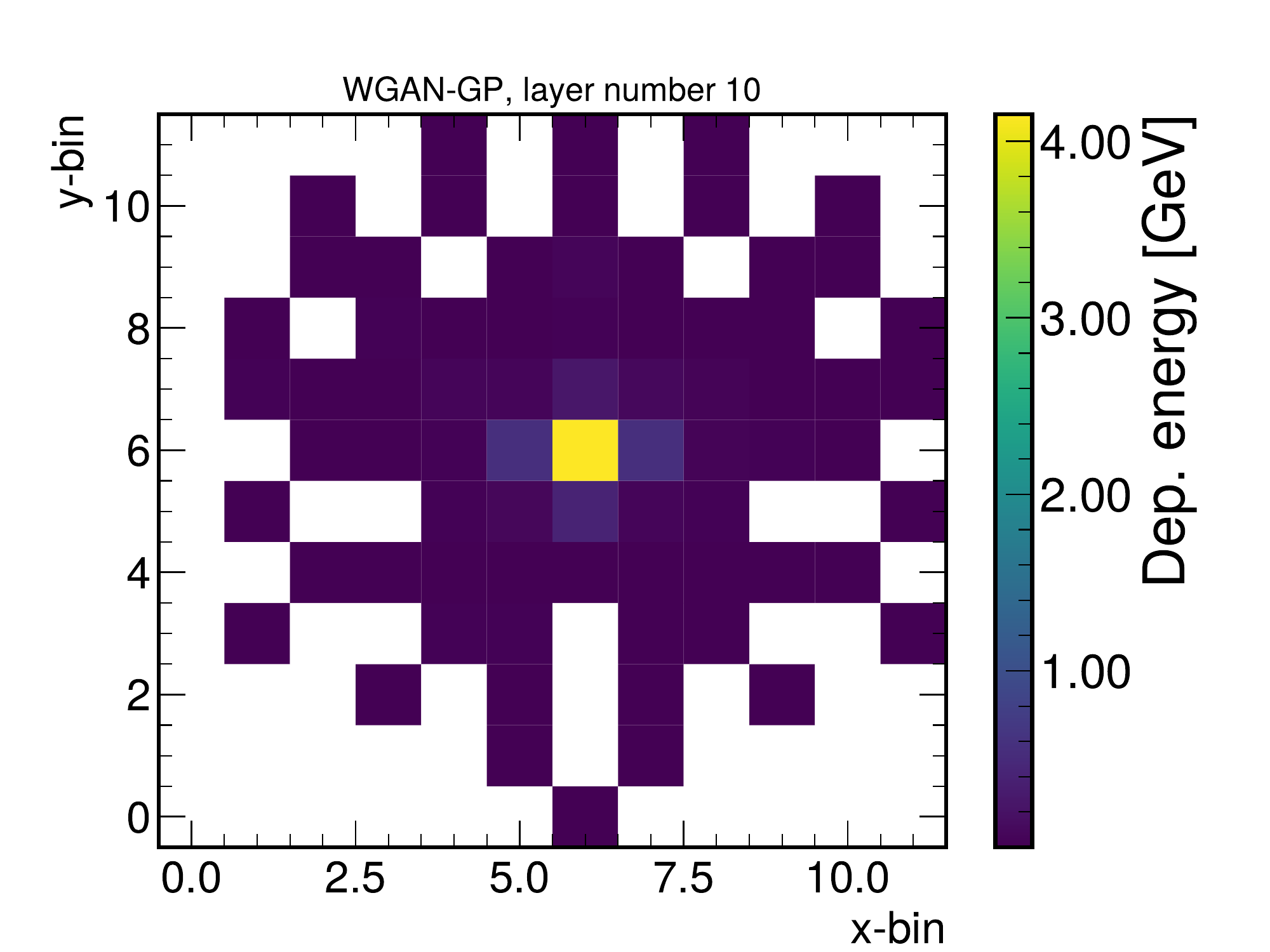}
\includegraphics[width=0.18\textwidth]{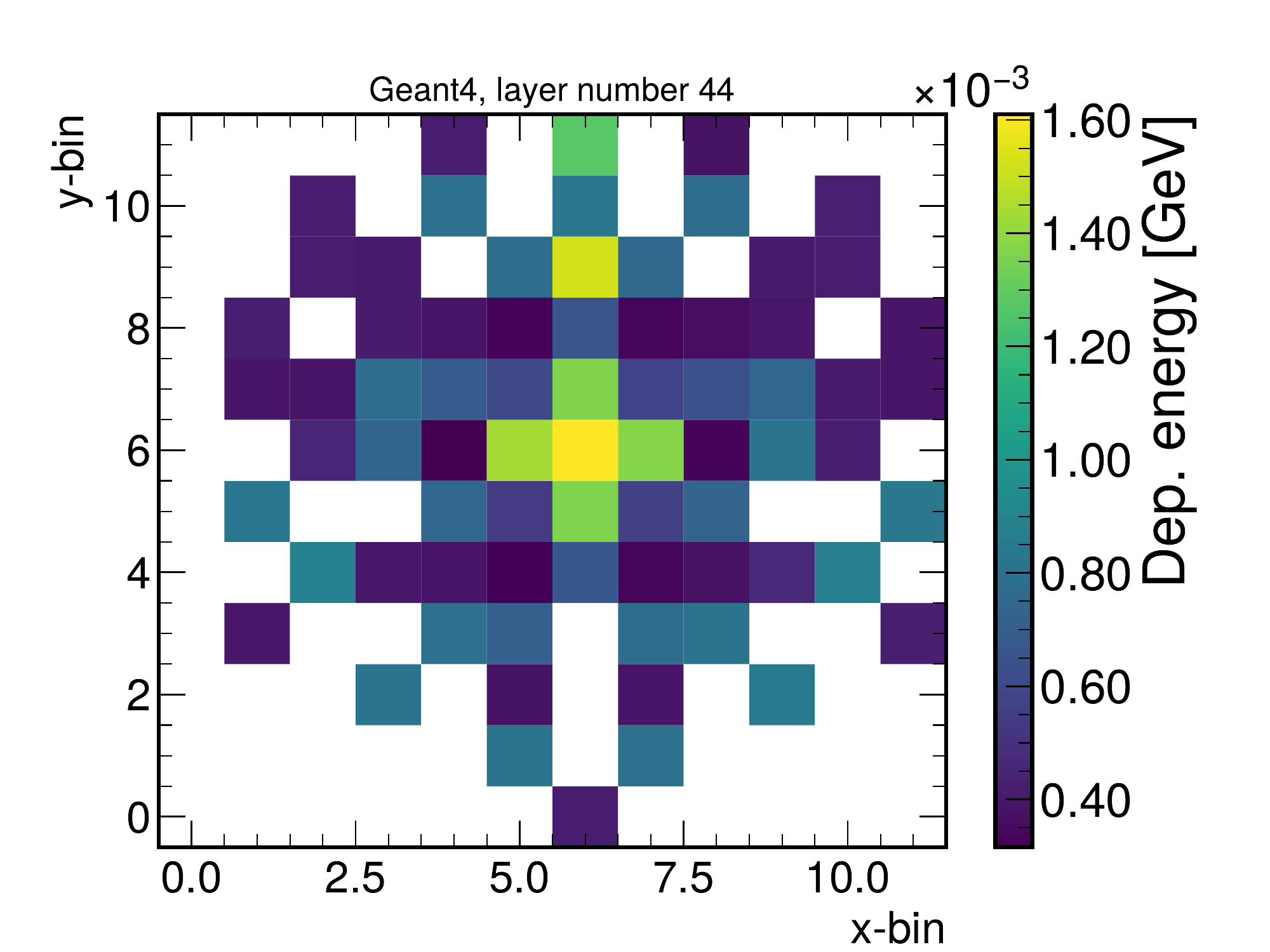}
\includegraphics[width=0.18\textwidth]{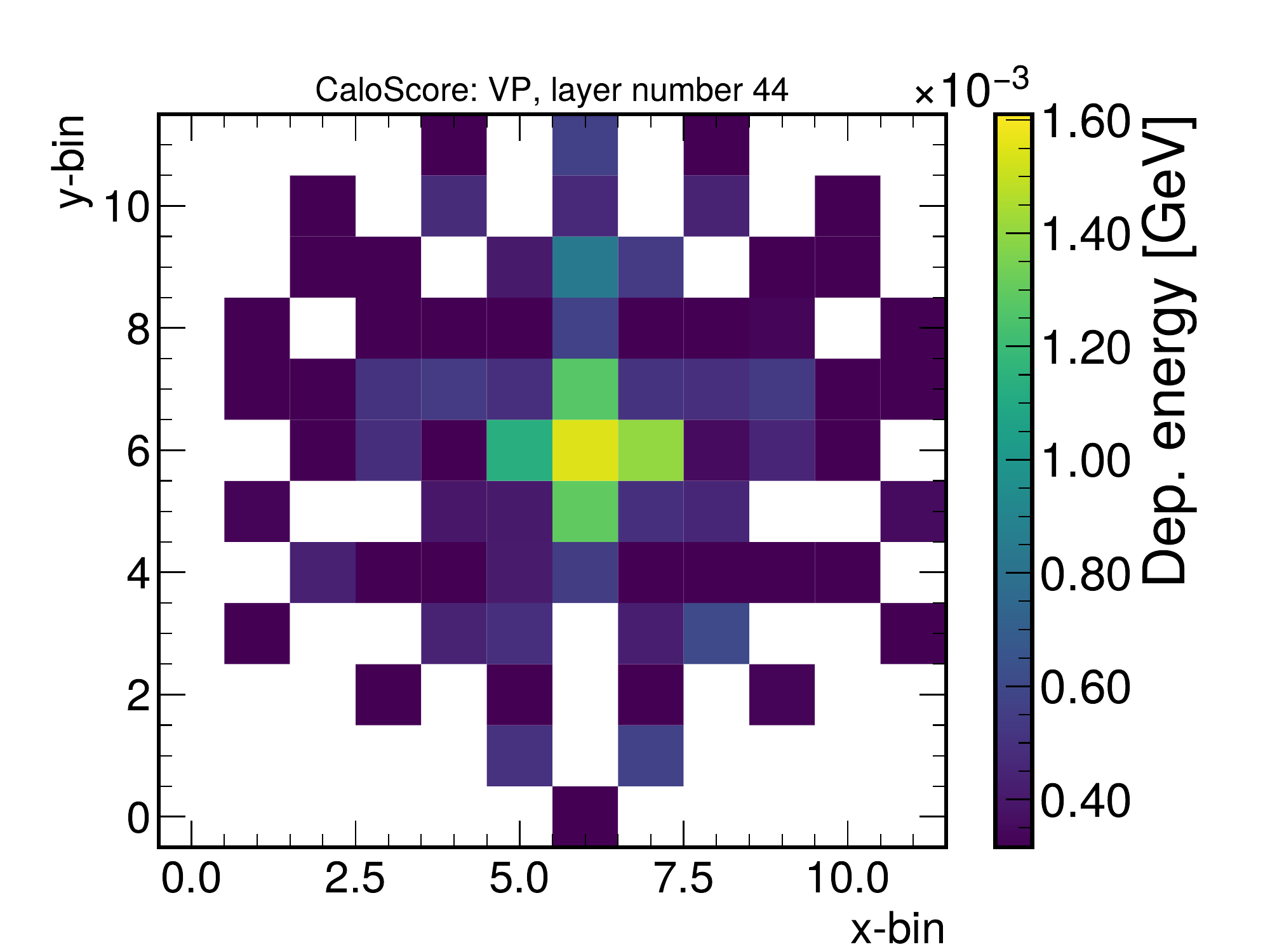}
\includegraphics[width=0.18\textwidth]{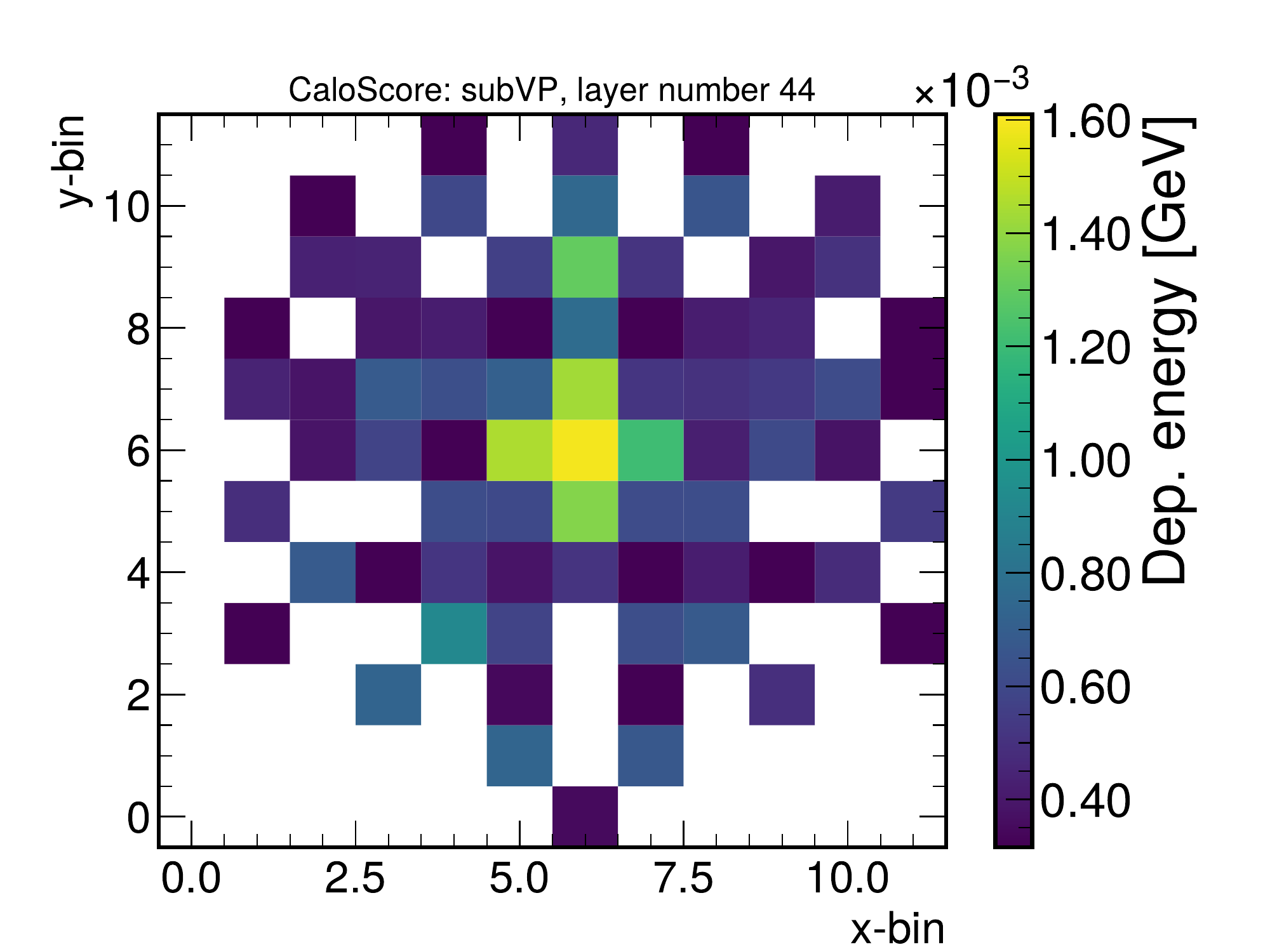}
\includegraphics[width=0.18\textwidth]{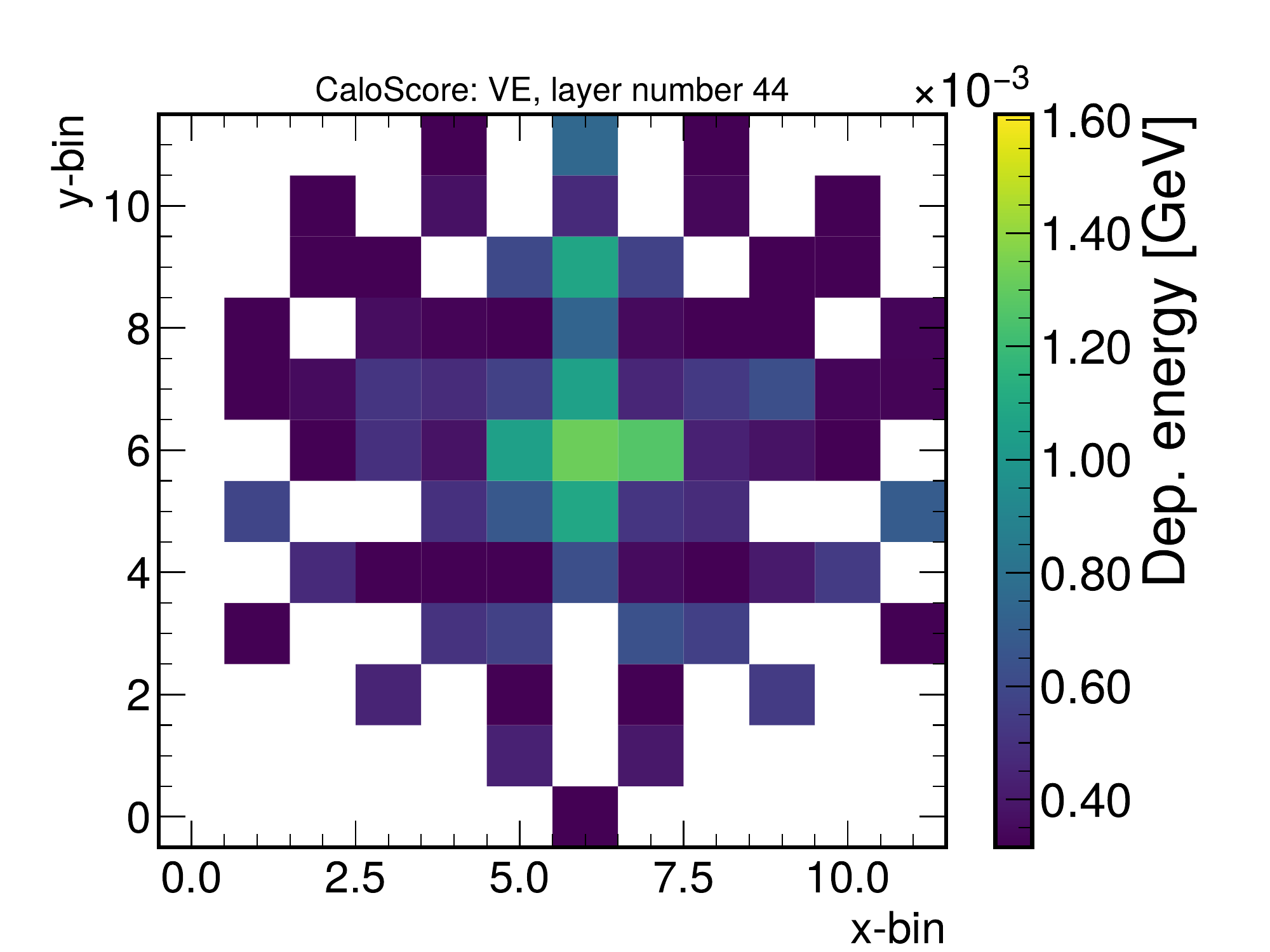}
\includegraphics[width=0.18\textwidth]{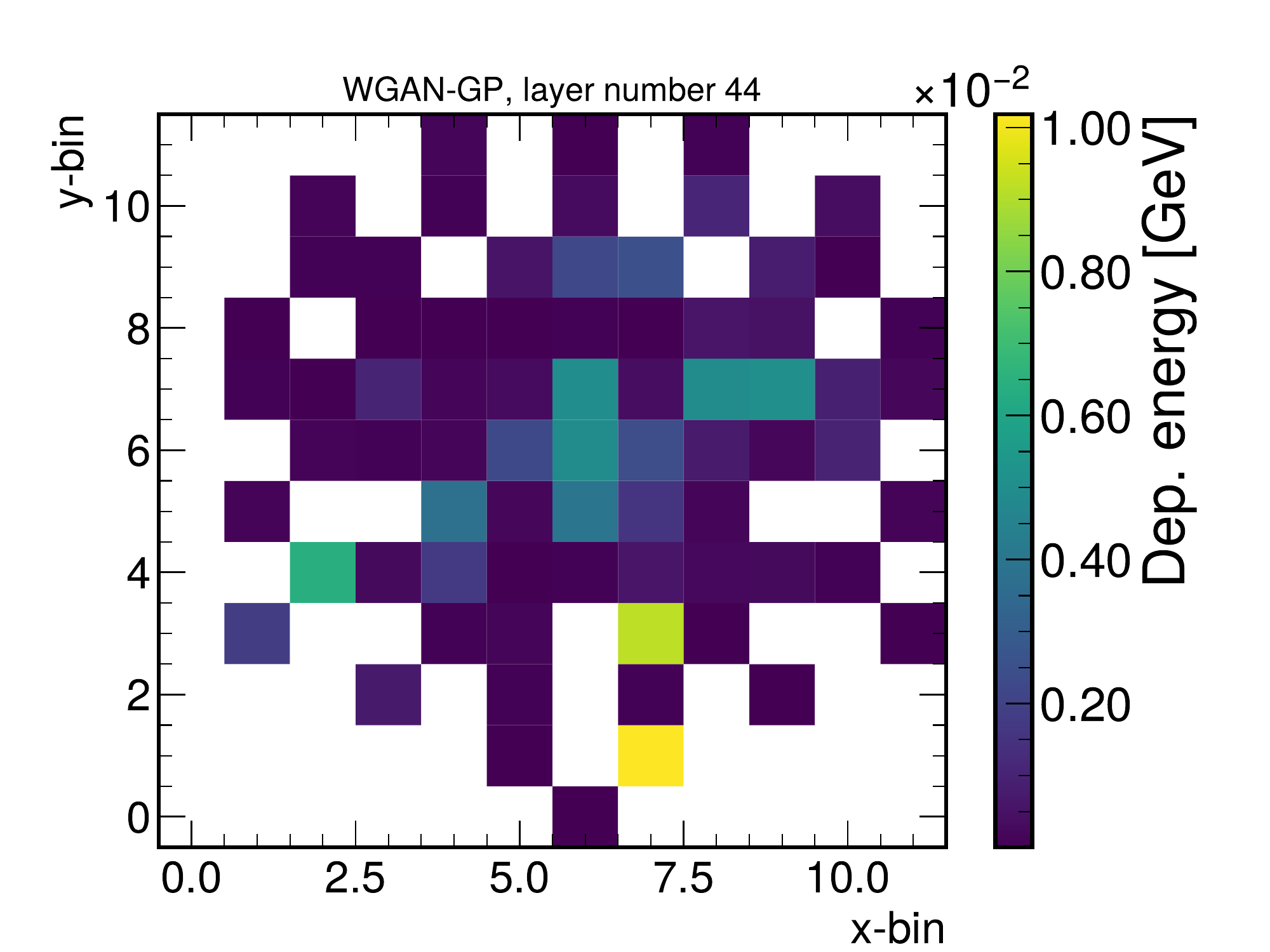}
\includegraphics[width=0.18\textwidth]{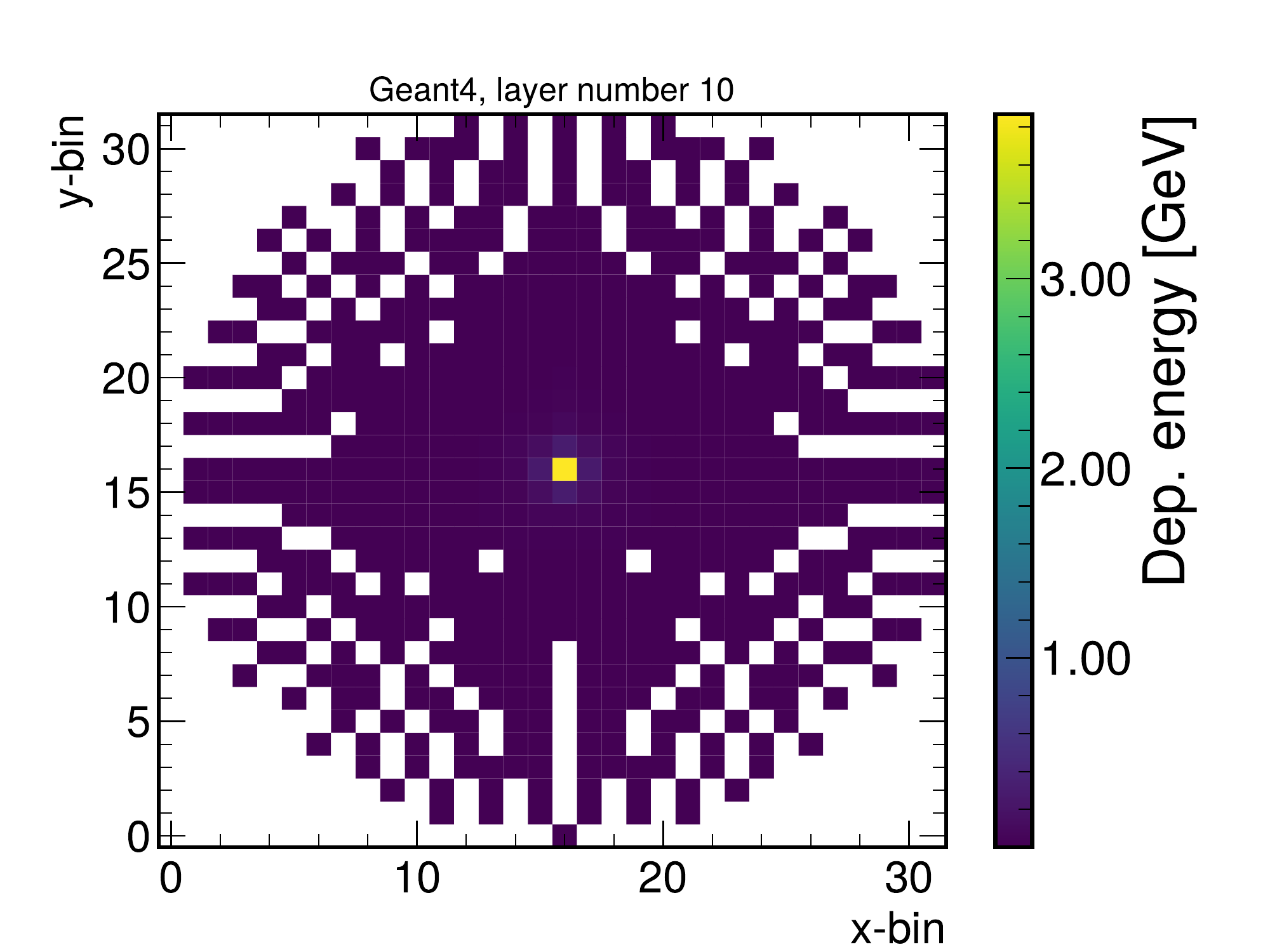}
\includegraphics[width=0.18\textwidth]{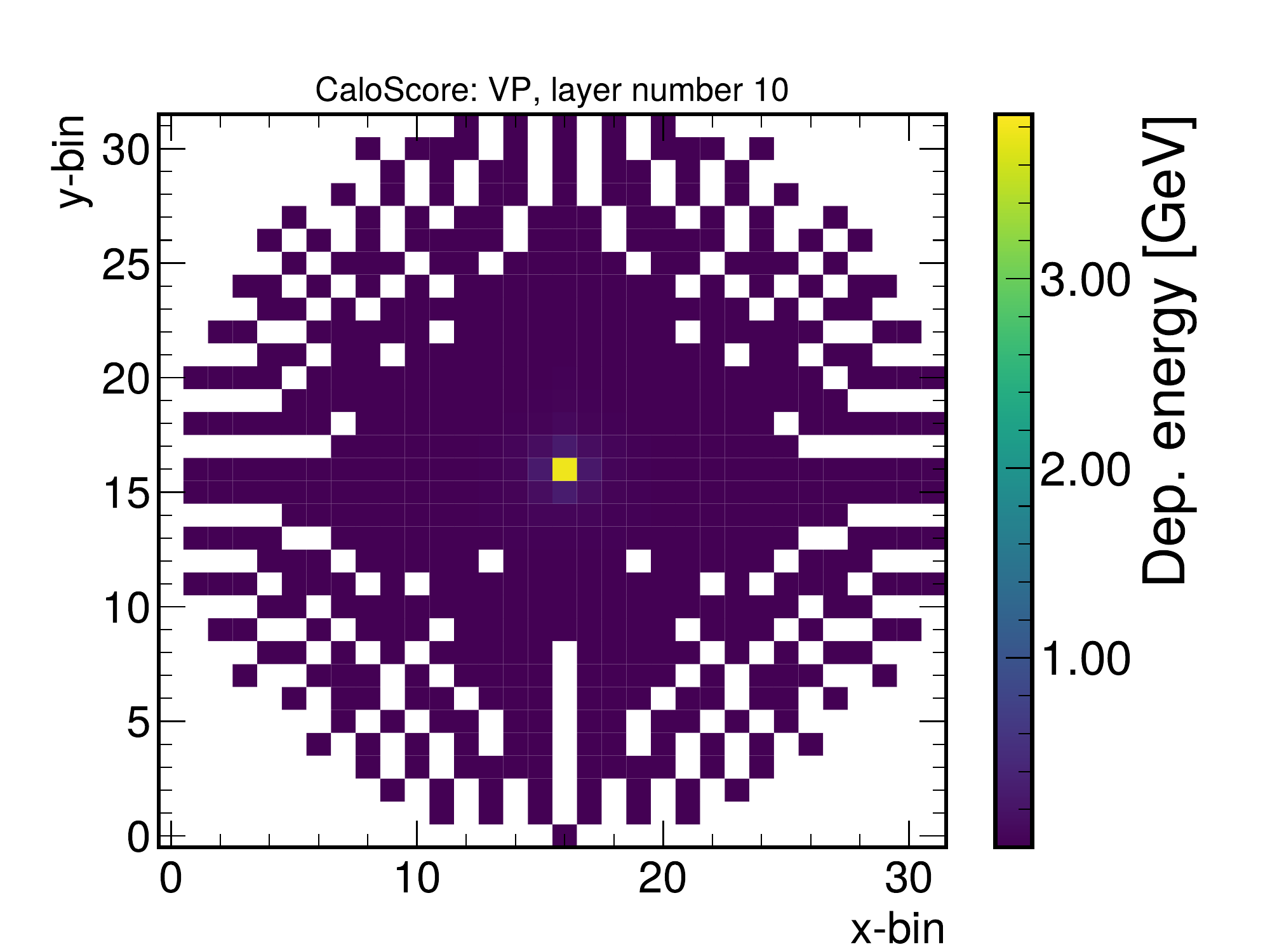}
\includegraphics[width=0.18\textwidth]{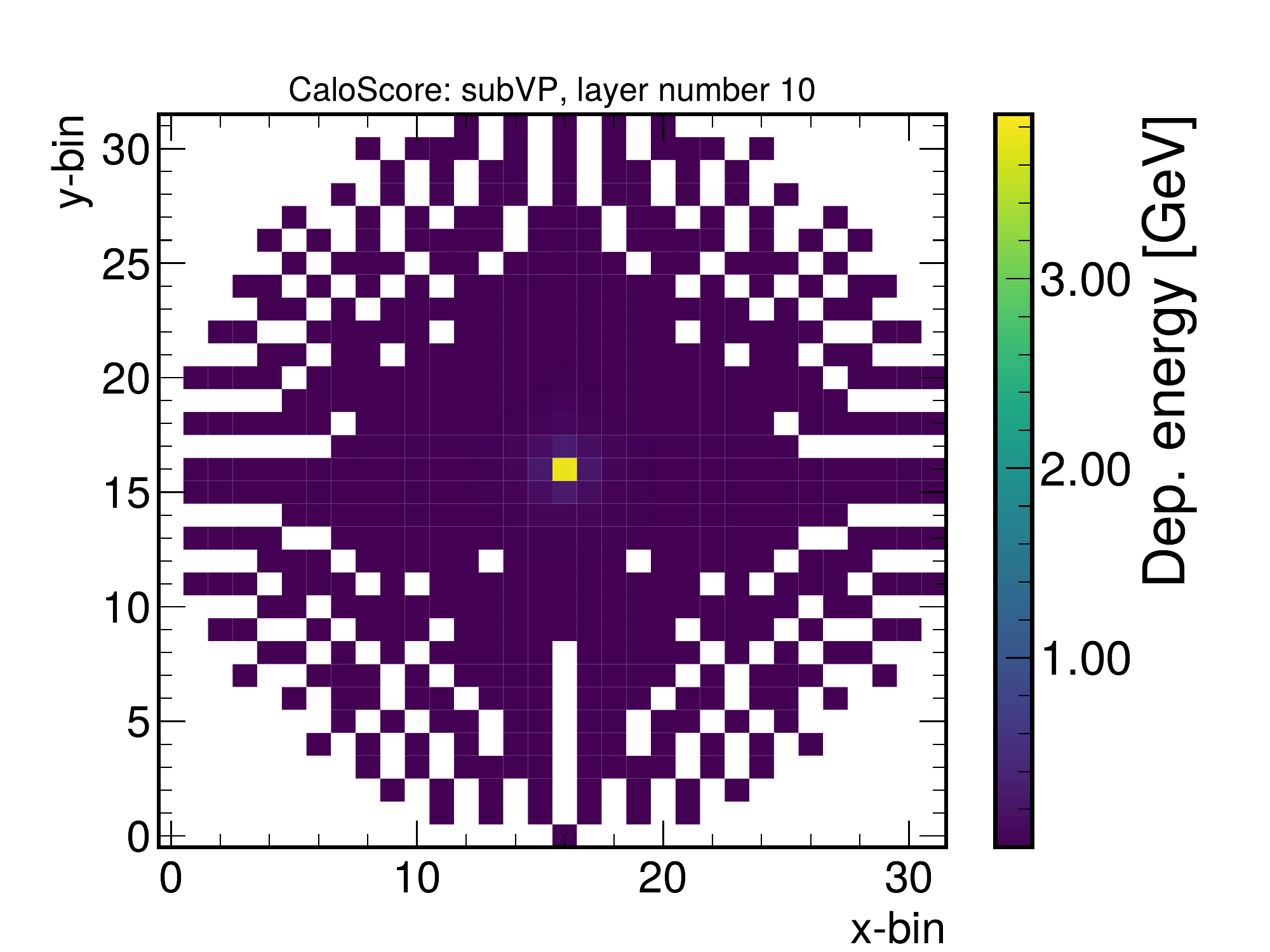}
\includegraphics[width=0.18\textwidth]{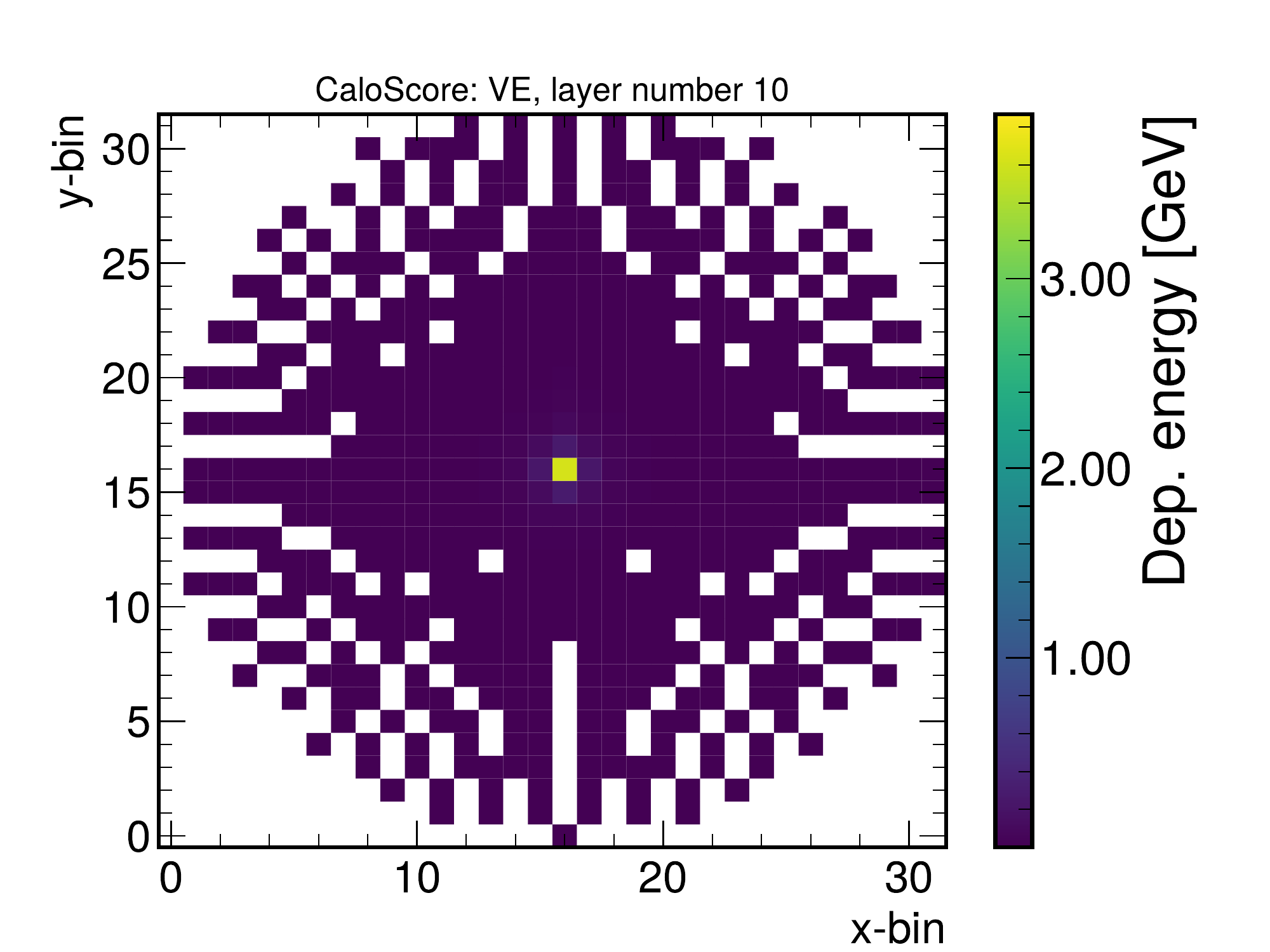}
\includegraphics[width=0.18\textwidth]{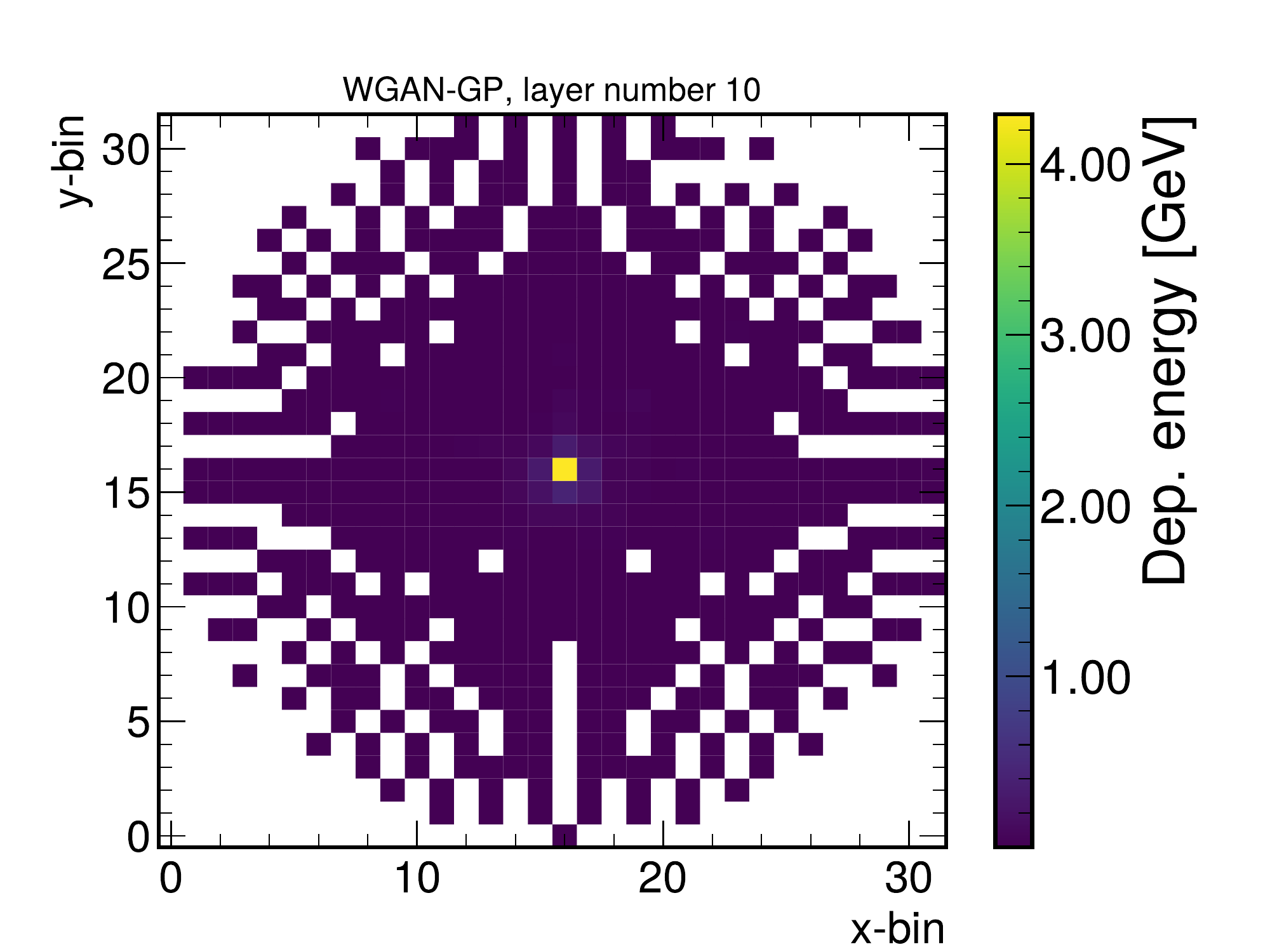}
\includegraphics[width=0.18\textwidth]{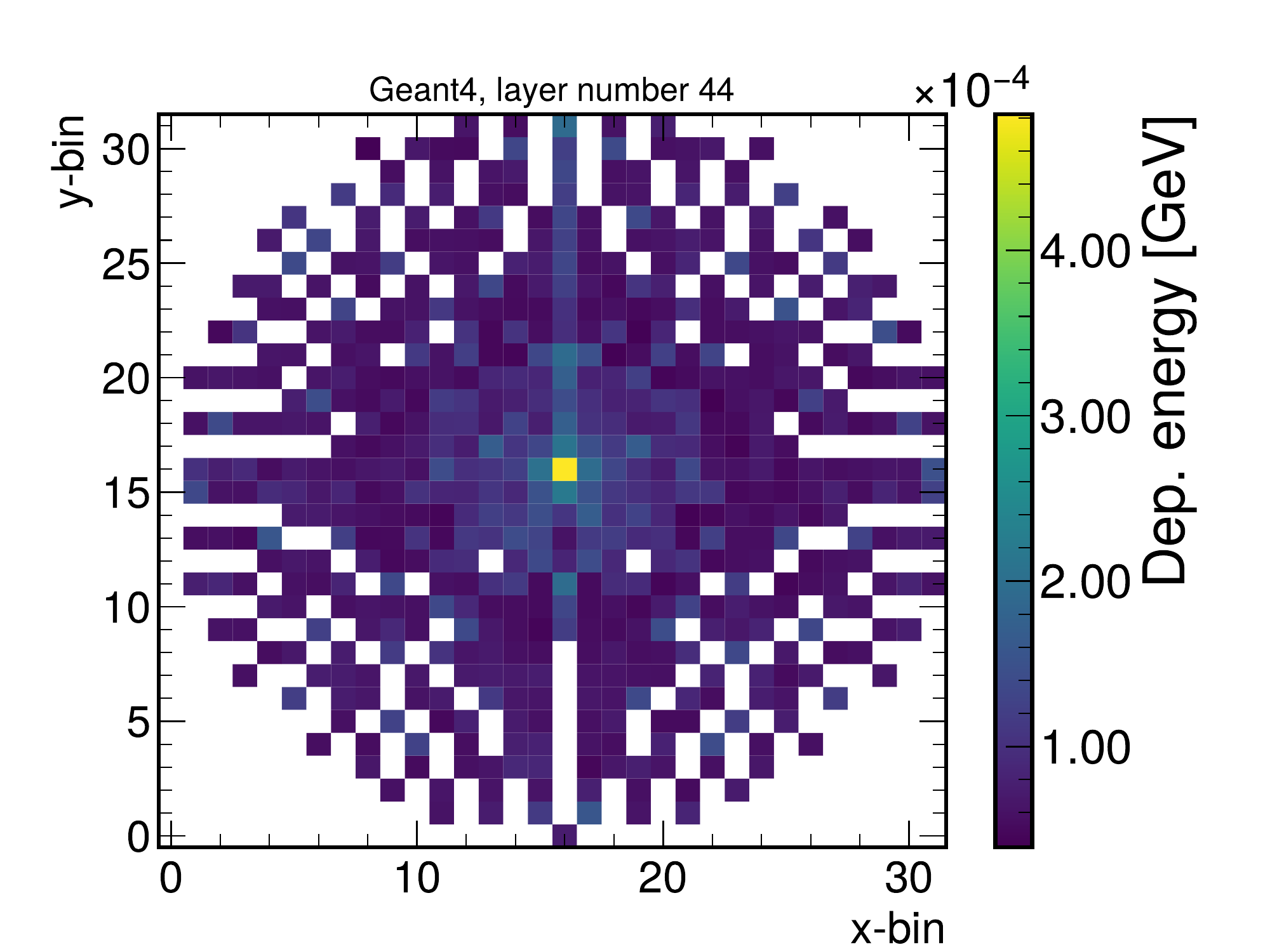}
\includegraphics[width=0.18\textwidth]{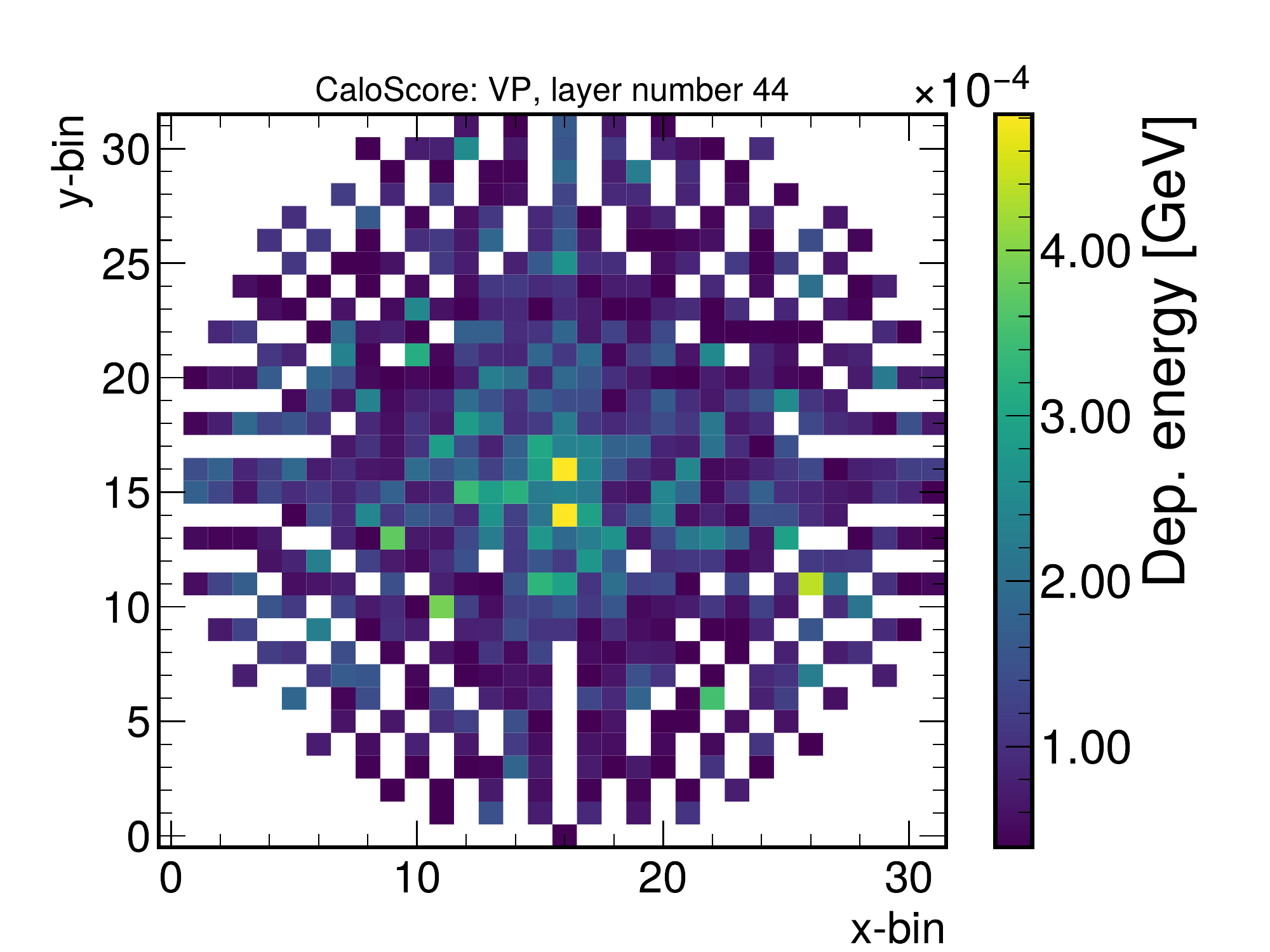}
\includegraphics[width=0.18\textwidth]{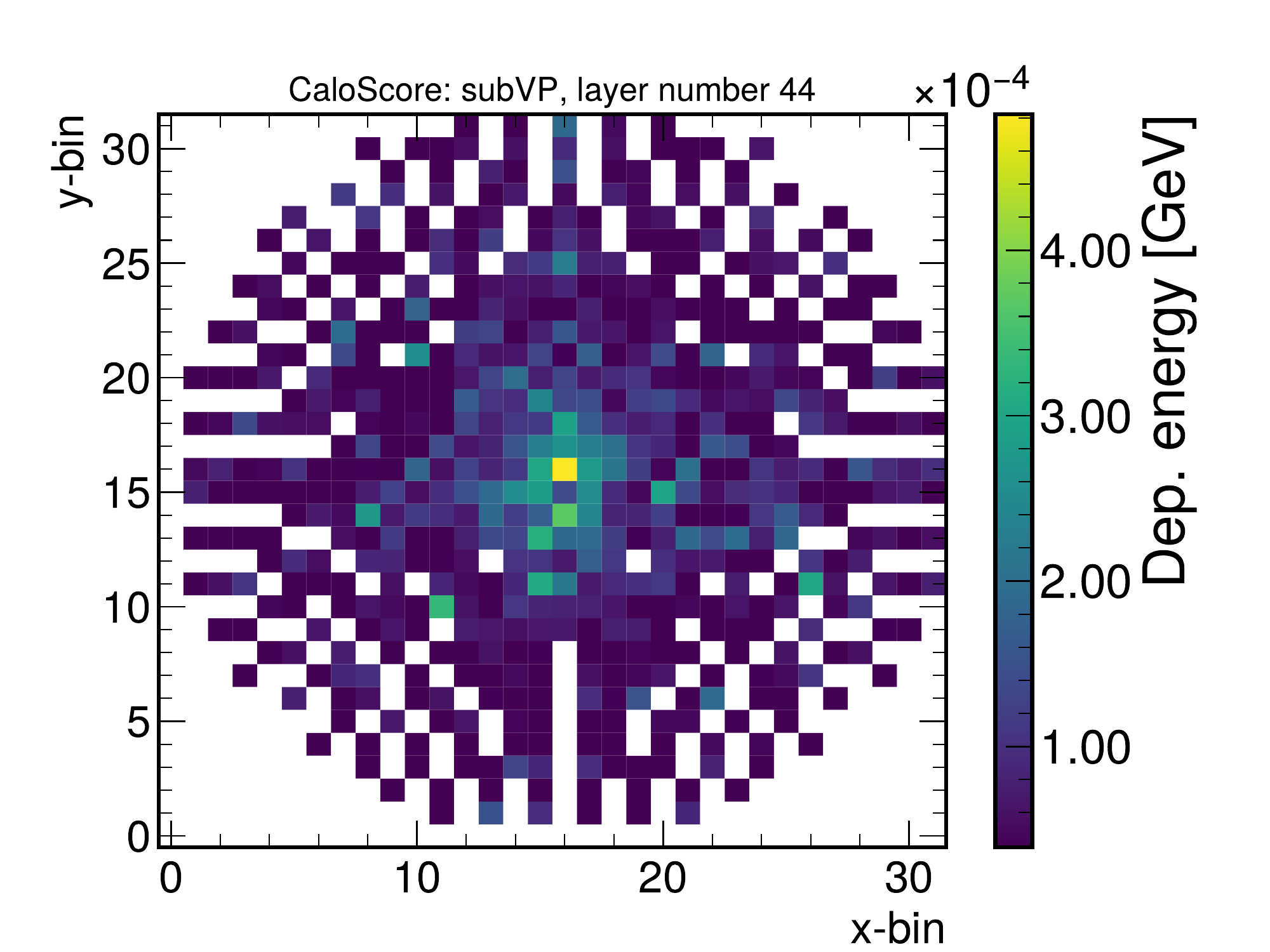}
\includegraphics[width=0.18\textwidth]{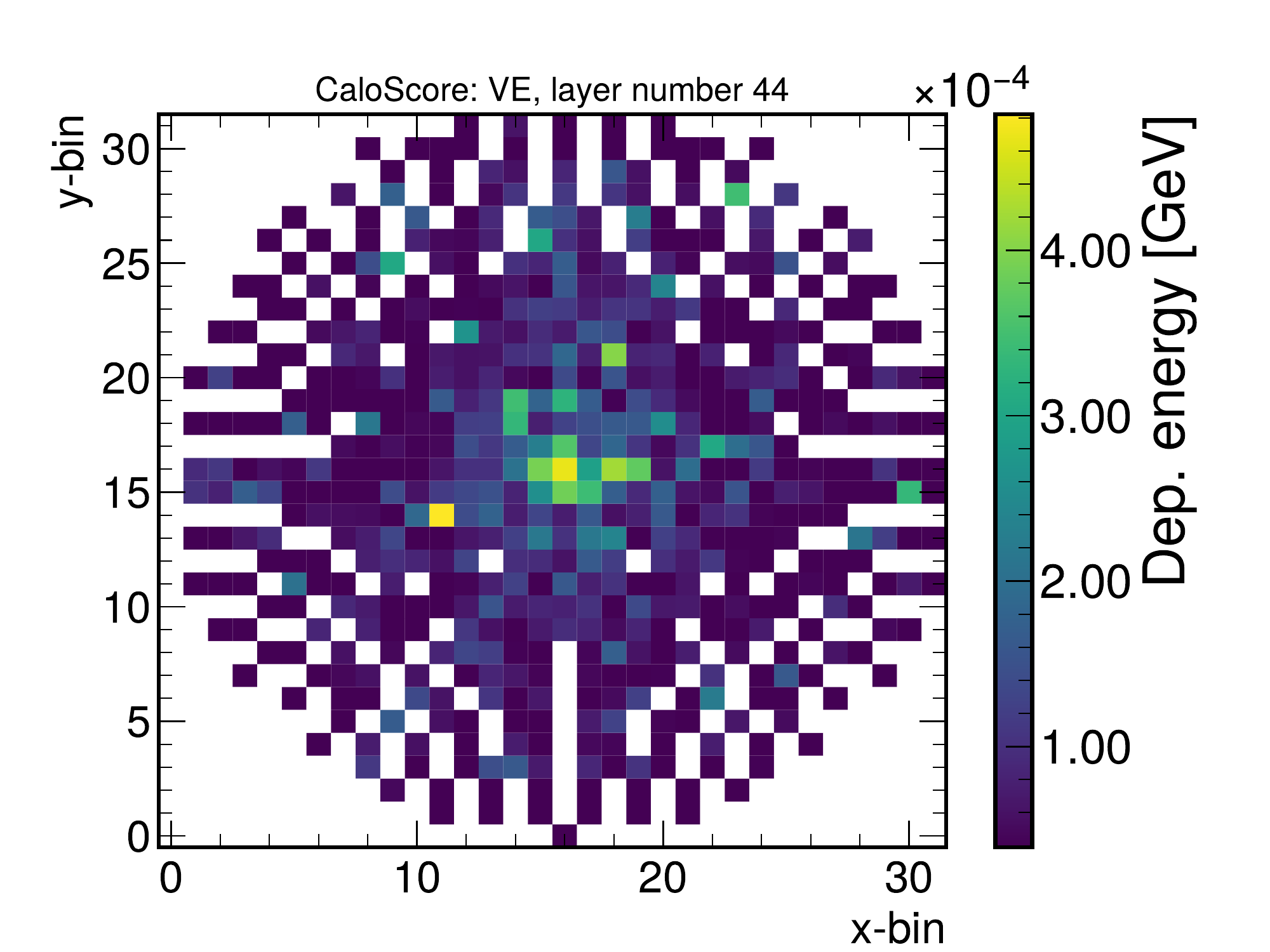}
\includegraphics[width=0.18\textwidth]{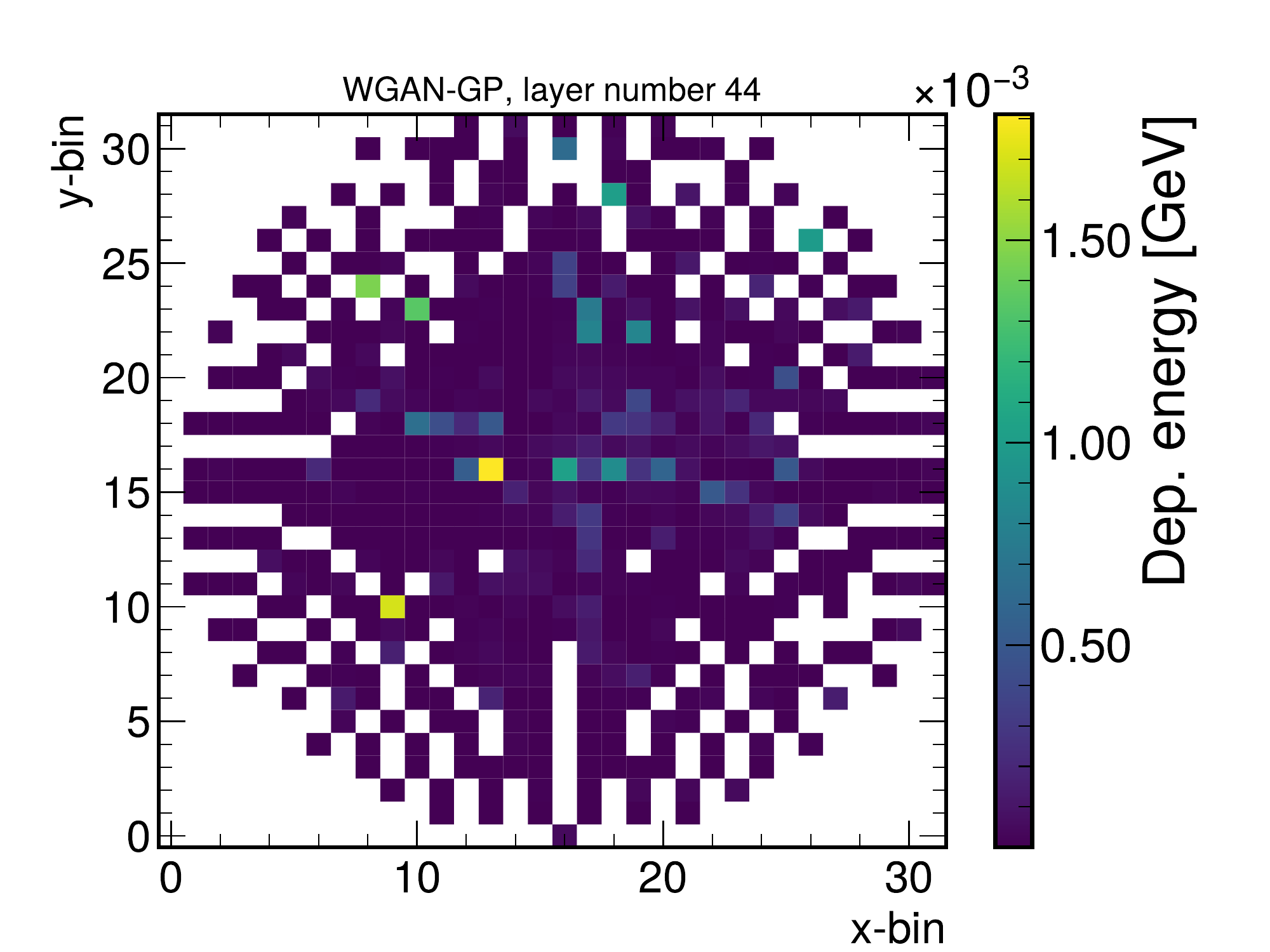}

\caption{The 2-dimensional distribution of the mean deposited energy in layers with highest (first and third rows) and lowest (second and fourth rows) mean energy depositions in datasets 2 (first two rows) and 3 (last two rows). Simulated samples from \geant~are shown in the first column, compared with different diffusion models: VP (second column), subVP(third column), and VE (fourth column). The WGAN-GP results are shown in the fourth column.}
\label{fig:2D_showers}
\end{figure*}

A qualitative assessment of the generation is shown in Fig.~\ref{fig:2D_showers} for datasets 2 and 3. The 2-dimensional distribution of the average energy deposition is shown in the detector layers with highest (layer 10) and lowest (layer 44)  mean energy depositions. Empty entries in the \geant~simulation are a result of the initial voxelization combined with the following transformation to Cartesian coordinates. All voxels with an expected energy deposition above 0 are populated in all \calosc~ and WGAN-GP implementations, an indication that the models are able to reproduce the shower diversity from the training set. Images at layer 10 are identical for all generative models, dominated by the central voxel. Layer 44; however, has more voxels sharing a significant fraction of the layer energy. The subVP implementation shows a visually similar average to \geant~compared to the other diffusion implementations, capturing the high energy depositions along the y-axis in dataset 2 and the isotropic pattern around the center in dataset 3. The WGAN-GP implementation, on the other hand, shows higher energy fractions away from the center of the image, where lower energy fractions are expected.

\begin{figure*}[ht]
\centering
\includegraphics[width=0.23\textwidth]{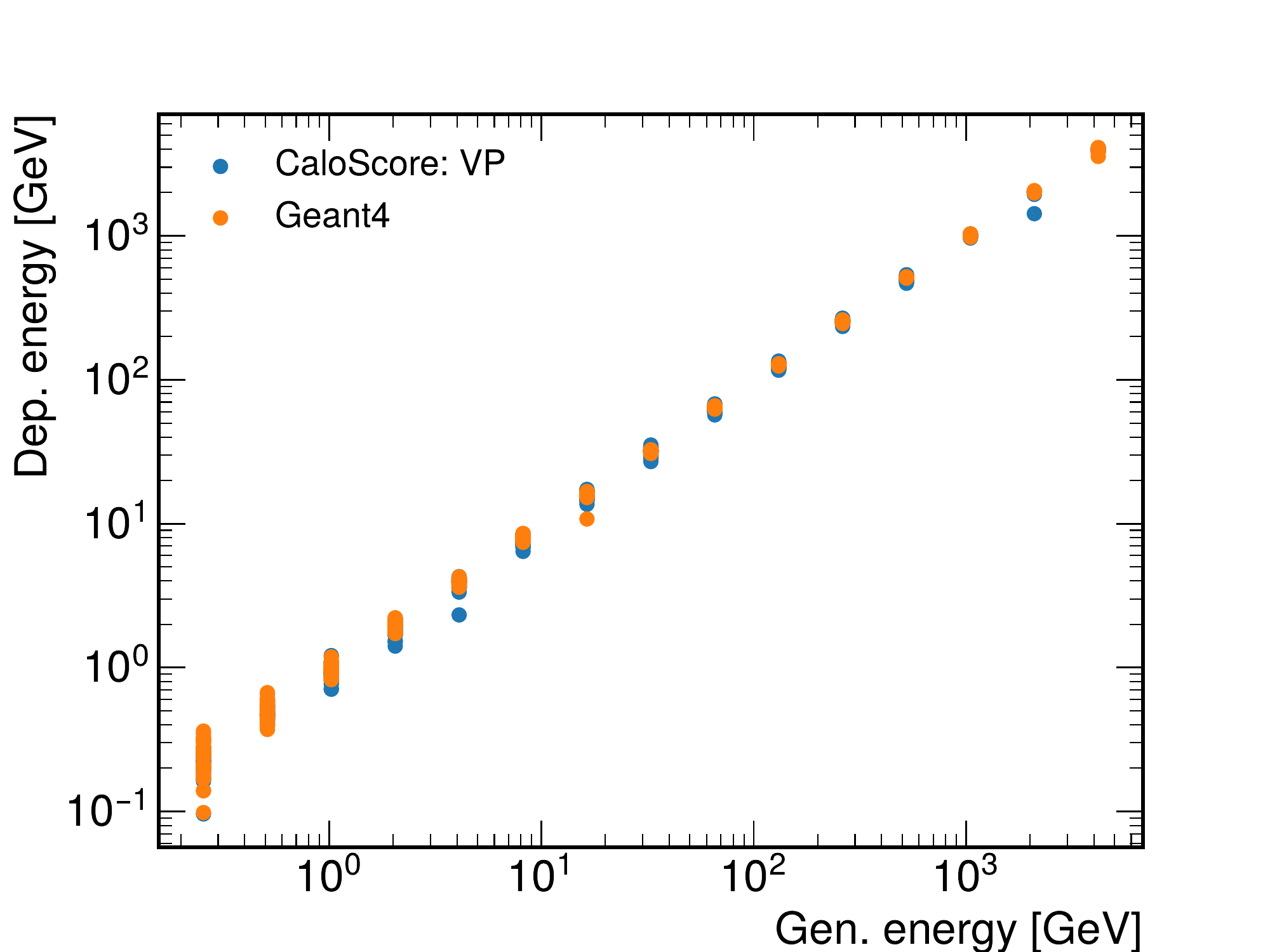}
\includegraphics[width=0.23\textwidth]{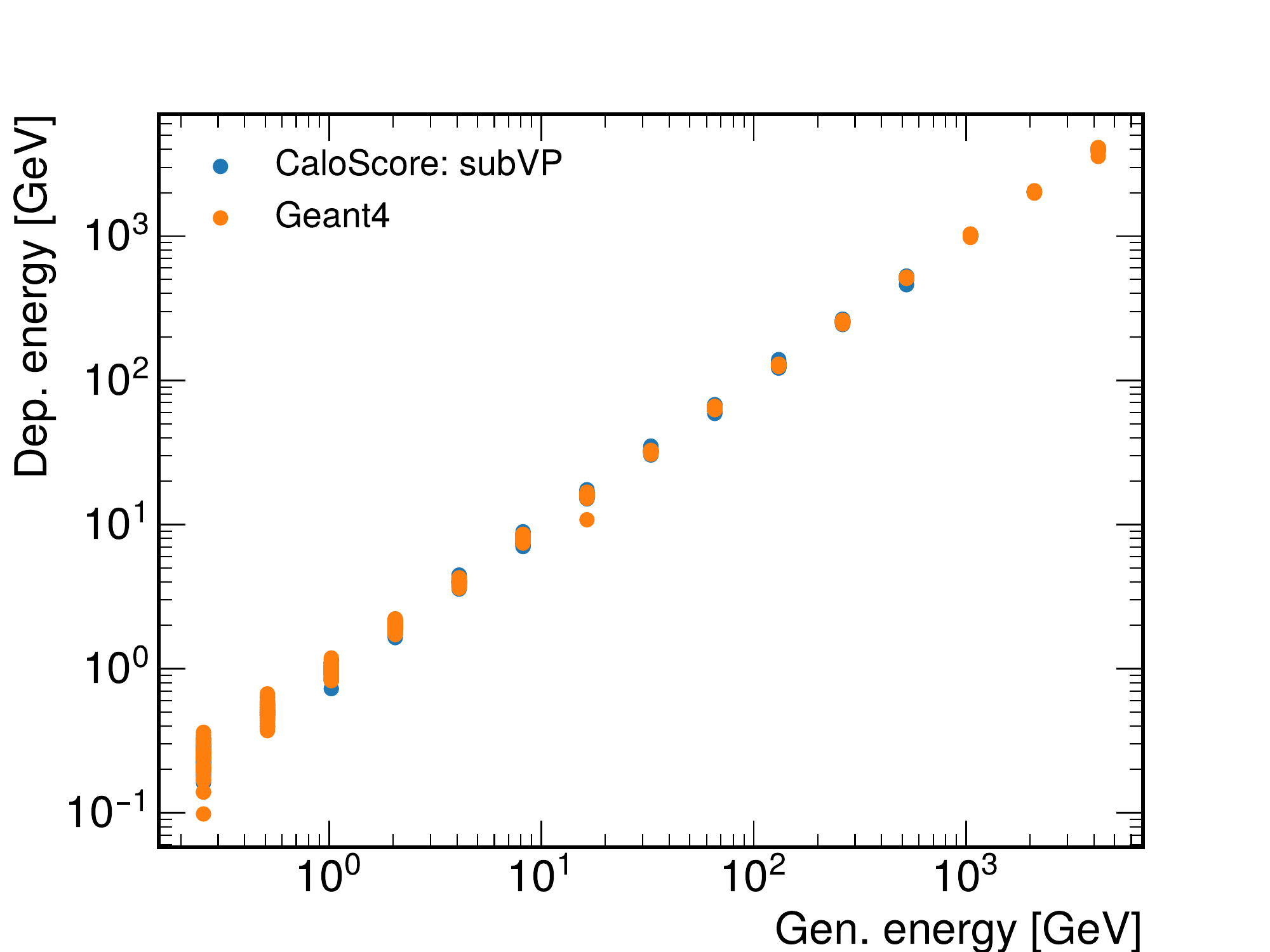}
\includegraphics[width=0.23\textwidth]{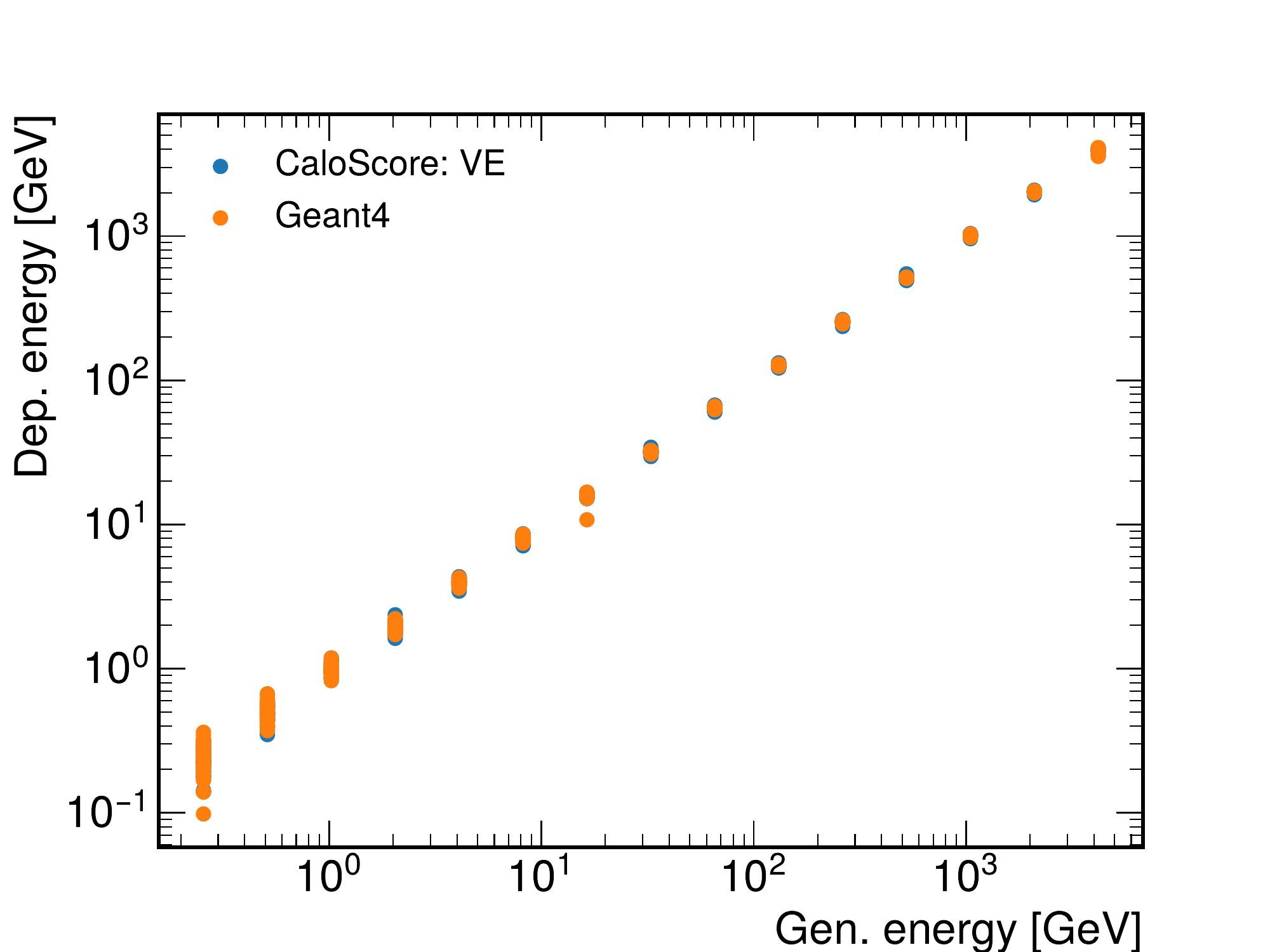}
\includegraphics[width=0.23\textwidth]{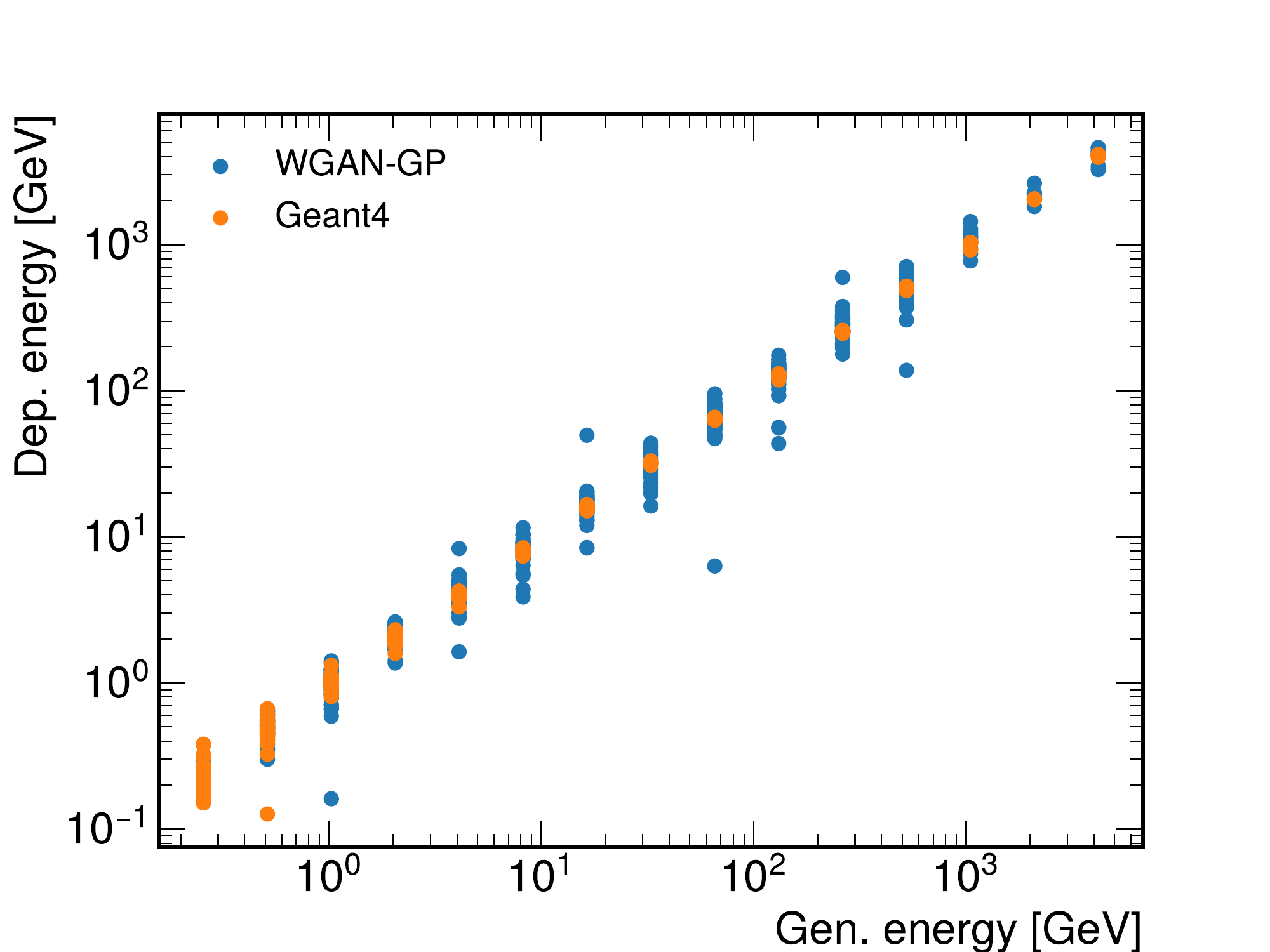}
\includegraphics[width=0.23\textwidth]{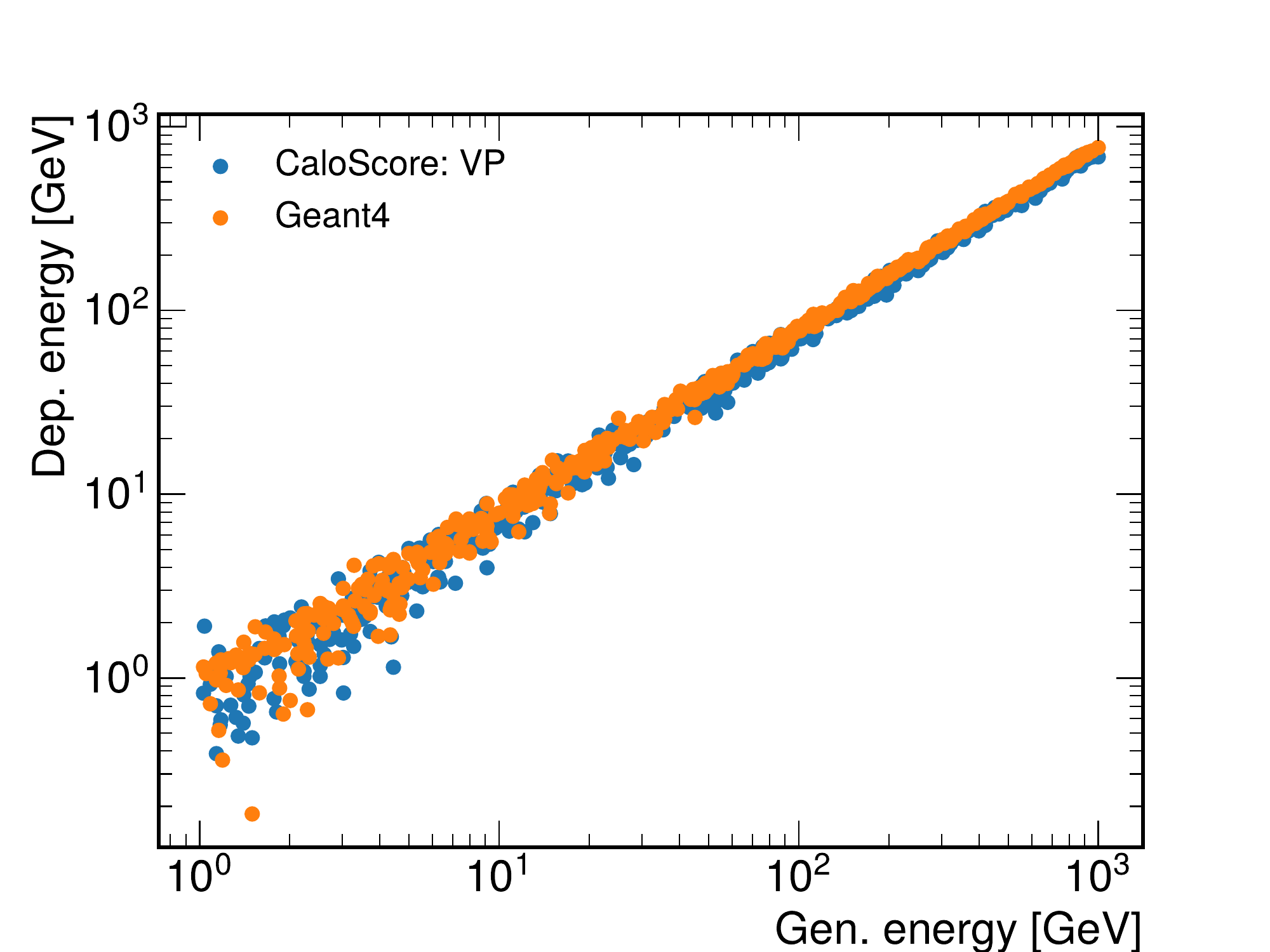}
\includegraphics[width=0.23\textwidth]{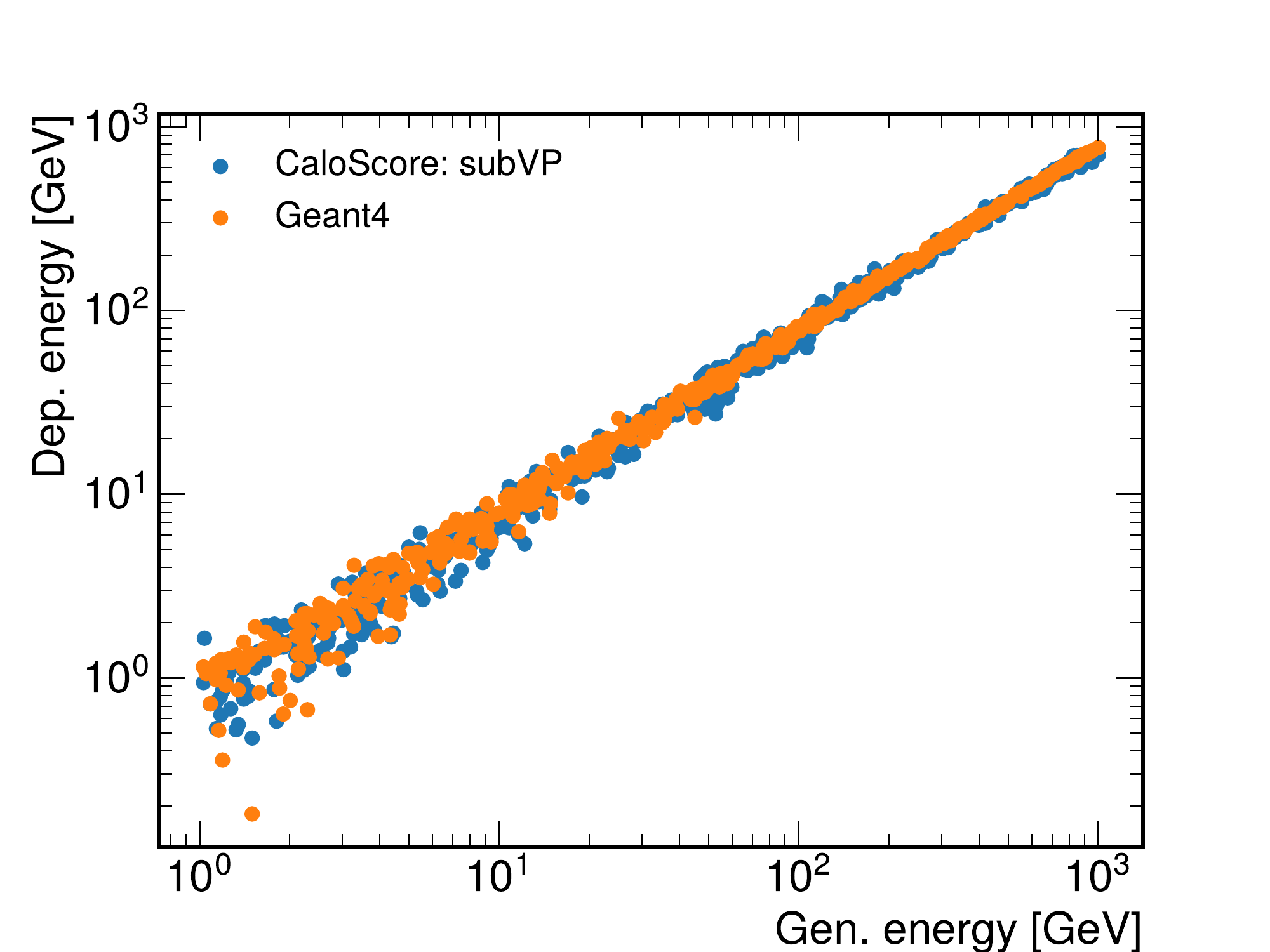}
\includegraphics[width=0.23\textwidth]{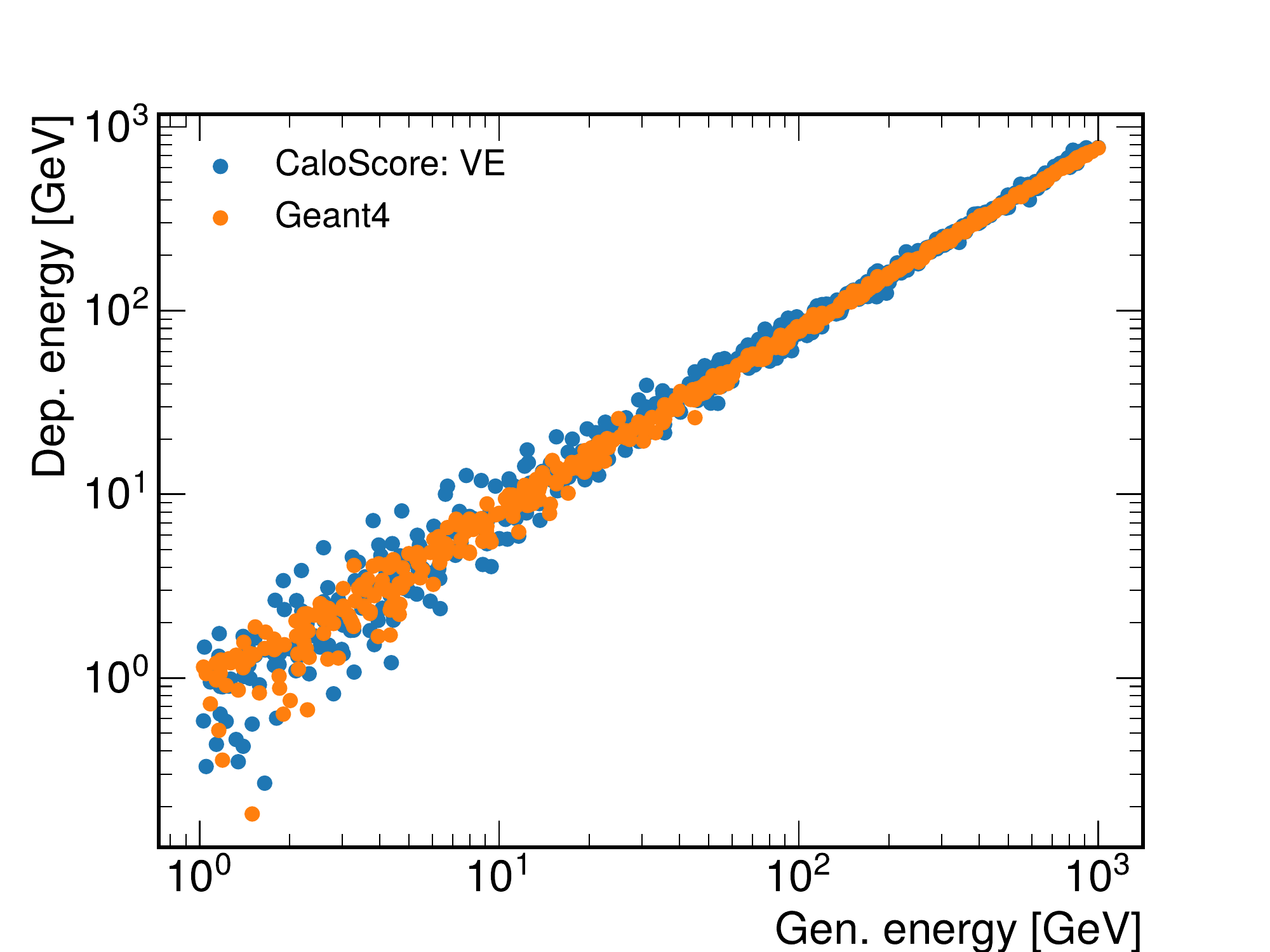}
\includegraphics[width=0.23\textwidth]{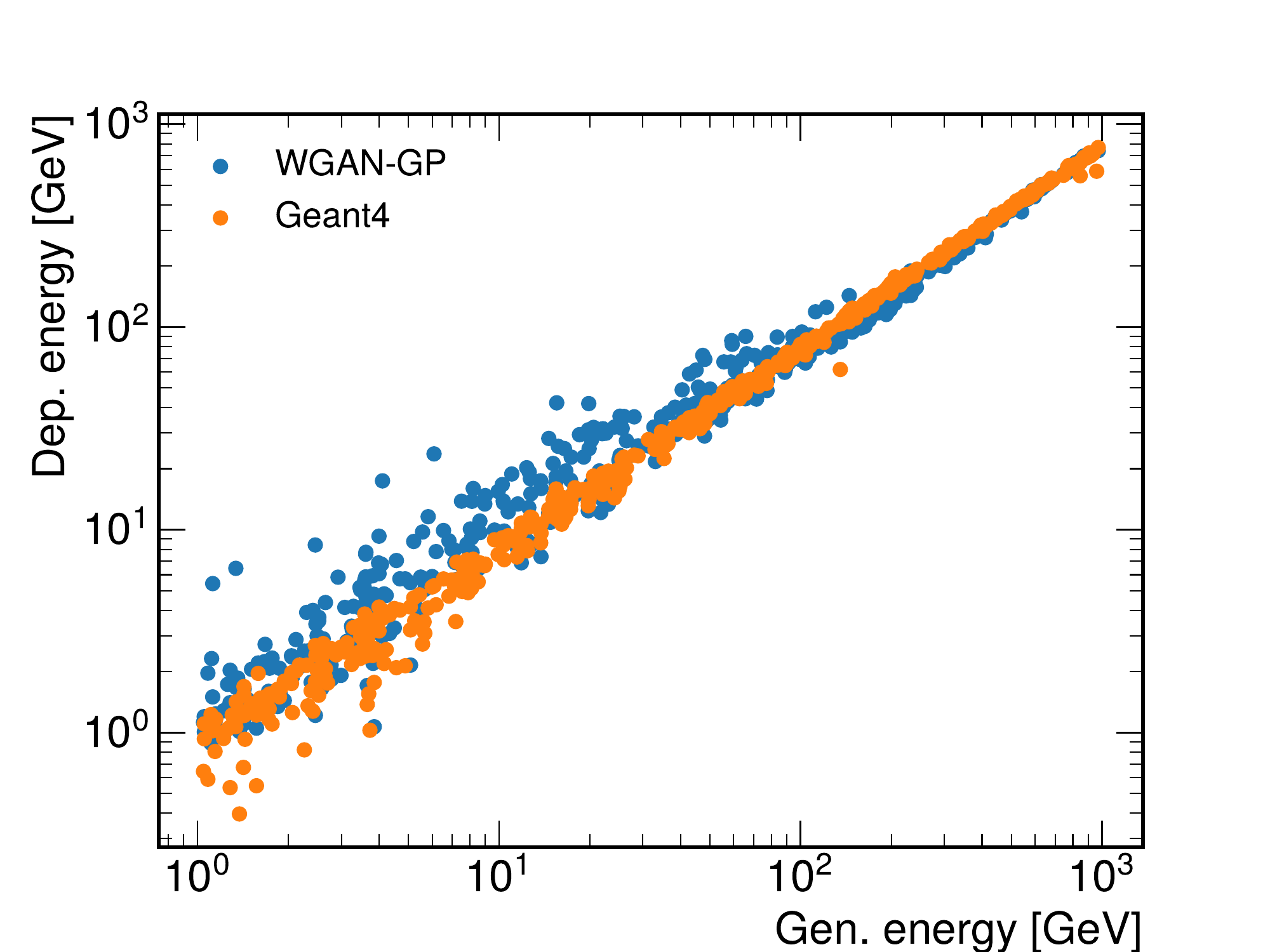}
\includegraphics[width=0.23\textwidth]{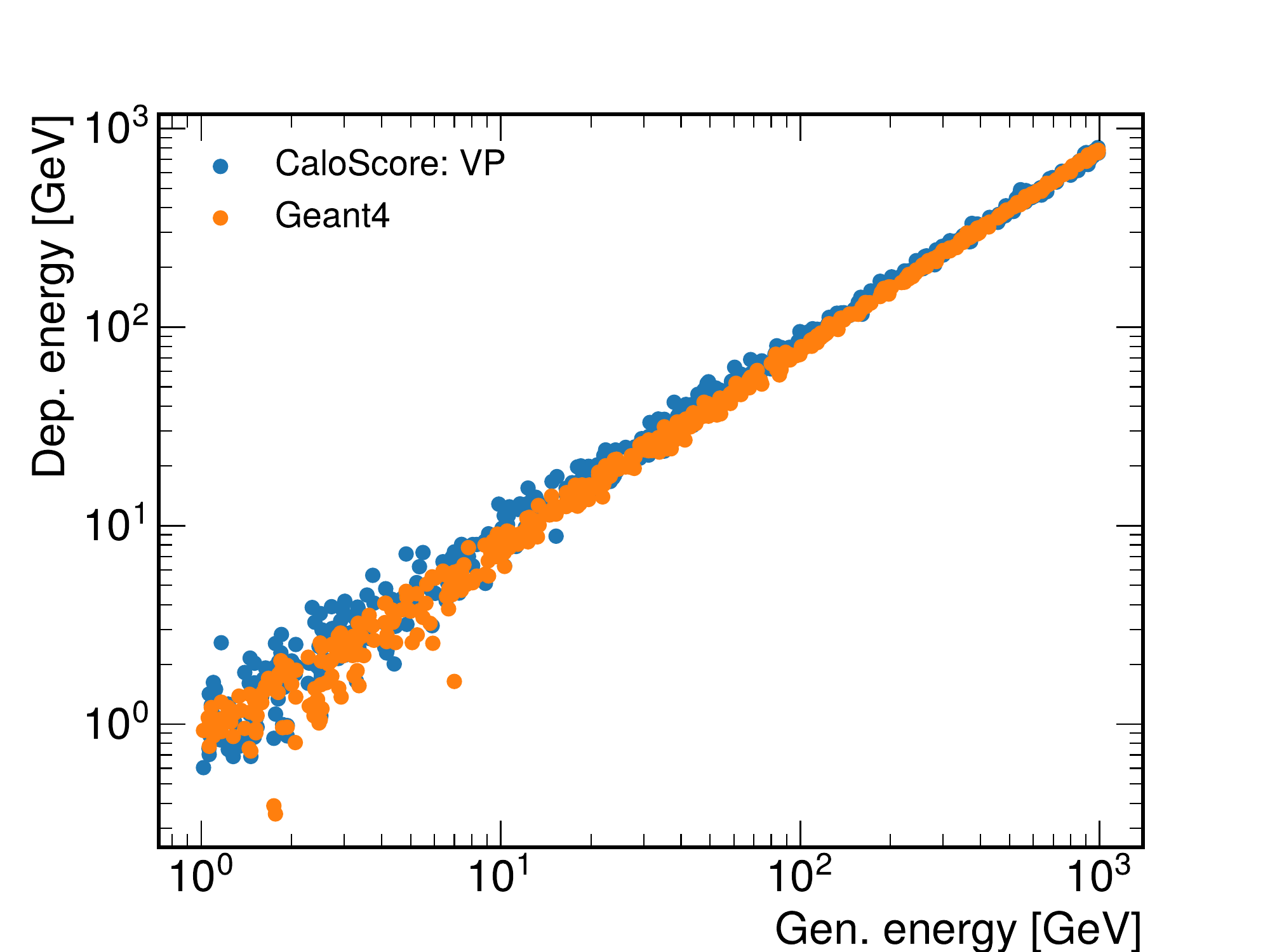}
\includegraphics[width=0.23\textwidth]{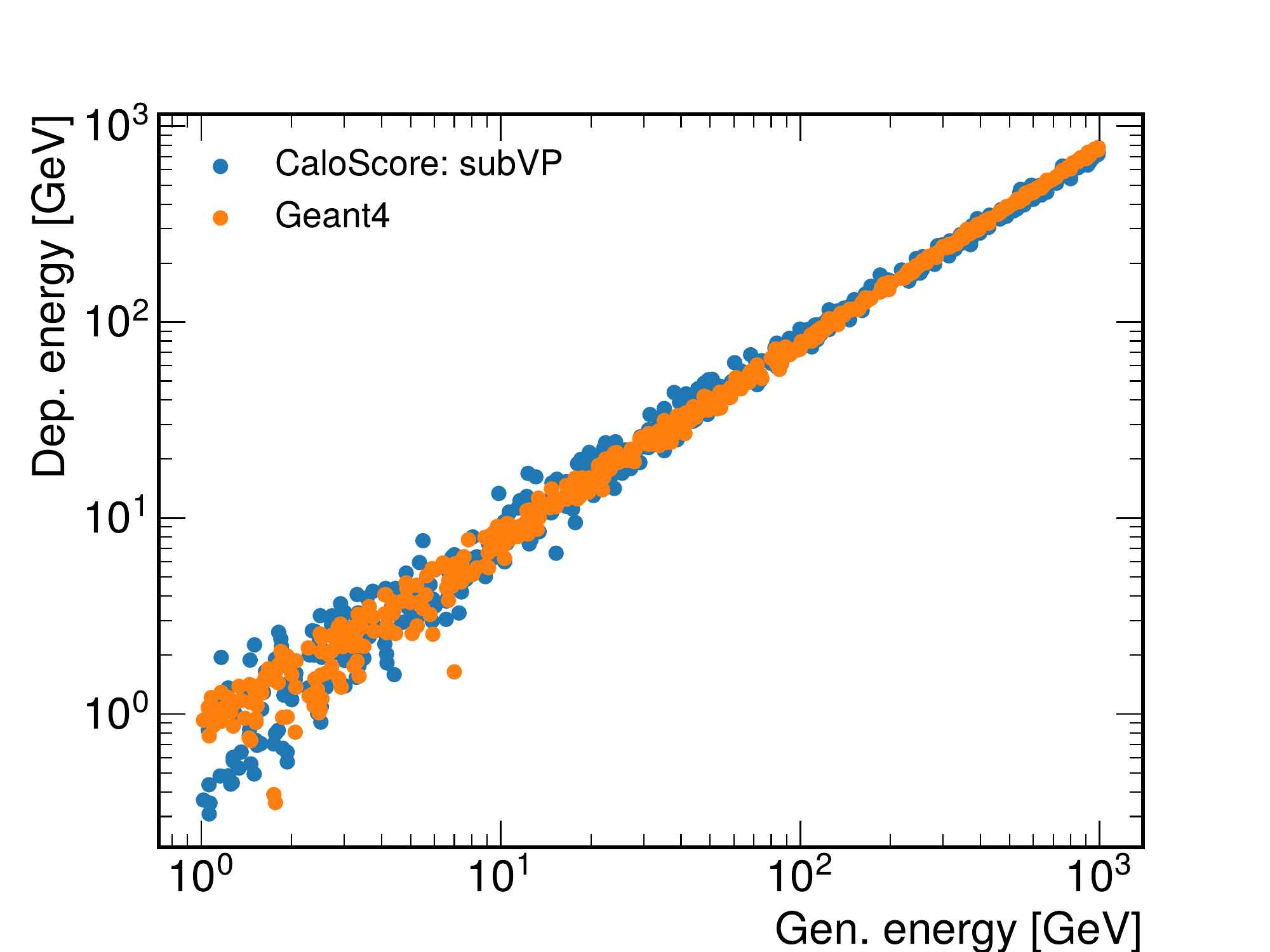}
\includegraphics[width=0.23\textwidth]{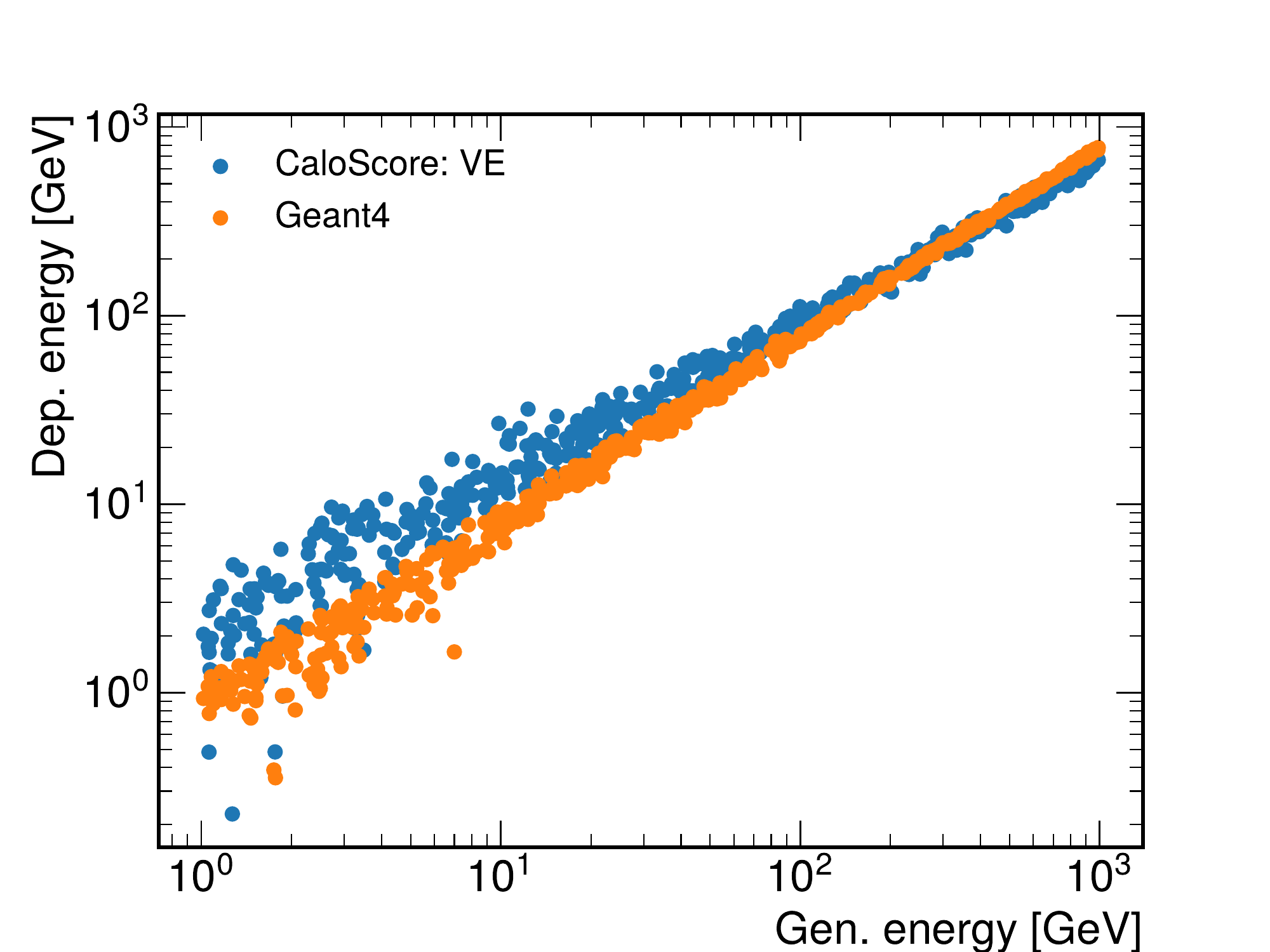}
\includegraphics[width=0.23\textwidth]{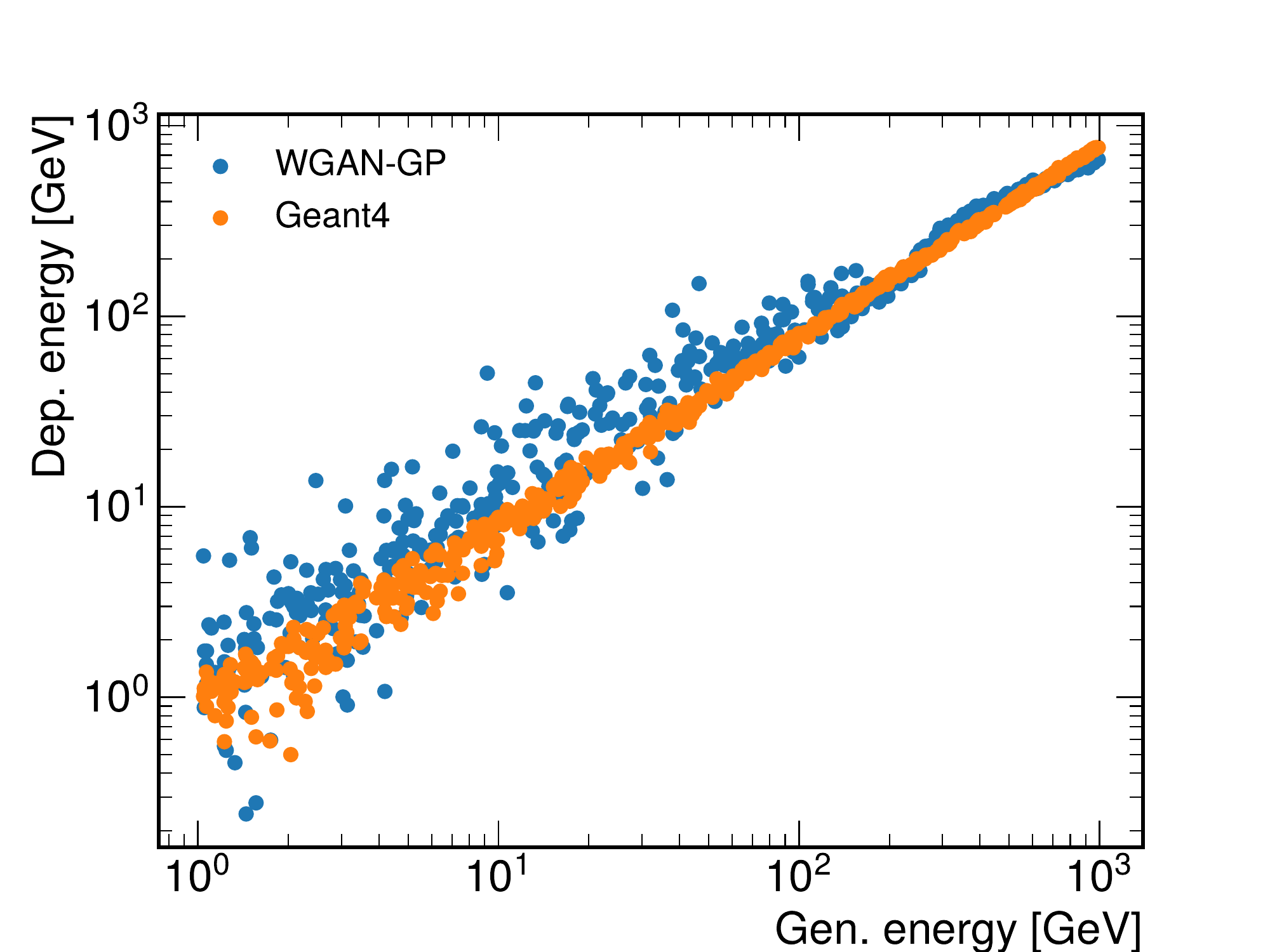}

\caption{Deposited energy versus generated energy in \calosc~(blue) and \geant~(orange) for the three different diffusion models: VP (first column), subVP(second column), and VE (third column). WGAN-GP results are shown in the fourth column. First row samples are generated using energies from dataset 1, second row from dataset 2 and third row from dataset 3.}
\label{fig:econd}
\end{figure*}

Finally, the assessment of generated samples using different conditional energies is investigated in Fig.~\ref{fig:econd}, by comparing the total deposited energy versus the generated particle energy.

All \calosc~models show similar mean and spread compared to \geant, with the exception of the VE implementation that shows a wider spread for dataset 2 and higher mean in dataset 3. The WGAN-GP model shows a wider spread in all datasets compared to all \calosc besides the VE implementation.

We have also explored the classifier metric introduced in \textsc{CalowFlow} whereby a post-hoc classifier is trained to distinguish generated showers from \textsc{Geant}~4 examples.  While the classifier could not perfectly identify fake from real showers, it did have an area under the receiver operating characteristic curve (AUC) of about 0.98 for all three models. The classifier was trained using the generated calorimeter images. While this suggests that further (hyperparameter)optimization would be beneficial, it already serves as an important baseline for other methods. It should also be noted that compared to the dataset introduced in the \textsc{CaloGAN} paper, the datasets considered in this work are either more realistic (dataset 1) or higher dimensional (a factor 10 for dataset 2 and 100 for dataset 3) and would be interesting to see the classifier metric obtained from models such as \textsc{CaloFlow}. 

\section{Conclusions and Outlook}
\label{sec:conclusions}

In this paper we introduced \textsc{CaloScore}, a novel generative model for calorimeter shower simulation based on score-matching, applied for the first time in the context of collider physics. 

The performance of \calosc~is studied using the Fast Calorimeter Simulation Challenge 2022 datasets and compared to three different model implementations based on different drift and diffusion coefficients. An additional comparison is also provided using a Wasserstein GAN with gradient penalty. This is the first method to produce results for all three datasets of the challenge and we look forward to comparisons with other models as they become available.

For lower number of voxels, all models are capable of producing realistic calorimeter showers, showing a good agreement with the \geant~simulation in all datasets and for a variety of observables. At the highest dimensional dataset, \calosc~with diffusion process described by the subVP stochastic differential equation is able to produce realistic calorimeter showers while VP and VE SDEs show larger deviations. \calosc also shows overall better results compared to the WGAN-GP implementation in all distributions investigated in this work with exception to generation time, where the WGAN-GP implementation is 3-16 times faster than \calosc. \calosc~is also shown to be scalable, with number of trainable parameters sensitive to the overall model architecture rather than the total dimensionality of the dataset. 

While the voxelization strategy used in dataset 1 minimizes the number of empty voxels, the irregular voxelization reduces the geometrical information present in the calorimeter shower. The geometrical information is included as an important inductive bias for datasets 2 and 3 and leads to an order of magnitude fewer trainable parameters in the model compared to the one used in dataset 1. Moreover, a regular voxelization amenable to convolutional operations is obtained in datasets 2 and 3 only after an additional transformation of coordinates. Since the transformation is applied on the voxelized inputs, nonphysical artifacts are introduced, such as empty voxels in regions that are not expected to be empty. All of these issues could be addressed if alternative voxelization schemes were available or if access to the datasets prior to any voxelization was possible.

The addition of inductive biases to the model is also expected to improve the generation capabilities of \calosc, possibly leading to better and even smaller model architectures. Energy conservation in particular is challenging to enforce, since generated samples are not produced during training time, but only at generation time when the reverse stochastic differential equation is solved. We partially address this issue by increasing the dimensionality of dataset 1, introducing an additional entry that stores an overall normalization. Since datasets 2 and 3 rely on the geometrical description of the voxels, this strategy is not readily applicable and would instead benefit from a two-step approach as used in \textsc{CaloFlow}, where the overall normalization is determined separately and used as a conditional input to a second model that learns the normalized detector response.

The major challenge to be addressed in \calosc~is the generation time, currently requiring hundreds of model evaluations to solve the reverse SDE. While the total generation time of \calosc~is still faster compared to the \geant~simulation, we envision future works targeting high fidelity generation with lower number of function evaluations. Indeed, since this limitation is also observed in applications of diffusion models in general, a number of different attempts are currently being proposed to accelerate the generation procedure~\cite{jolicoeur2021gotta,zhang2022fast,lyu2022accelerating,lu2022dpm,xiao2022DDGAN}, with feasibility for collider physics applications yet to be studied. 

Finally, \calosc~introduces a new generative paradigm to collider physics with scalable training strategy and able to generate realistic calorimeter showers consisting of tens of thousands of dimensions. While the generation time represents the main challenge to be overcome, \calosc~is able to incorporate different advantages from other generative models while addressing some of their limitations. These include scalable and stable training schedule, based on the minimization of the convex score-matching loss, and exact likelihood estimation, previously only available with methods such as normalizing flows.

\section*{Acknowledgments}

We thank Jean-Roch Vlimant, David Shih, and Daniel Britzger for feedback on the manuscript. VM and BN are supported by the U.S. Department of Energy (DOE), Office of Science under contract DE-AC02-05CH11231. The work of DS was supported by DOE grant DOE-SC0010008. This research used resources of the National Energy Research Scientific Computing Center, a DOE Office of Science User Facility supported by the Office of Science of the U.S. Department of Energy under Contract No. DE-AC02-05CH11231 using NERSC award HEP-ERCAP0021099.

\bibliography{HEPML,other}
\bibliographystyle{apsrev4-1}

\clearpage
\appendix

\section{Score function and the connection with continuous normalizing flows}
\label{app:cnf}
Full data likelihood access is obtained from the trained score-matching model by first identifying the deterministic ordinary differential equation (ODE) associated to the SDE in Eq.~\ref{eq:rsde} that reads:
\begin{equation}
    \mathrm{d}x = [f(x,t)-\frac{1}{2}g(t)^2s_\theta(x,t)]\mathrm{d}t = \tilde{f}_\theta(x_t,t)\mathrm{d}t .
    \label{eq:ode}
\end{equation}

This ODE, named probability flow ODE by \cite{Song2021ScoreBasedGM}, has the property of having all trajectories sharing the same marginal probability densities as the SDE in Eq.~\ref{eq:rsde} and is fully determined once the score function is estimated by the generative model. 
The time evolution of the density is given by the instantaneous change of variables defined in \cite{DBLP:journals/corr/abs-1806-07366}:
\begin{equation}
    \log p_0(x_0) = \log p_T(x_T) + \int_0^T \nabla.\tilde{f}_\theta(x_t,t)\mathrm{d}t. 
    \label{eq:cnf}
\end{equation}
Eq.~\ref{eq:cnf} is equivalent to the change of variables often used in continuous normalizing flows. This expression can be estimated efficiently by first noticing that
\begin{equation}
    \nabla.\tilde{f}_\theta(x_t,t) = \mathrm{Tr}\left (  \nabla\tilde{f}_\theta(x_t,t) \right ),
\end{equation}
where $\nabla\tilde{f}_\theta(x_t,t)$ represents the Jacobian of $\tilde{f}_\theta$ and using algorithms such as the Skilling-Hutchinson trace estimator \cite{skilling1989eigenvalues,hutchinson1990stochastic} to approximate the trace calculation. 

\section{Cylindrical to Cartesian coordinate transformation}
\label{app:transform}
The initial voxelization provided for datasets 2 and 3 are in Cylindrical coordinates ($r$,$\alpha$,$z'$). While this set of coordinates reflect the detector symmetry, we found beneficial to convert the voxelization to Cartesian coordinates ($x$,$y$,$z$).  A voxel initially described in Cylindrical coordinates ($r_i$,$\alpha_i$,$z_i$) is then converted as:
\begin{align}
    x_i &= r_i\cos{\alpha_i}\\
    y_i &= r_i\sin{\alpha_i}\\
    z_i &= z_i',
\end{align}
where for simplicity we assume $r_i\in[0,1]$ and $\alpha_i\in[0,2\pi]$, resulting in $x\in[-1,1]$ and $y\in[-1,1]$. Since the overall transformation is not linear, some voxels of the new set of coordinates will always be empty, while others contain the sum of multiple voxels in the initial set of coordinates.

\section{Generation quality for different sampling parameters}
\label{app:corrector}

\begin{figure*}[ht]
\centering
\includegraphics[width=0.3\textwidth]{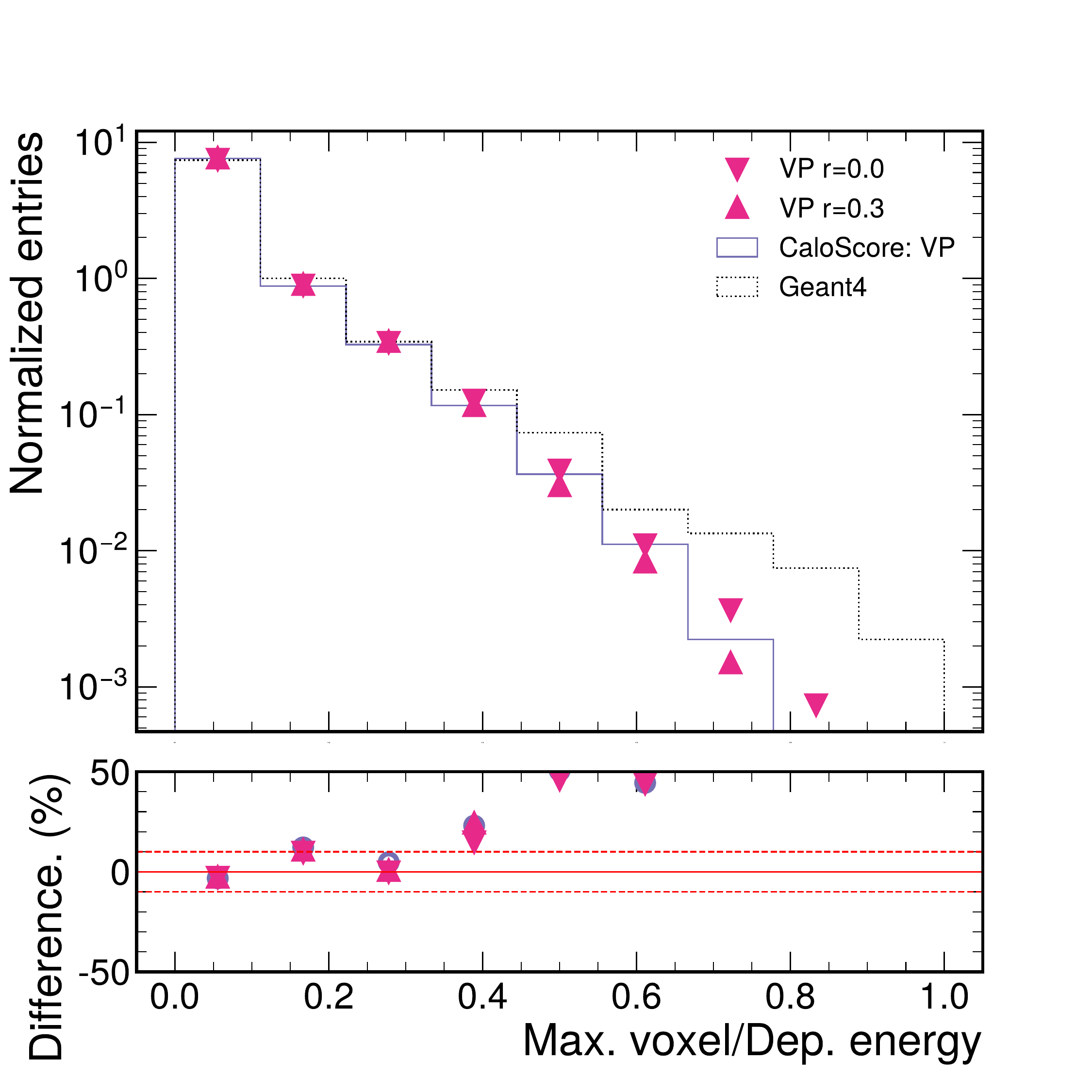}
\includegraphics[width=0.3\textwidth]{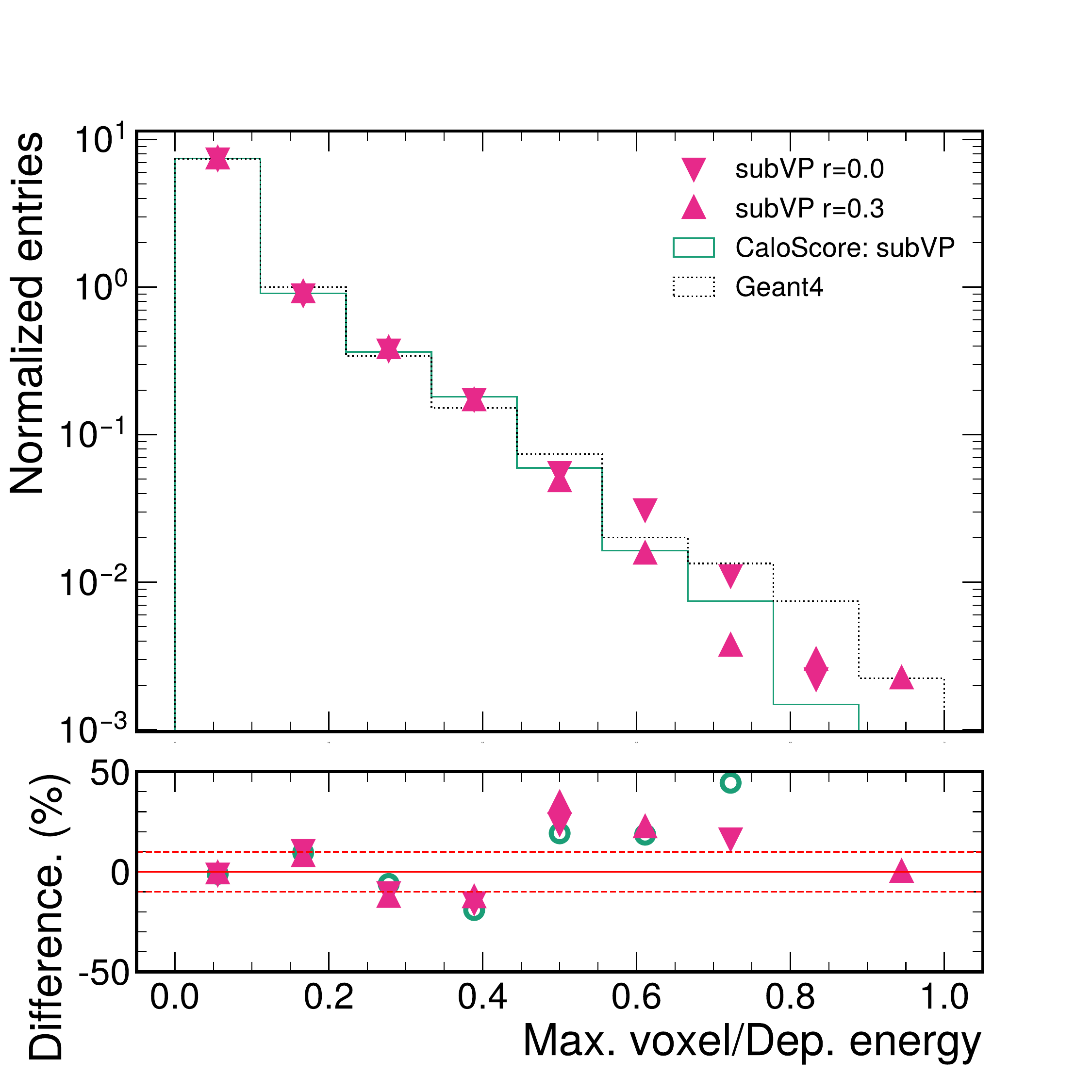}
\includegraphics[width=0.3\textwidth]{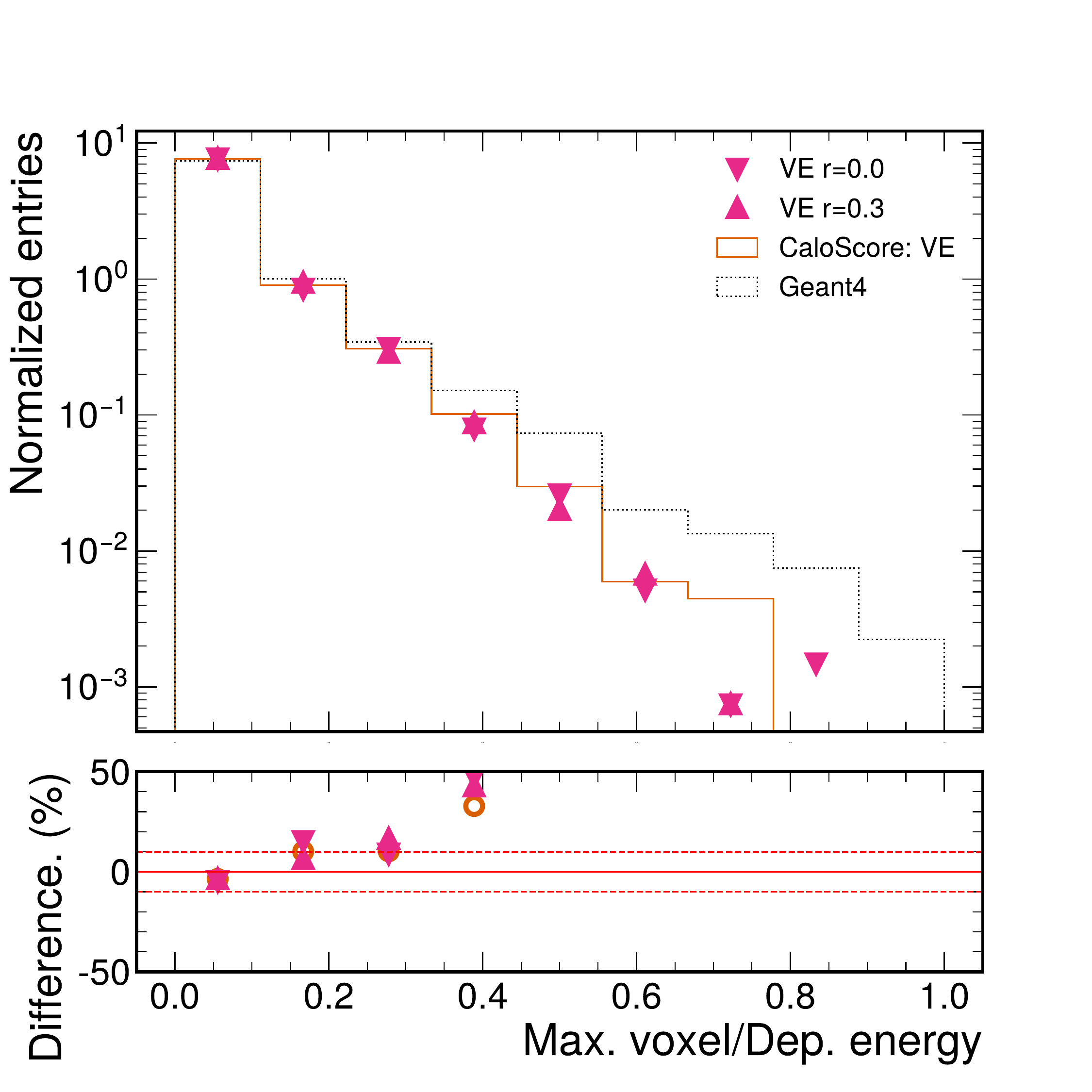}
\includegraphics[width=0.3\textwidth]{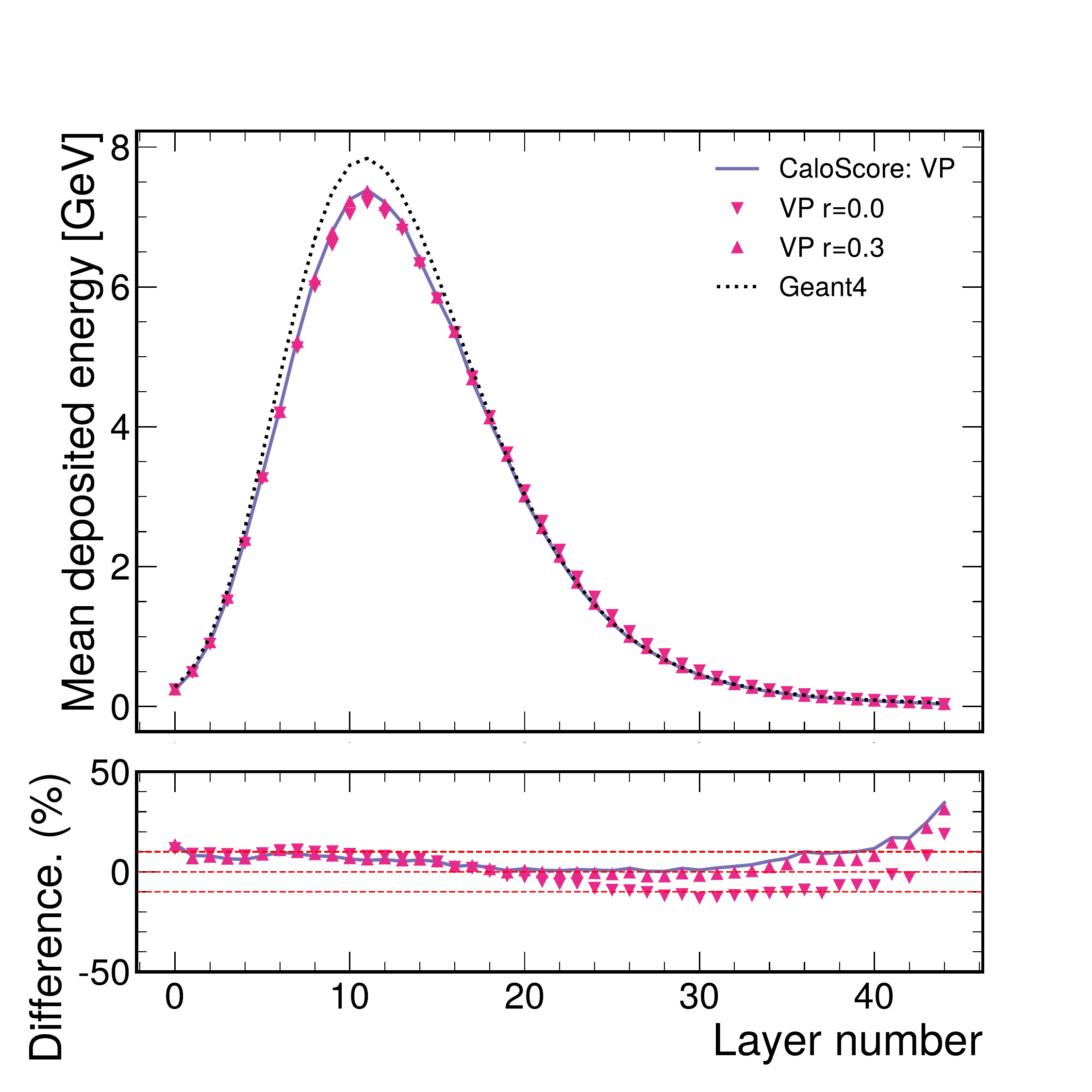}
\includegraphics[width=0.3\textwidth]{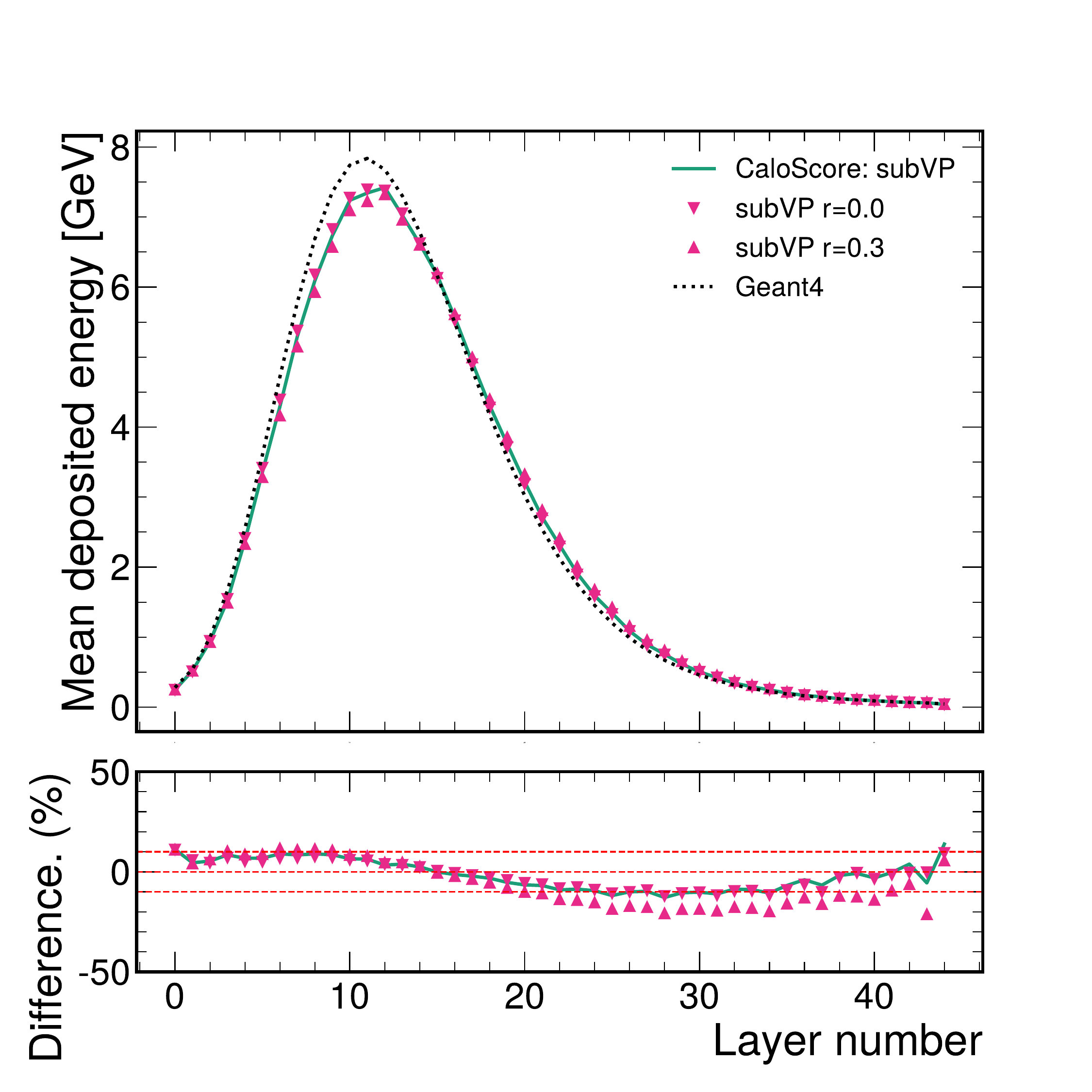}
\includegraphics[width=0.3\textwidth]{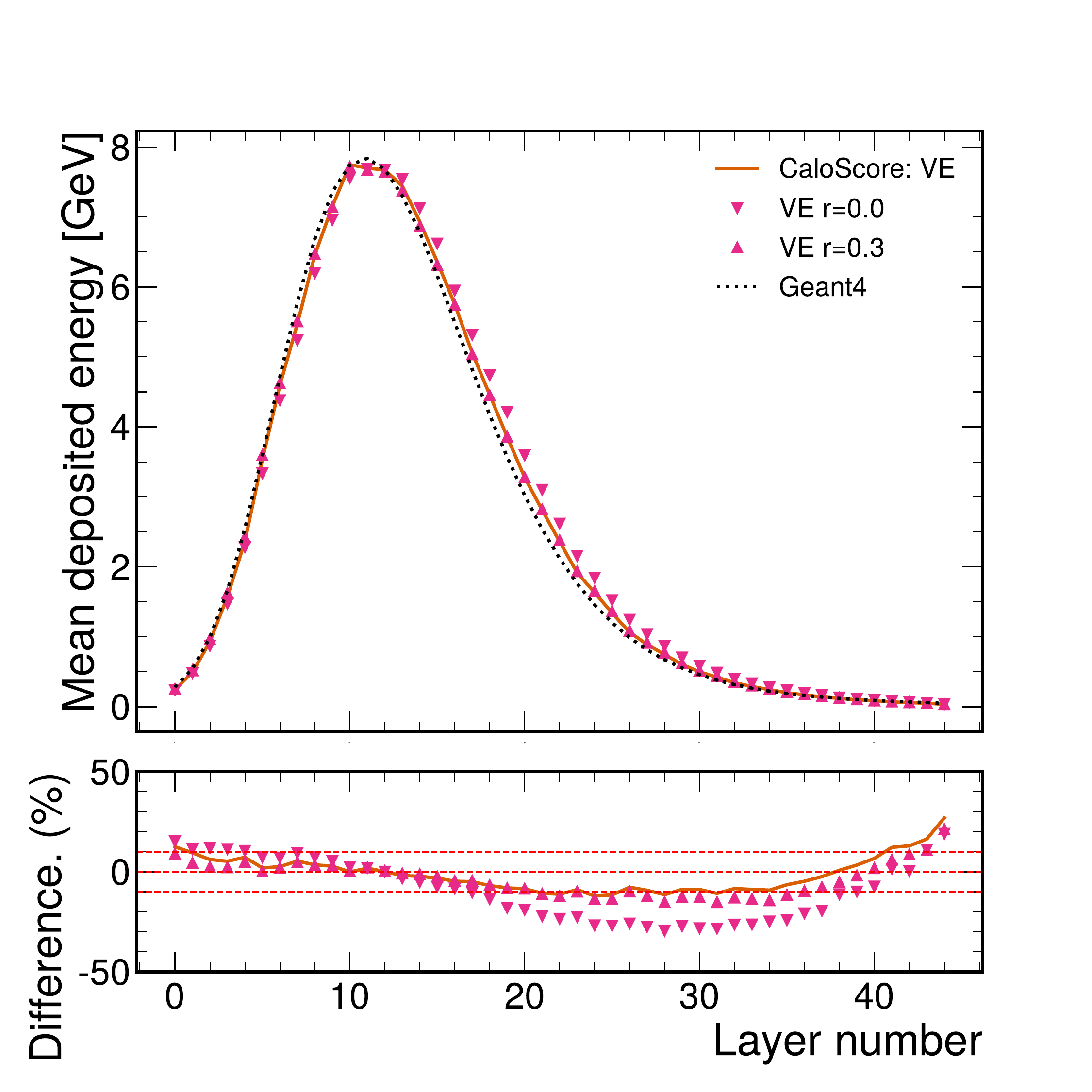}
\includegraphics[width=0.3\textwidth]{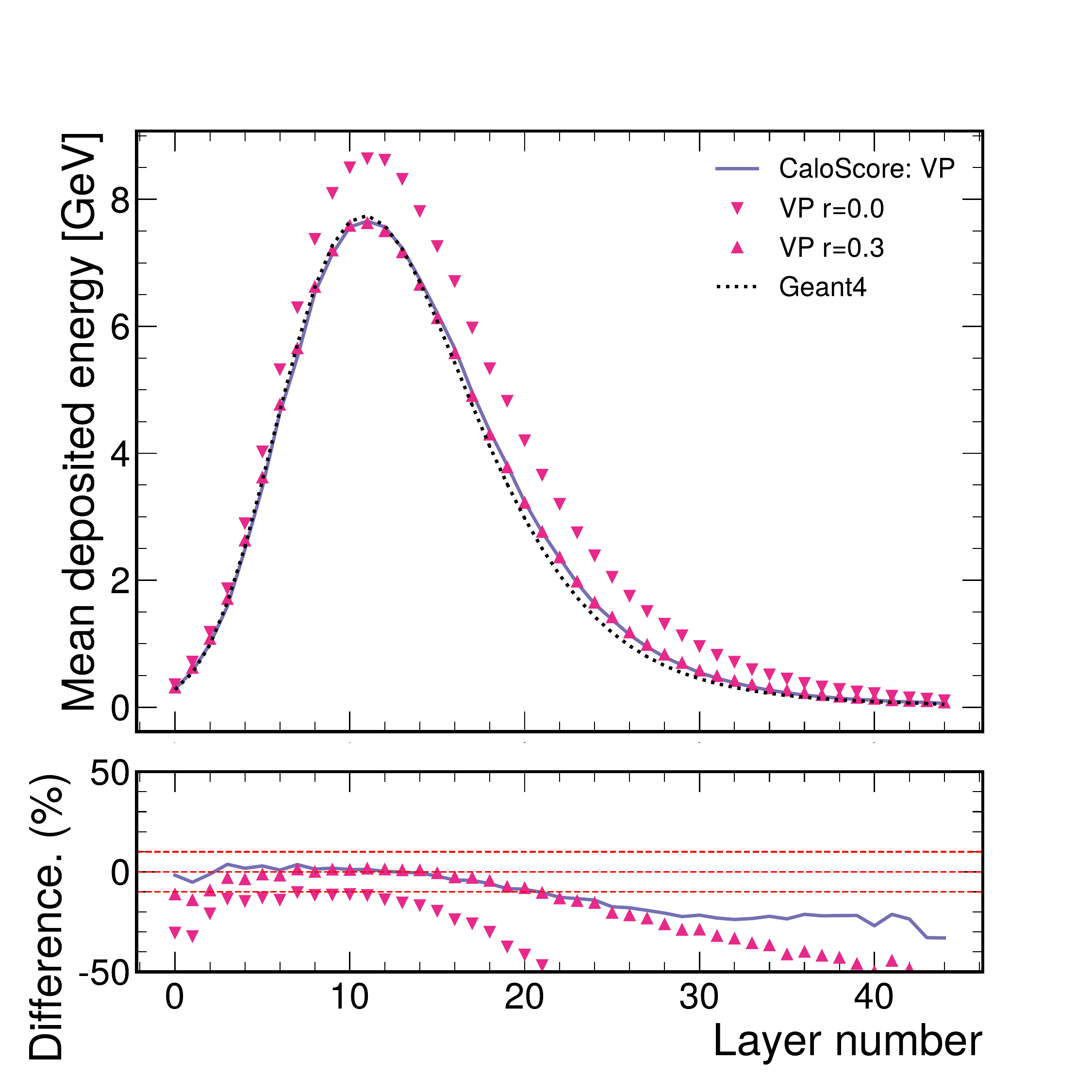}
\includegraphics[width=0.3\textwidth]{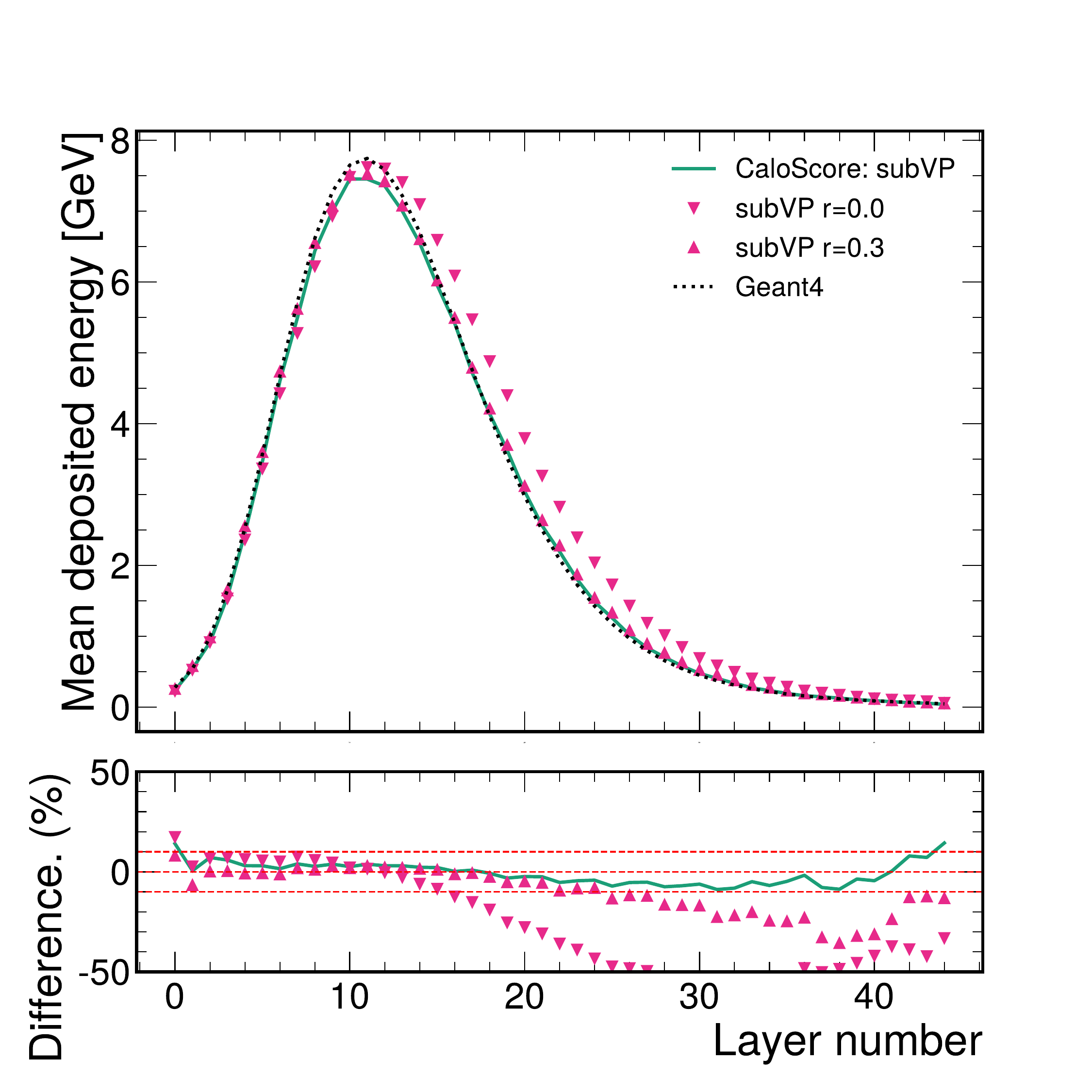}
\includegraphics[width=0.3\textwidth]{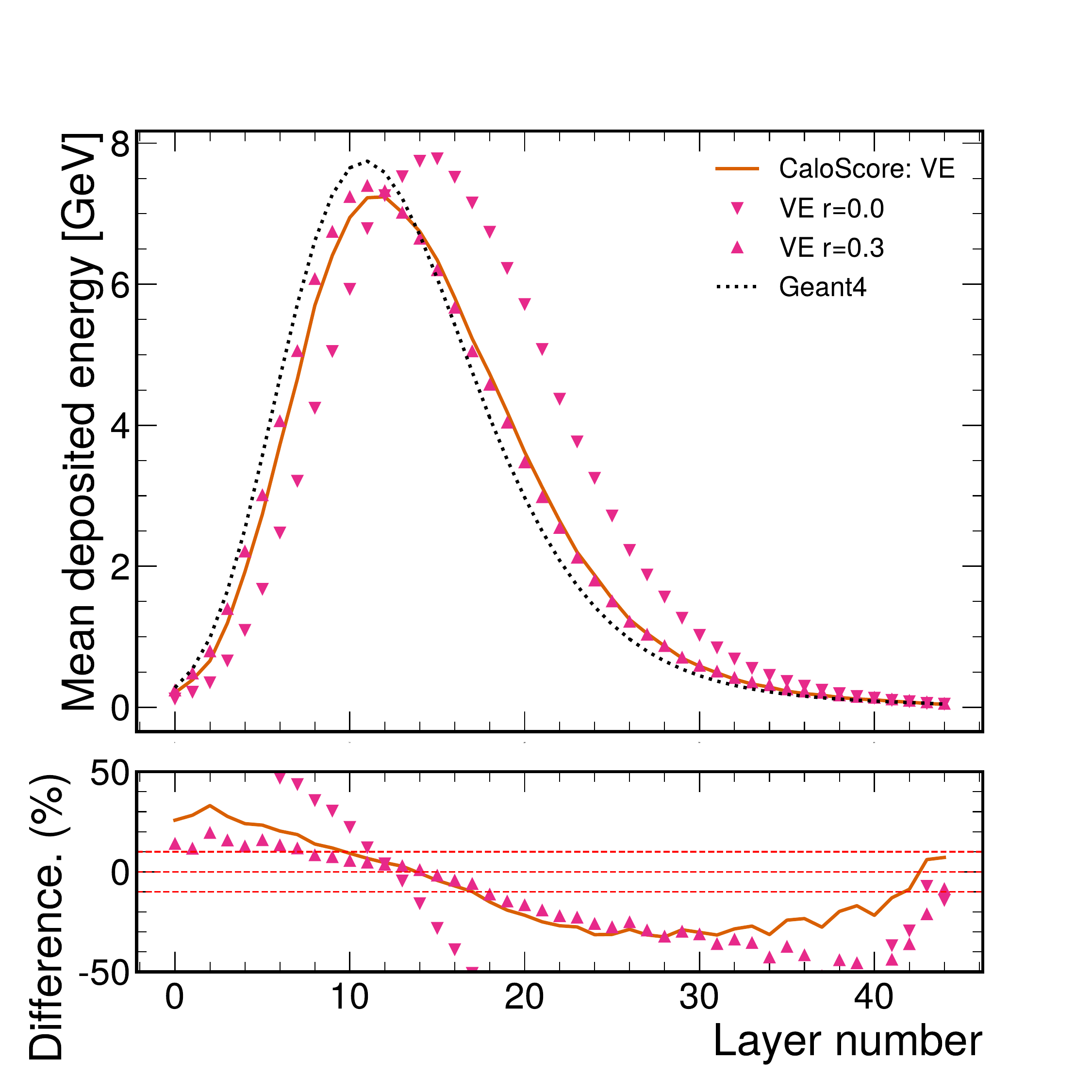}

\caption{Comparison of different signal-to-noise choices on the maximum energy fraction deposited in a single voxel for dataset 1 (top) and average energy deposition per layer for datasets 2 (middle) and 3 (bottom) for the three different diffusion models: VP (left), subVP(middle), and VE (right).}
\label{fig:app:snr}
\end{figure*}

Results presented in this work are derived using the value of 0.2 for signal-to-noise-ratio $r$ of the Langevin corrector (Eq.~\ref{eq:corrector}) and fixed number of times steps of 100 for dataset 1 and 200 for datasets 2 and 3. This choice of parameters are used to balance the generation quality and the generation time. In Fig.~\ref{fig:app:snr} different choices of $r$ are compared while maintaining the same number of steps as before for all datasets and diffusion models, evaluated on the distribution of maximum energy fraction in a single voxel for dataset 1 and average energy deposition per layer for datasets 2 and 3, the distributions that show were observed to be more sensitive to the choice of generation parameters used. 

Different choices of $r$ yield similar results for all diffusion models in both datasets 1 and 2. On the other hand, the corrector step has a stronger effect on dataset 3 and is crucial to achieve good generation quality with minimal additional computational complexity.

\begin{figure*}[ht]
\centering
\includegraphics[width=0.3\textwidth]{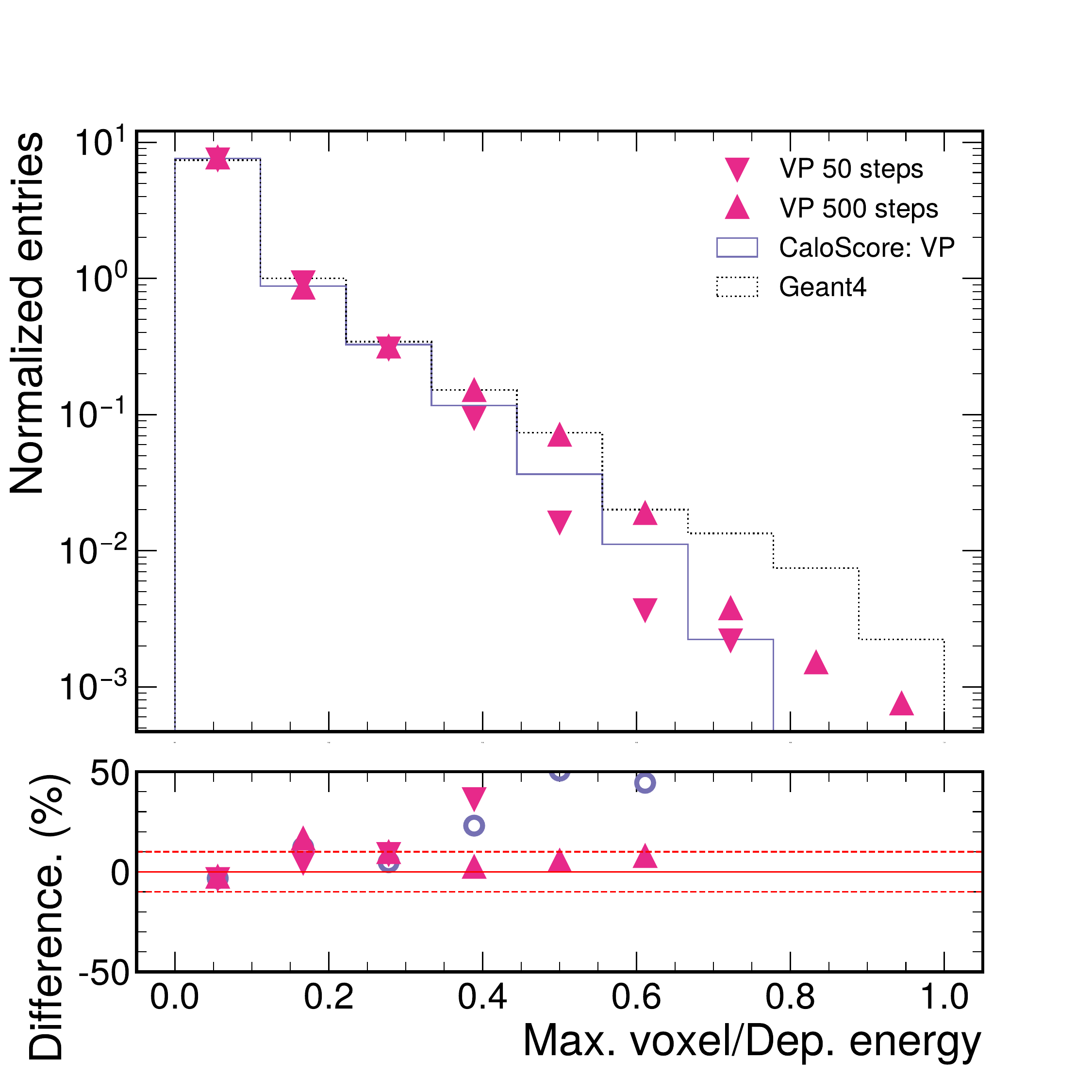}
\includegraphics[width=0.3\textwidth]{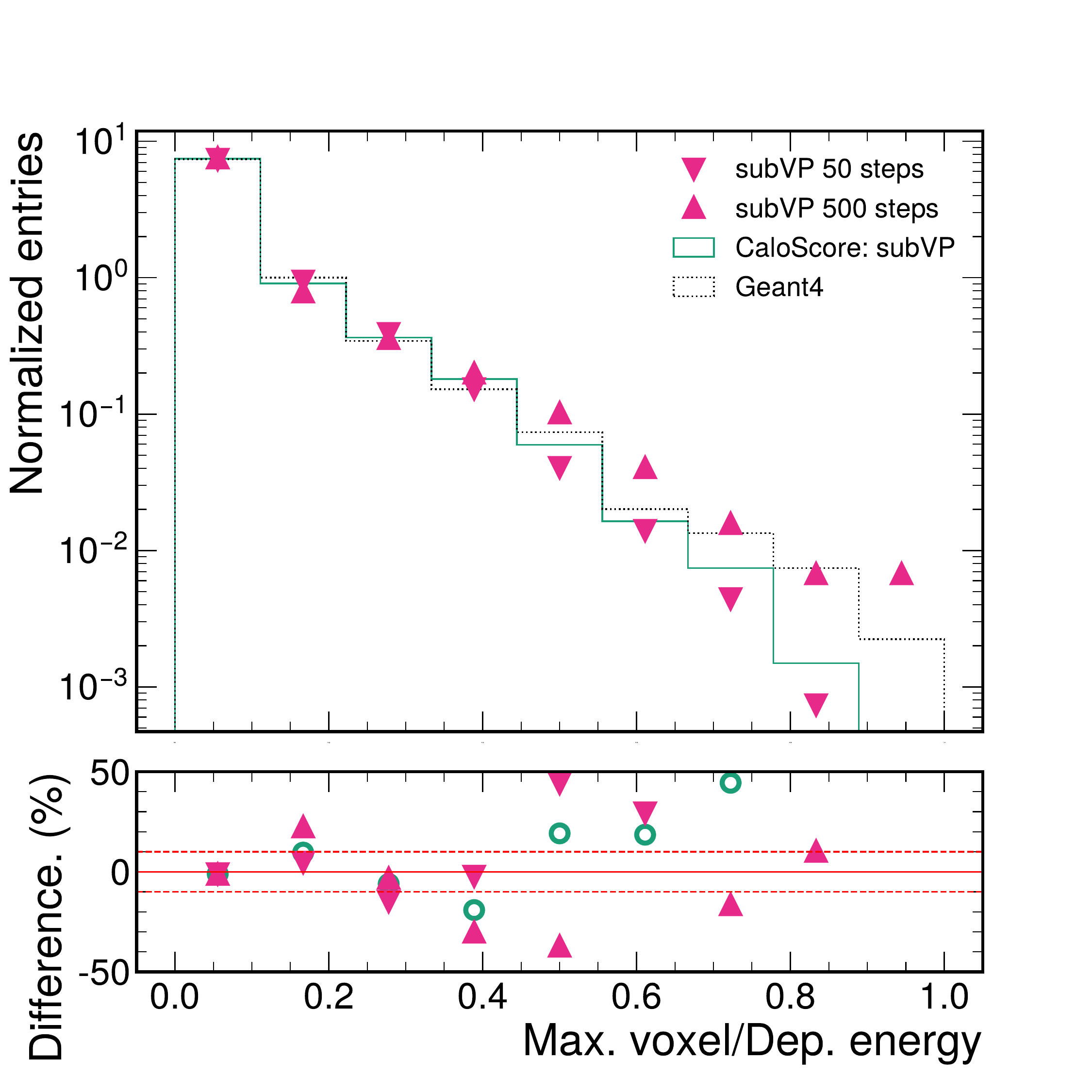}
\includegraphics[width=0.3\textwidth]{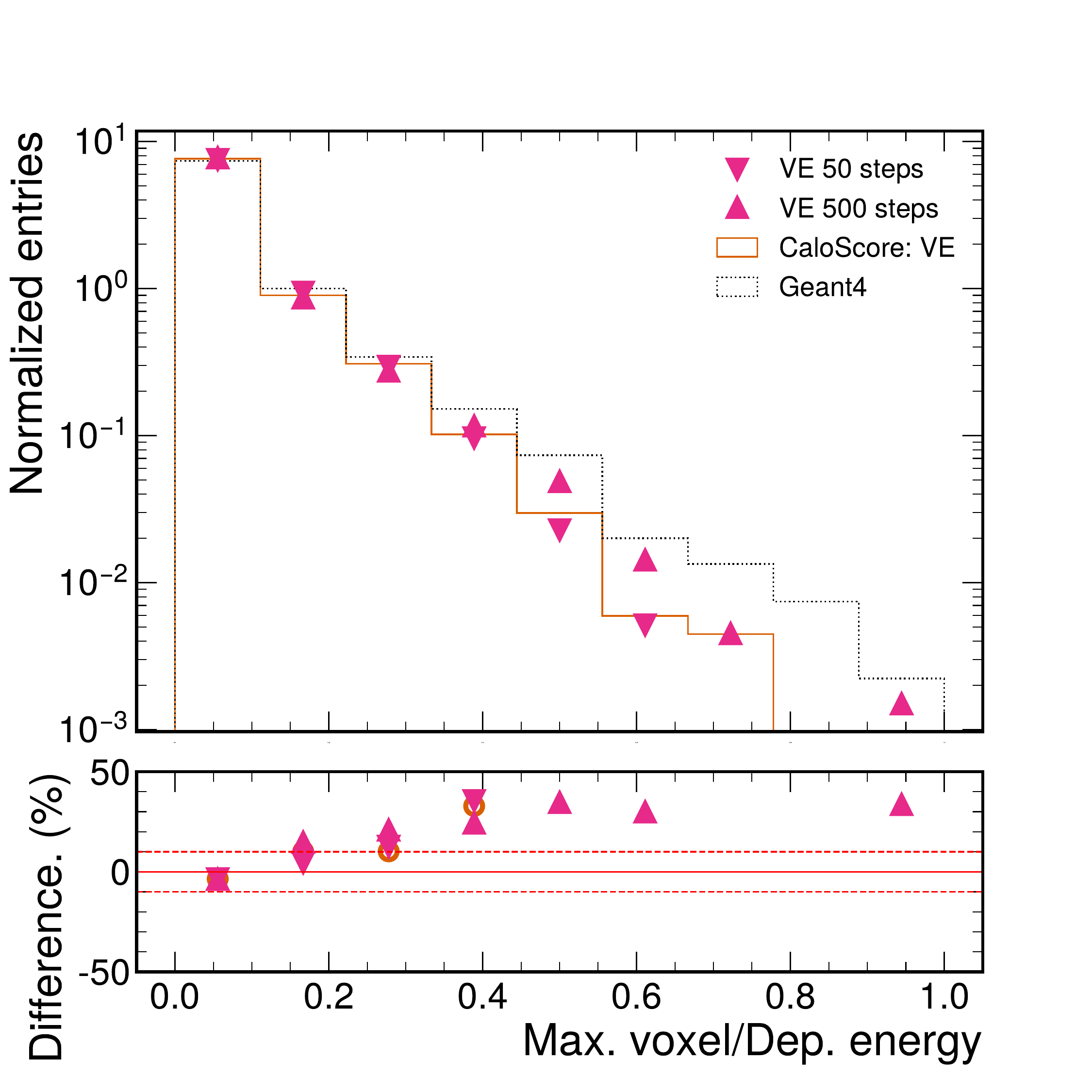}
\includegraphics[width=0.3\textwidth]{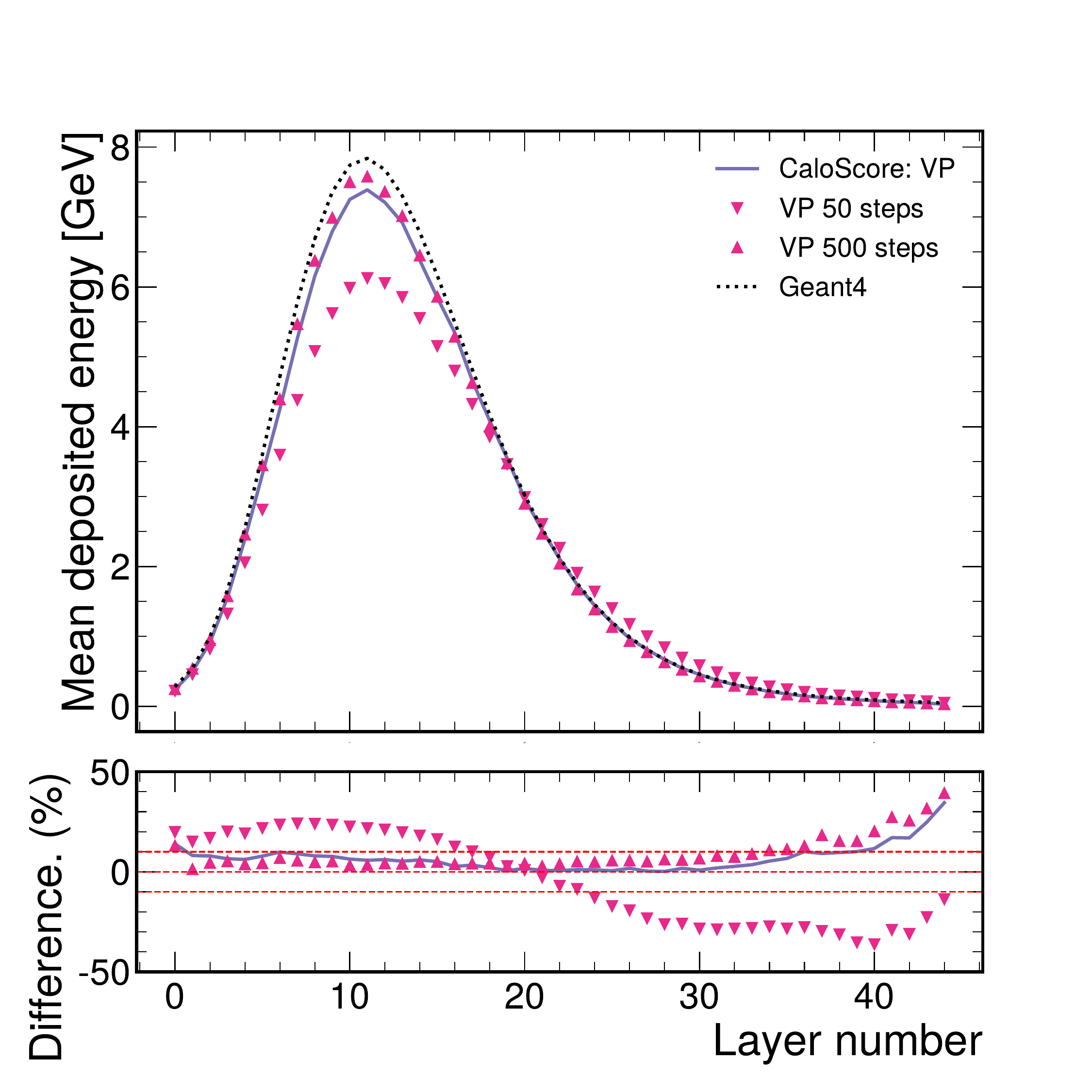}
\includegraphics[width=0.3\textwidth]{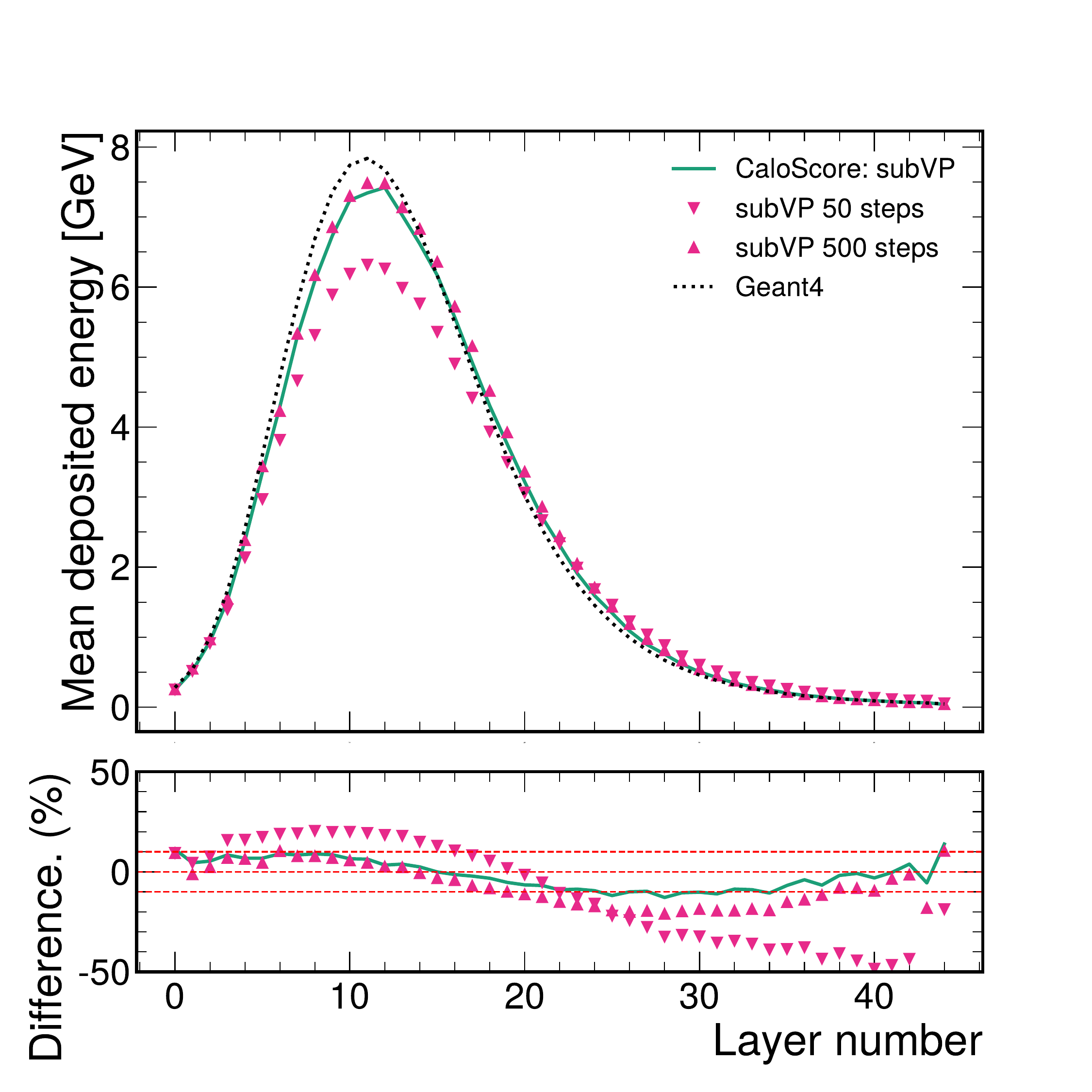}
\includegraphics[width=0.3\textwidth]{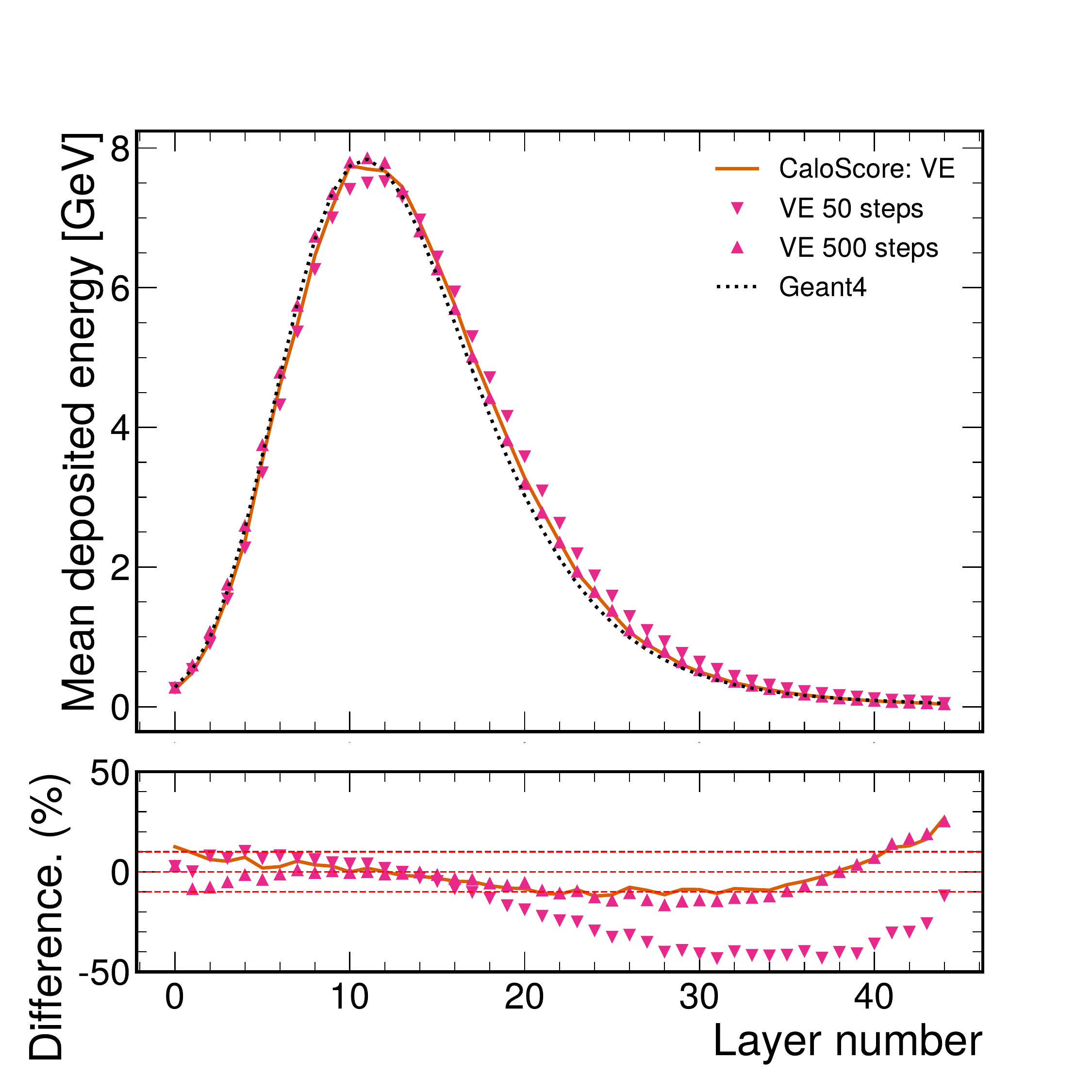}
\includegraphics[width=0.3\textwidth]{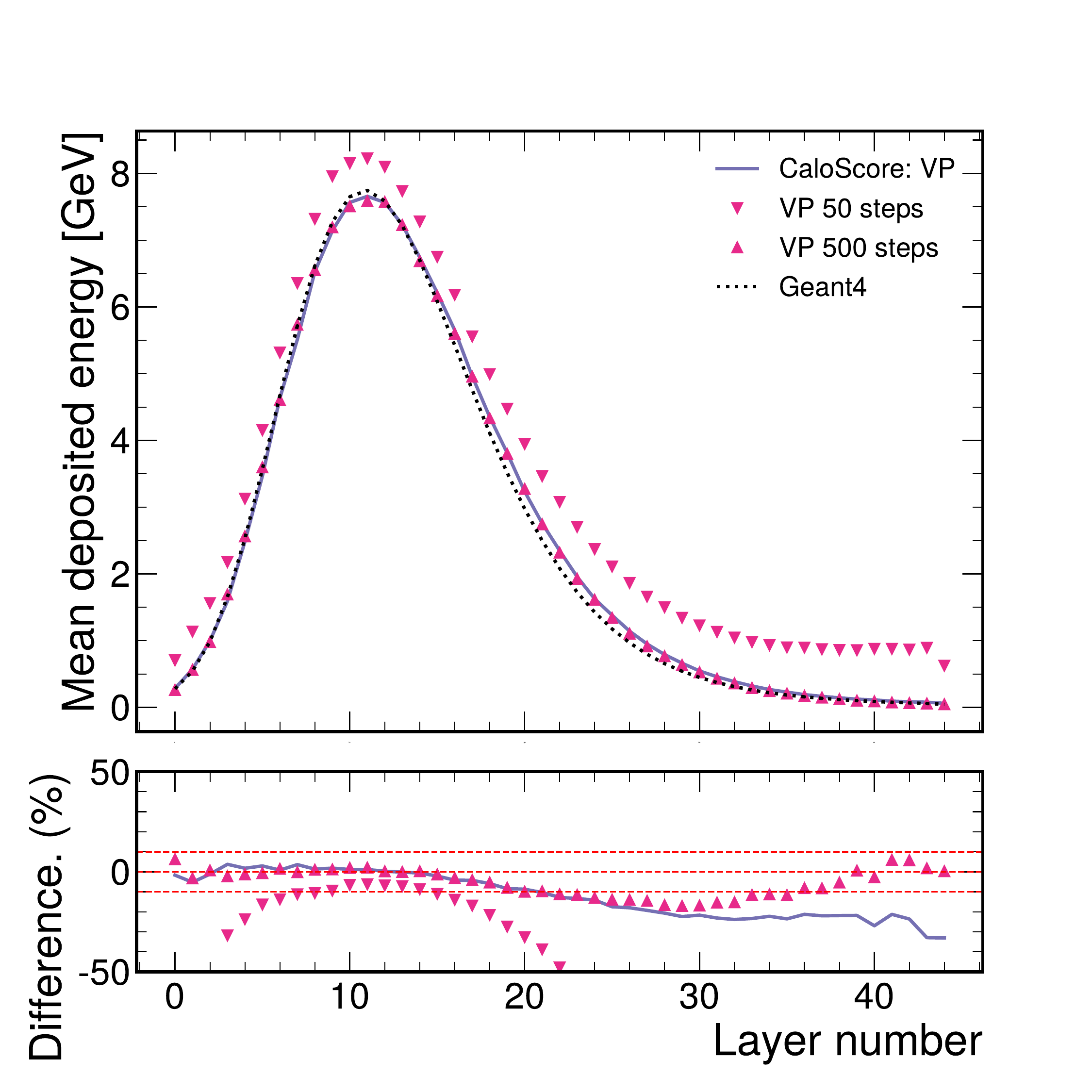}
\includegraphics[width=0.3\textwidth]{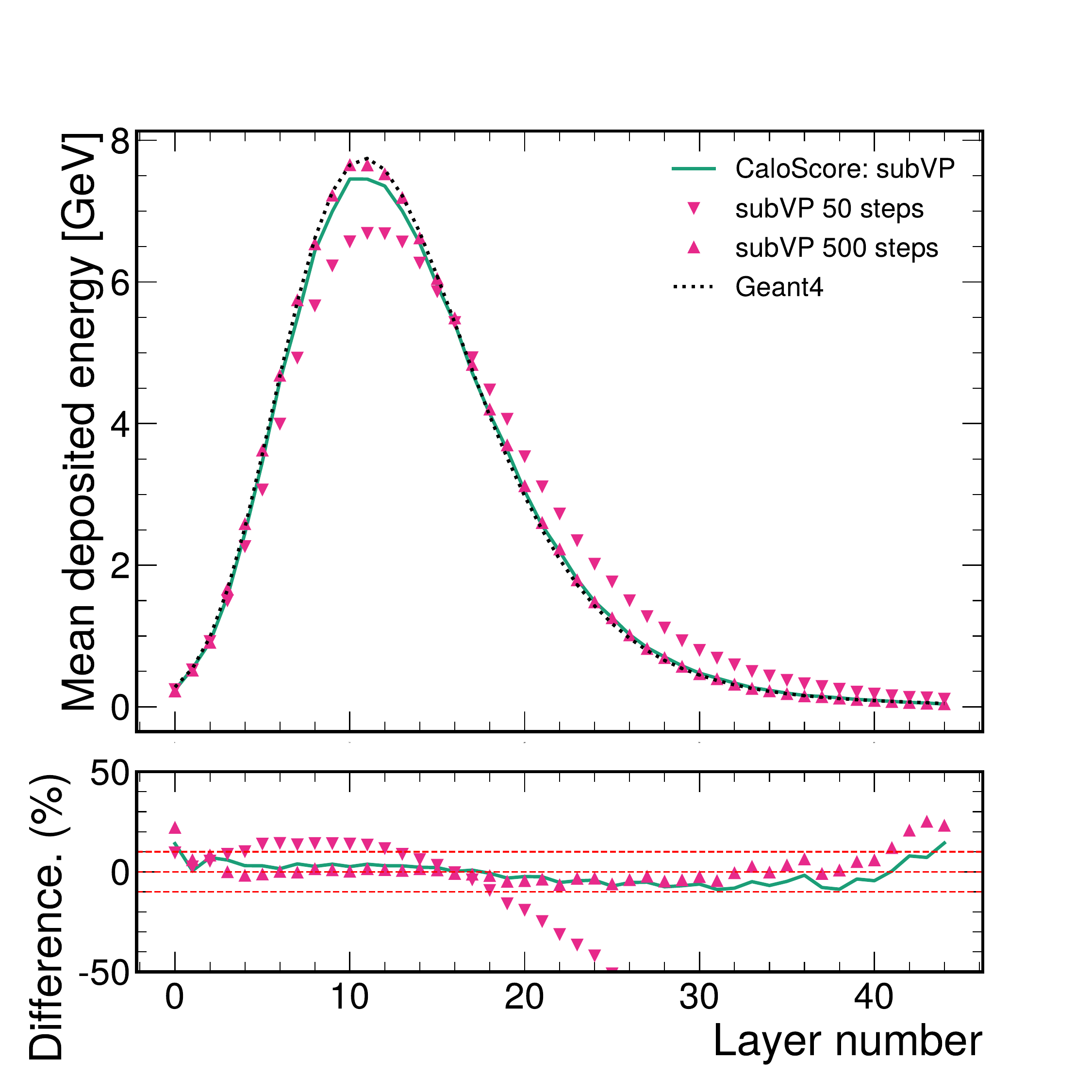}
\includegraphics[width=0.3\textwidth]{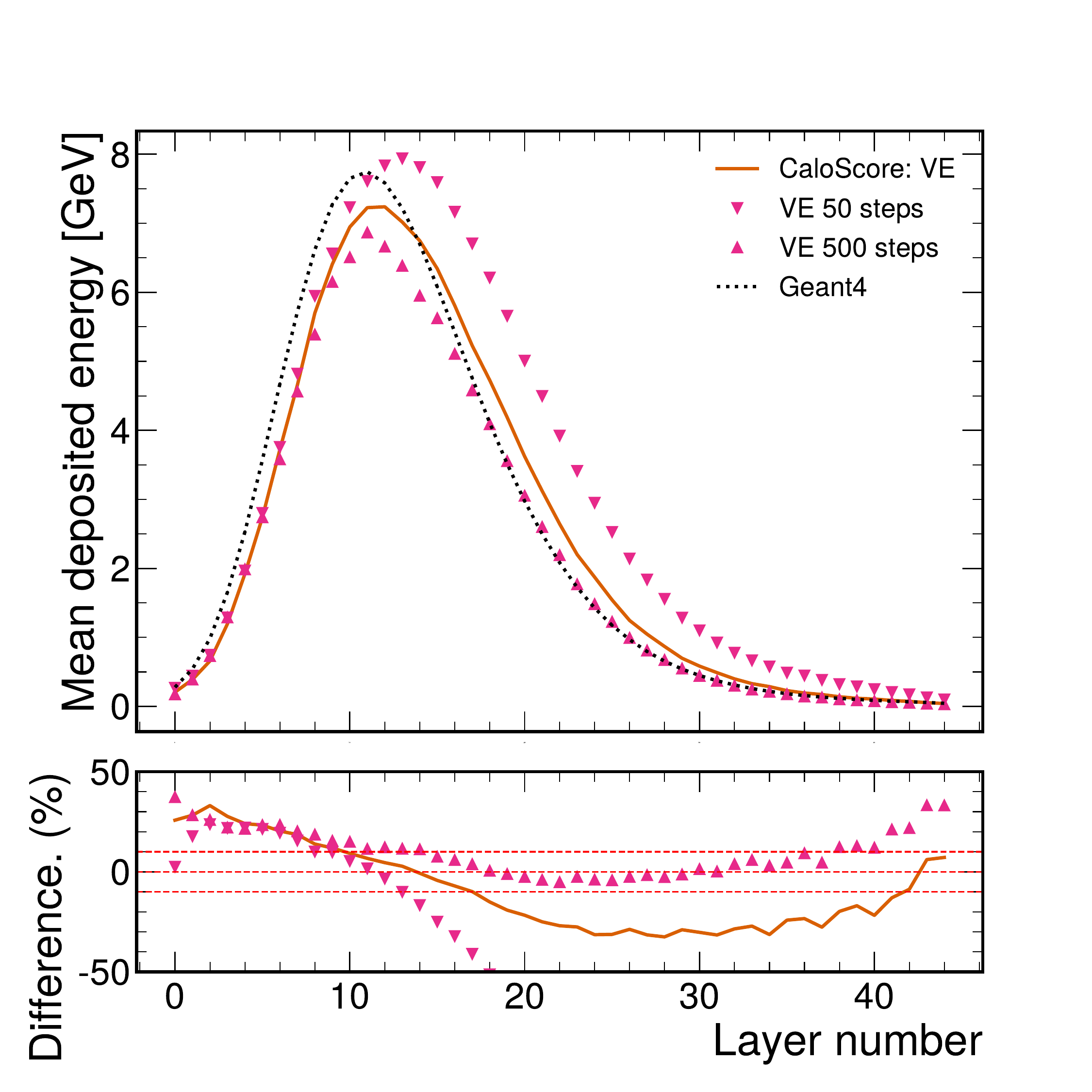}

\caption{Comparison of different number of time steps on the maximum energy fraction deposited in a single voxel for dataset 1 (top) and average energy deposition per layer for datasets 2 (middle) and 3 (bottom) for the three different diffusion models: VP (left), subVP(middle), and VE (right).}
\label{fig:app:nsteps}
\end{figure*}

Contrary to the corrector step, increasing the number of time steps directly affect the generation time, dominated by the number of score function evaluations. Different choices of number of times steps are shown in Fig.~\ref{fig:app:nsteps} with $r$ value fixed to the baseline value.

In all cases using fewer time steps deteriorate the agreement with \geant~while additional steps are able to improve the agreement in both datasets 1 and 3 while dataset 2 shows similar results compared to the baseline. In Tab.~\ref{tab:app:gen}, the time required to generate 100 calorimeter showers using different number of time steps is listed. Even though the additional time steps improve the simulation quality, the time to generate the same amount of new observations increase by more than a factor 2.

\begin{table}[ht]
    \centering
	\small
    \caption{Time comparison to generate 100 calorimeter showers using the baseline model and different number of time steps}
    \label{tab:app:gen}
	\begin{tabular}{lccccccc}
        Dataset & Baseline [s] & N=50 [s] & N=500 [s]\\
        \hline
        dataset 1 & 4.0 & 2.9 & 14.8\\
        dataset 2 & 5.8 & 2.7 & 13.1\\
        dataset 3 & 33.4 & 10.3 & 80.2\\
	\end{tabular}
\end{table}

\end{document}